\documentclass[10pt,a4paper]{article}

\usepackage{graphicx}
\usepackage{subfigure}
\usepackage{latexsym}
\usepackage{amsmath}
\usepackage{amssymb}
\usepackage{theorem}
\usepackage[section]{algorithm}
\usepackage{algorithmic}
\usepackage{enumerate}
\usepackage{comment}
\usepackage{multirow}
\usepackage{xcolor}

\newcommand{\argmax}[1]{\underset{#1}{\operatorname{arg}\operatorname{max}}\;}
\newcommand{\argmin}[1]{\underset{#1}{\operatorname{arg}\operatorname{min}}\;}

\title{Quantitative recovery conditions for tree-based compressed sensing}
\author{Coralia Cartis\footnote{Mathematical Institute, University of Oxford, Oxford, UK. E-mail: \texttt{cartis@maths.ox.ac.uk}.}\quad and Andrew Thompson\footnote{Mathematical Institute, University of Oxford, Oxford, UK. E-mail: \texttt{thompson@maths.ox.ac.uk}.}}

\newcommand{\e}{\epsilon}
\newcommand{\al}{\alpha}
\newcommand{\au}{\underline{\alpha}}
\newcommand{\sg}{\sigma}
\newcommand{\gm}{\gamma}
\newcommand{\lm}{\lambda}

\newcommand{\rr}{\rho}
\newcommand{\dd}{\delta}
\newcommand{\xh}{\hat x}
\newcommand{\xb}{\bar x}

\newcommand{\RR}{\mathbb R }
\newcommand{\PP}{\mathbb P }
\newcommand{\EE}{\mathrm{I\!E\!}}
\newcommand{\Gm}{\Gamma}
\newcommand{\Lm}{\Lambda}
\newcommand{\sm}{\setminus}

\newcommand{\TIL}{\mathcal{TIL}}
\newcommand{\TIU}{\mathcal{TIU}}
\newcommand{\LL}{\mathcal{L}}
\newcommand{\FF}{\mathcal{F}}
\newcommand{\UU}{\mathcal{U}}
\newcommand{\TU}{\mathcal{TU}}
\newcommand{\TL}{\mathcal{TL}}
\newcommand{\TIF}{\mathcal{TIF}}
\newcommand{\astar}{a^{\ast}(\rr;\zeta)}
\newcommand{\ra}{\rightarrow}

\newcommand{\rh}{\hat{\rho}}

\newcommand{\Tk}{\mathcal{T}_k}
\newcommand{\Ts}{\mathcal{T}_s}

\newcommand{\bN}{\overline{NSP_{\al}}}
\newcommand{\bNN}{\overline{NSP_{\au(\dd,\rr;\zeta)}}}
\newcommand{\Sg}{\Sigma}

\newcommand{\mP}{\mathcal{P}}
\newcommand{\mH}{\mathcal{H}}
\newcommand{\mN}{\mathcal{N}}

\newcommand{\xs}{x^{\ast}}
\newcommand{\ys}{y^{\ast}}
\newcommand{\ao}{\overline{\al}}
\newcommand{\astart}{a^{\ast}(\rr;\zeta)}
\newcommand{\Ta}{\Theta^1_n}
\newcommand{\Tb}{\Theta^2_n}
\newcommand{\ha}{\hat{\al}}
\newcommand{\mZ}{\mathcal{Z}}

\newtheorem{ass}{Assumption}
\newtheorem{thm}{Theorem}[section]
\newtheorem{cor}[thm]{Corollary}
\newtheorem{lem}[thm]{Lemma}
\newtheorem{defn}[thm]{Definition}
\newtheorem{prob}{Problem}

\begin{document}

\maketitle

\begin{abstract}

\noindent As shown in~\cite{union_subspaces,modelbased}, signals whose wavelet coefficients exhibit a rooted tree structure can be recovered using specially-adapted compressed sensing algorithms from just $n=\mathcal{O}(k)$ measurements, where $k$ is the sparsity of the signal. Motivated by these results, we introduce a simplified proportional-dimensional asymptotic framework which enables the quantitative evaluation of recovery guarantees for tree-based compressed sensing. In the context of Gaussian matrices, we apply this framework to existing worst-case analysis of the Iterative Tree Projection (ITP) algorithm~\cite{union_subspaces,modelbased} which makes use of the tree-based Restricted Isometry Property (RIP). Within the same framework, we then obtain quantitative results based on a new method of analysis, recently introduced in~\cite{stablepoint}, which considers the fixed points of the algorithm. By exploiting the realistic average-case assumption that the measurements are statistically independent of the signal, we obtain significant quantitative improvements when compared to the tree-based RIP analysis. Our results have a refreshingly simple interpretation, explicitly determining a bound on the number of measurements that are required as a multiple of the sparsity. For example we prove that exact recovery of binary tree-based signals from noiseless Gaussian measurements is asymptotically guaranteed for ITP with constant stepsize provided $n\geq 50k$. All our results extend to the more realistic case in which measurements are corrupted by noise.

\end{abstract}

\section{Introduction}\label{intro}

Compressed sensing is motivated by the observation that many signals have an approximately sparse representation in some basis. Under this assumption, it has been proven that, to guarantee signal recovery, the sampling rate need only be proportional to the sparsity of the signal's approximation, rather than the signal dimension~\cite{donoho,candes1}. Given an unknown signal $\xs$ of dimension $N$, our aim is to recover $\xs$ from $n<N$ undersampled linear measurements of the form $b=A\xs+e$, where $e$ is sampling noise. Many signals have additional structure that can be exploited in the recovery process, and one such example occurs when a wavelet basis is used to represent the signal. Wavelet representations are now widely used in a variety of signal processing contexts, most notably image processing, due to the fact that piecewise smooth signals have sparse representations in wavelet bases~\cite{mallat}. Wavelet representations have a multi-scale tree structure, in which signals are decomposed from coarse to fine scales, with the nested support properties of wavelets inducing a parent/child relationship between wavelet coefficients at different scales. One-dimensional wavelets, for example, have a binary tree structure, in which almost all coefficients have precisely two children. Section~\ref{prelim} gives a precise characterization of the tree structures we consider here.

Since wavelets essentially work as local discontinuity detectors, signal discontinuities give rise to a chain of large coefficients along a single branch~\cite{modelbased}. For this reason, if a particular wavelet coefficient is large, its parent wavelet coefficient is also likely to be large, which means that the large wavelet coefficients of many signals can be modelled as forming a connected subset of the whole tree which is itself a \textit{rooted tree}. This motivates an alternative model of data simplicity: assume that the image is supported on some rooted tree of cardinality $k$, for some sparsity parameter $k$.

Several algorithms have been proposed which approximately perform the Euclidean projection onto the set of vectors supported on a rooted tree of given cardinality~\cite{CSSA,fast_proj,nearly_linear}. Algorithms guaranteed to exactly calculate the projection were proposed in~\cite{exact,tractability}. Consequently, certain iterative projection algorithms for compressed sensing can be adapted to the tree-based setting. One such algorithm proposed in~\cite{modelbased}, and also in~\cite{union_subspaces}, is an adaptation of the well-known Iterative Hard Thresholding (IHT) algorithm~\cite{thresh}, which we choose to call Iterative Tree Projection (ITP). Section~\ref{stepsize} gives precise details on the ITP algorithm and associated stepsize variants.

ITP is one of several algorithms that have been proposed for the tree-based compressed sensing problem. Also relying upon tree projection, an adaptation of the CoSaMP algorithm~\cite{cosamp} was proposed in~\cite{modelbased}. Tree-based variants of matching pursuit algorithms were proposed in~\cite{treeMP,treeOMP}. Convex relaxations of the tree-based compressed sensing problem have also been considered~\cite{hidden_markov,clash,bach,discrete_convex,sparse_regression}.

Worst-case recovery guarantees for ITP (with exact tree projection) were obtained in~\cite{union_subspaces,modelbased} in the case of binary trees, by extending the notion of the ubiquitous Restricted Isometry Property~\cite{candes} to the tree-based setting. More recently, worst-case recovery guarantees based on tree-based RIP have been proved for approximate versions of ITP and tree-based CoSaMP in which the tree projections are computed to a given accuracy~\cite{approx_alg}. 

Bounds on tree-based RIP for random matrices with subgaussian entries were obtained in~\cite{union_subspaces,modelbased} in terms of the ratio $k/n$. The bounds imply that it suffices to take only $n=C\cdot k$ measurements to guarantee recovery, for some implicitly quantified constant $C$. The value of the constant $C$ is an issue of crucial importance to practitioners since it essentially determines how many measurements must be taken as a multiple of the signal sparsity. The main contribution of this paper is to determine explicit bounds on the constant $C$ guaranteeing recovery. While our bounds are likely to be pessimistic compared to observed behaviour, they makes clear the extent of current theory in explicit quantitative terms. We obtain results in the context of one particular family of measurement matrices, the Gaussian ensemble, in which each entry of the matrix is i.i.d. Gaussian.

Since a Gaussian matrix is stochastic by nature, it is not possible to obtain deterministic results. However, by exploiting the remarkable concentration of measure properties of Gaussian matrices, it is possible to obtain limiting results as one lets the matrix dimensions grow. In the context of simple sparsity, Donoho introduced a \textit{proportional-dimensional asymptotic framework} as a way of quantifying results for recovery using $l_1$ minimization~\cite{neighborliness}. More precisely, let $(k,n,N)\ra\infty$ such that $n/N\ra\dd\in(0,1]$ and $k/n\ra\rr\in(0,1]$, where $\dd$ is the undersampling ratio and $\rr$ is the oversampling ratio. Following this framework, limiting results were obtained in~\cite{greedy} for three state-of-the-art greedy algorithms including IHT, the algorithm on which ITP is based. These results, which are worst-case in nature, make use of analysis in~\cite{ihtCS} which relies upon the RIP. More recently, by introducing a new method of analysis and by switching to an average-case framework, the present authors obtained improved quantitative results for IHT in~\cite{stablepoint}.

We now describe the main contributions of this paper.

{\bf 1)\,  We introduce a simplified proportional growth asymptotic to enable quantitative comparison of recovery guarantees for tree-based compressed sensing.}\quad The aforementioned results from~\cite{union_subspaces,modelbased} show that tree-based compressed sensing recovery depends only upon the ratio between $n$ and $k$, and is independent of $N$, the ambient signal dimension. This suggests that recovery results may be captured by a simplified proportional-growth asymptotic in which we dispense with the undersampling ratio $\dd$ and consider only the oversampling ratio $\rr$. 

\begin{defn}[\textbf{Simplified proportional-growth asymptotic}]\label{propdimdef2}
We say that a sequence of problem sizes $(k,n,N)$, where $0<k\le n\le N$, obeys the simplified proportional-growth asymptotic if, for some $\rr\in(0,1]$,
$$\frac{k}{n}\ra\rr\;\;\;\;\mbox{as}\;(k,n,N)\ra\infty.\footnote{Note that the only restriction that the simplified proportional-growth asymptotic places upon $N$ is that we must have $N\ra\infty$ such that $N\geq n$.}$$
\end{defn}

While the commmon two-variable asymptotic framework leads to recovery phase transitions in the $(\dd,\rr)$-plane, our recovery conditions take the refreshingly simple form of a threshold $\hat{\rr}$, such that stable recovery is asymptotically guaranteed provided the oversampling ratio satisfies $\rr<\hat{\rr}$. The framework allows a direct comparison of recovery conditions for different tree-based recovery algorithms, and for different methods of analysis. 

A possible objection to our claim that our results are of practical relevance is that they are asymptotic in nature. However, our recovery results take the form of asymptotic bounds which hold for a sequence of increasing problem sizes, except with probability which decays exponentially in the problem dimension. We therefore believe that it is reasonable to expect recovery behaviour in practice to rapidly approach asymptotic limits, or be even better (since our asymptotic bounds are likely to be pessimistic).

{\bf 2)\,  We obtain explicit quantitative recovery guarantees for ITP algorithms with Gaussian measurement matrices in this simplified asymptotic framework.} Our results are based upon a translation of the RIP analysis in~\cite{HTP} to the tree-based setting, and require the derivation of upper bounds on tree-based RIP constants for Gaussian matrices in the simplified proportional-growth asymptotic. We tighten the implicit bounds on tree-based RIP from~\cite{union_subspaces} (see discussion in Section 3.3). We quantify oversampling thresholds for ITP and Gaussian matrices, the precise recovery values being dependent on the ITP stepsize scheme variant used (see Section~\ref{stepsize}). In the case of zero noise, we have exact recovery of the original signal. In the case of noise, we derive \textit{stability factors} which bound the approximation error of the output of ITP as a multiple of the noise level. The analysis in the present paper broadly follows the approach used to analyze IHT in~\cite{thesis}, and deviates from it by tightening union bound arguments by exploiting the fact that only certain support sets (those corresponding to rooted trees) are permissible in the tree-based model. We compare our quantification with that obtainable from the existing analysis in~\cite{union_subspaces,modelbased} for binary trees, demonstrating a dramatic improvement in the value of the constant.

{\bf 3)\,  We obtain improved quantitative recovery guarantees for ITP algorithms by exploiting average-case assumptions.}\quad We obtain results in the same framework based upon a translation of the \textit{stable point} approach recently introduced by the present authors in~\cite{stablepoint} to the tree-based setting. Whereas the RIP is entirely worst-case, this alternative approach is more amenable to probabilistic analysis under the average-case (but realistic) assumption that the original signal and measurement matrix are statistically independent. Just as for the RIP analysis, the extension of the results in~\cite{stablepoint} involves the tightening of union bound arguments.  The stable point condition is especially amenable to probabilistic analysis for Gaussian matrices under the average-case (but realistic) assumption that  Central to the analysis are large deviations results for quantities related to Gaussian matrices, which are used to bound the constituent terms of the stable point condition, employing union bounds over all permissible support sets. We obtain oversampling thresholds for the same stepsize schemes, enabling a quantitative comparison with those derived from tree-based RIP analysis. For both stepsize schemes, the incorporation of average-case assumptions leads to a significant quantitative improvement in recovery guarantees for ITP and Gaussian matrices. We also extend our stable point recovery analysis to the case of noisy measurements, obtaining stability factors that show a substantial quantitative improvement over those derived from tree-based RIP analysis.

{\bf Outline of the paper.}\quad The rest of the paper is structured as follows: In Section~\ref{prob_alg}, we give full technical details of the tree-based compressed sensing problem, describe in more detail the generic ITP algorithm along with two possible stepsize schemes, and give a brief roadmap to the proofs. We describe our main results in Sections~\ref{statement1} and~\ref{statement2}, first for those derived from tree-based RIP analysis (Section~\ref{statement1}), followed by the results derived from our stable point analysis (Section~\ref{statement2}. A discussion of all our main results then follows in Section~\ref{discussion}. All proofs can be found in the appendix. We present the tree-based RIP analysis in Appendix~\ref{RIP_proofs}, and the stable point analysis in Appendix~\ref{stable_proofs}. Both analyses rely crucially upon large deviations results for quantities related to Gaussian matrices (including bounds on tree-based RIP constants), and proofs of these subsidiary results can be found in Appendix~\ref{large_dev}.

\section{Problems and algorithms}\label{prob_alg}

\subsection{Problem statement}\label{prelim}

Suppose we have a signal $\ys\in\RR^N$ which has a sparse rooted-tree representation $\xs\in\RR^N$ in some orthogonal wavelet basis, so that $\xs=\Psi\ys$ where $\Psi\in\RR^{N\times N}$ is an orthogonal discrete wavelet transform matrix. We obtain the measurements $b=\Phi\ys+e\in\RR^n$, where $\Phi\in\RR^{n\times N}$, where $e$ is sampling noise, and where we assume $n<N$. Referring to $A=\Phi\Psi^{-1}\in\RR^{n\times N}$ from now on as the measurement matrix, we have 
\begin{equation}\label{measurements}
b=A\xs+e.
\end{equation}

We say that a vector $\xs$ is $k$\textit{-tree sparse} if it is supported on a rooted tree of cardinality $k$, and denote by $\Tk$ the set of supports permitted by the model. Denoting by $\|\cdot\|$ the Euclidean norm $\|\cdot\|_2$, and defining
\begin{equation}\label{psidef}
\Psi(x):=\frac{1}{2}\|b-Ax\|^2,
\end{equation}
we can formulate signal recovery as the following optimization problem. 
\begin{equation}\label{treeprog}
\displaystyle\min_{x\in\RR^N}\Psi(x)\;\;\;\;\;\mbox{subject to}\;\;\;\mbox{supp}(x)\in\Tk,
\end{equation}
where $\mbox{supp}(x)$ denotes the support of the signal $x$. We write $\mathcal{P}_k$ for the (exact) Euclidean projection onto the set $\{x:\mbox{supp}(x)\in\Tk\}$, namely
\begin{equation}\label{Pk_proj}
\mP_k(z):=\argmin{\mathrm{supp}(x)\in\Tk}{\|x-z\|}.
\end{equation}

Our analysis will consider arbitrary tree structures, characterized only by the existence of a root coefficient (that is, a coefficient with no parents) a \textit{tree order} $d$, defined to be the maximum number of children of any coefficient in the tree. We will at times refer to a tree of order $d$ as a $d$-ary tree. The coefficients of one-dimensional wavelet transforms typically have a binary tree structure, that is tree order $d=2$. The two-dimensional wavelet transforms often used in image processing typically form quad-trees ($d=4$). Orthogonal discrete wavelet transforms often have a particular canonical tree structure, in which every coefficient essentially has the same number of children, but this condition is never enforced in our analysis.

Our challenge, then, is to recover the wavelet representation $\xs$ (and therefore the original signal $\ys$) from the measurements (\ref{measurements}), which we formally state as the following two problems.

\begin{prob}[\textbf{Tree-sparse recovery from exact measurements}]\label{treesparseprob}
Recover exactly a $k$-tree sparse $\xs\in\RR^N$ from the noiseless measurements $b=A\xs\in\RR^n$, where $k\le n\le N$.
\end{prob}

\begin{prob}[\textbf{Tree-sparse recovery from noisy measurements}]\label{treenoiseprob}
Recover a $k$-tree sparse $\xs\in\RR^N$ from the noisy measurements $b=A\xs+e\in\RR^n$, where $k\le n\le N$.
\end{prob}

We consider the case where $\Phi$ is chosen to be a Gaussian matrix with entries distributed i.i.d. as $\{\Phi_{ij}\}\sim\mN(0,1/n)$. The orthogonality assumption on the wavelet transform $\Psi$ then implies that the entries of $A$ are also distributed i.i.d. as $\{A_{ij}\}\sim\mN(0,1/n)$, i.e. $A$ is also i.i.d. Gaussian. Assuming $\Phi$ to be Gaussian is therefore equivalent to placing the same assumption on $A$, which we formalize as follows.

\begin{ass}\label{gaussianass}
The measurement matrix $A$ has i.i.d. $\mathcal{N}(0,1/n)$ entries. 
\end{ass}

It can be shown that $\xs$ is the unique global solution to problem (\ref{treeprog}) whenever $A$ is a Gaussian matrix~\cite[Sections 3 and 4.1]{stablepoint}.\\
\\
{\bf Notation.} Given some index set $\Gm\subseteq\{1,2,\ldots N\}$, we define the complement of $\Gm$ to be $\Gm^C=\{1,2,\ldots
N\}\setminus\Gm$. We write $x_{\Gm}$ for the restriction of the vector $x$ to the coefficients indexed by the elements of $\Gm$, and we write $A_{\Gm}$ for the restriction of the matrix $A$ to those columns indexed by the elements of $\Gm$.

\subsection{ITP algorithms and stepsize schemes}\label{stepsize}

In this section, we describe in more detail the ITP algorithm along with two possible stepsize schemes. Generically, on each iteration $m$, a steepest descent step, possibly with linesearch, is calculated
for the objective $\Psi$ in (\ref{treeprog}), namely, a move is performed from the current iterate
$x^m$ along the negative gradient
of $\Psi$,
$$-\nabla \Psi(x^m)=-A^T(Ax^m-b).$$
Recalling the definition of $\mP_k$ from Section~\ref{prelim}, the resulting step is then projected onto the (nonconvex) constraint in (\ref{treeprog}) which defines the set of all vectors supported on rooted trees of cardinality $k$.

\begin{algorithm}\caption{\textbf{Generic ITP~\cite{union_subspaces,modelbased}}}
Inputs: $A,b,k$.\\
Initialize $x^0=0;\;m=0$.\\
While some termination criterion is not satisfied, do:
\begin{enumerate}
\item $x^{m+1}:=\mathcal{P}_k\left\{x^m+\al^m A^T(b-Ax^m)\right\}$, where $\mP_k(\cdot)$ is defined in (\ref{Pk_proj}) and $\al^m>0$ is a stepsize.
\item $m:=m+1$
\end{enumerate}
End; output $\xh=x^m$.
\end{algorithm}

To avoid a situation in which the support set $\Gamma$ is not uniquely defined, if for instance some of the coefficients are equal in magnitude, then a support set for the identical components can be selected either randomly or according to some predefined ordering. In our analysis, we will consider the possibly infinite sequence of iterates generated by ITP, though in practice a useful termination criterion such as requiring the residual to be sufficiently small, would need to be employed. 

Two stepsize choices will be addressed in this paper: {\it constant stepsize} $\alpha^m=\alpha \in (0,1)$ for all $m$, which we will hereafter refer to simply as ITP \cite{union_subspaces,modelbased}, and a {\it variable stepsize} scheme which we will call Normalised ITP (NITP), which adopts the same stepsize scheme as prescribed in the Normalised IHT variant of IHT algorithms proposed in~\cite{normalized}. The constant stepsize ITP variant can be summarized as follows.

\begin{algorithm}\label{constantITP}\caption{\textbf{ITP~\cite{union_subspaces,modelbased}}}
Given some $\alpha>0$, on \textbf{step 1} of each iteration $m\geq 0$ of generic ITP, set
\begin{equation}\label{ITPstep}
\al^m:=\al.
\end{equation}
\end{algorithm}

The NITP variant defined below follows \cite{normalized}, having the stepsize $\al^m$ chosen according to an {\it exact linesearch} \cite{nocedal_wright} when the support set of consecutive iterates stays the same, and using a {\it shrinkage} strategy when the support set changes, in order to ensure sufficient decrease in the objective of (\ref{treeprog}).

\begin{algorithm}\label{normalisedITP}\caption{\textbf{NITP}}
Given some $c\in (0,1)$ and $\kappa>1/(1-c)$, on \textbf{step 1} of each iteration $m\geq 0$ of generic ITP, do:
\begin{description}
\item[1.1.] \textbf{Exact linesearch.} 
\begin{enumerate}[(a)]
\item Set $\Gm^m:=\mbox{supp}(x^m)$. 
\item Compute 
\begin{equation}\label{exactchoice_tree}
\al^m:=\displaystyle\frac{\|A_{\Gm^m}^T(b-Ax^m)\|^2}{\|A_{\Gm^m}A^T_{\Gm^m}(b-Ax^m)\|^2}.
\end{equation}
\item Let $\tilde{x}^{m+1}:=\mathcal{P}_k\left\{x^m+\al^m A^T(b-Ax^m)\right\}$.
\end{enumerate}
\item[1.2.] \textbf{Backtracking.} If $\mbox{supp}(\tilde{x}^{m+1})=\mbox{supp}(x^m)$, end; output $\alpha^m$.\\
Else, while
  $\al^m\geq(1-c)\frac{\|\tilde{x}^{m+1}-x^m\|^2}{\|A(\tilde{x}^{m+1}-x^m)\|^2}$,
  do:
\begin{enumerate}[(a)]
\item $\al^m:=\al^m/(\kappa(1-c))$.
\item $\tilde{x}^{m+1}:=\mathcal{P}_k\left\{x^m+\al^m A^T(b-Ax^m)\right\}$.
\end{enumerate}
End; output $\alpha^m$.
\end{description}
\end{algorithm}

In practice, the choice of $\kappa$ constitutes a trade-off between recovery performance and computational efficiency: for optimal performance, $\kappa$ close to $1$ should be chosen, while increasing $\kappa$ will lead to fewer shrinkage steps, making the algorithm more computationally efficient. The shrinkage strategy ensures a potentially desirable property of the NITP algorithm, namely that, provided the measurement matrix satisfies mild linear independence assumptions, it is guaranteed to converge (see Section~\ref{conv}). A practical scheme similar to the one in~\cite{normalized} was proposed in~\cite{recipes} which does not employ a shrinkage strategy.

An important property of the operator $\mP_k$ is that it preserves the value of selected coefficients.
\begin{equation}\label{treeproj_def2}
\left\{\mP_k(x)\right\}_i:=\left\{\begin{array}{ll}
x_i&i\in\Gm\\
0&i\notin\Gm
\end{array}\right.\;\;\;\;
\mbox{where}\;\;\;\;\Gm:=\argmax{\Gm\in\Tk}\left\|x_{\Gm}\right\|.
\end{equation}
See \cite[Lemma 6.1]{thesis} for a proof of (\ref{treeproj_def2}) given its definition. It follows from (\ref{treeproj_def2}) that $\mP_k$ can be framed as an integer program with $\{0,1\}$ decision variables. This problem can either be solved exactly using dynamic programming~\cite{exact} or approximately by solving its linear programming or Lagrangian relaxations~\cite{CSSA,cartbob}. We refer the reader to~\cite{exact} for further details on methods for performing the projection onto rooted trees. 

\section{Recovery results for tree-based RIP analysis}\label{statement1}

\subsection{Results for deterministic matrices}\label{det_results}

Our first analysis relies upon a deterministic recovery condition originally given in \cite{HTP}. Our contribution is to extend it to the tree-based setting and then obtain from it quantitative results for Gaussian matrices. We consider an extension of the ubiquitous (asymmetric) Restricted Isometry Property (RIP)~\cite{candes,lqphase} to the tree-based setting.

\begin{defn}[\textbf{Tree-based RIP~\cite{union_subspaces,modelbased}}]\label{treeRIPdef} 
For a given matrix $A$, define $TL_s$ and $TU_s$, the lower and upper tree-based RIP constants of order $s$, to be, respectively,
\begin{equation}\label{treeRIPeqn}
TL_s:=1-\displaystyle\min_{\emptyset\neq\mbox{supp}(y)\subseteq\Gm\in\Ts}\displaystyle\frac{\|Ay\|^2}{\|y\|^2}\;\;\;\;\mbox{and}\;\;\;\;TU_s:=\displaystyle\max_{\emptyset\neq\mbox{supp}(y)\subseteq\Gm\in\Ts}\displaystyle\frac{\|Ay\|^2}{\|y\|^2}-1.
\end{equation}
\end{defn}

We obtain deterministic recovery results of the following form for both ITP and NITP.

\begin{thm}[\textbf{Deterministic recovery result for ITP variants}]\label{ALG_result}
Consider Problem~\ref{treenoiseprob}. Let $\mu^{ALG}$ and $\xi^{ALG}$ be defined as in Definition~\ref{muxiITP_det}. Then, there exists functions $\mu^{ALG}$ and $\xi^{ALG}$ such that, provided $\mu^{ALG}<1$, the output, $\hat{x}$, at iteration $m$ of variant ALG of ITP satisfies
\begin{equation}\label{ALGerror}
\|\hat{x}-\xs\|\le\left(\mu^{ALG}\right)^m\|\xs\|+\frac{\xi^{ALG}}{1-\mu^{ALG}}\|e\|.
\end{equation}
\end{thm}

\textbf{Proof:} See Appendix~\ref{RIP_proofs}.\hfill$\Box$\\
\\
The functions $\mu^{ALG}$ and $\xi^{ALG}$ will play the role of a convergence factor and a factor controlling stability to noise. Though Theorem~\ref{ALG_result} gives a limiting bound on the approximation error, it does not necessarily imply convergence of the algorithm. In the simplified noiseless case however, the result implies convergence to $\xs$ at a linear rate.

Specifically, Theorem~\ref{ALG_result} holds for ITP with stepsize $\alpha$ if $\mu^{ALG}:=\mu^{ITP_{\al}}$ and $\xi^{ALG}:=\xi^{ITP_{\al}}$, while Theorem~\ref{ALG_result} holds for NITP with shrinkage parameter $\kappa$ if $\mu^{ALG}:=\mu^{NITP_{\kappa}}$ and $\xi^{ALG}:=\xi^{NITP_{\kappa}}$, defined as follows.

\begin{defn}[\textbf{Deterministic convergence and stability factor for ITP}]\label{muxiITP_det}
Provided $3k\le n$, define
\begin{equation}\label{CIHTmudef}
\mu^{ITP_{\al}}:=\sqrt{3}\max\{\al(1+TU_{3k})-1,1-\al(1-TL_{3k})\},
\end{equation}
\begin{equation}\label{CIHTxidef}
\xi^{ITP_{\al}}:=\al\sqrt{3(1+TU_{2k})},
\end{equation}
\begin{equation}\label{NIHTmudef}
\mu^{NITP_{\kappa}}:=\sqrt{3}\max\left\{\frac{1+TU_{3k}}{1-TL_k}-1,1-\frac{1-TL_{3k}}{\kappa[1+TU_{2k}]}\right\},
\end{equation}
\begin{equation}\label{NIHTxidef}
\xi^{NITP_{\kappa}}:=\frac{\sqrt{3(1+TU_{2k})}}{1-TL_k},
\end{equation}
where $TU$ and $TL$ are defined in Definition~\ref{treeRIPdef}.
\end{defn} 

\subsection{Asymptotic results for Gaussian matrices}\label{asympt_results} 

We derive quantitative recovery conditions for Gaussian matrices by means of upper bounds on tree-based RIP constants in the simplified proportional-growth asymptotic of Definition~\ref{propdimdef2}. We follow the broad approach used for the standard notion of RIP in \cite{lqphase,support_sizes,stablepoint}, in which a union bound was performed over the maximum/minimum singular values of all $\binom{N}{k}$ submatrices of $A$ of size $n\times k$. In the present work, however, the assumed tree structure means that the number of permissible support sets for iterates of the algorithm is much diminished, which means that union bound arguments can be tightened, leading to improved quantitative results. 
 
The number, $|\Tk|$, of permissible support sets in the $d$-ary tree-based framework, is bounded above by $T(k)$, the total number of ordered, rooted $d$-ary trees of cardinality $k$. Fortunately, a formula for $T(k)$ is known.

\begin{lem}[\textbf{Tree counting result~\cite{concrete}}]\label{tree_count}
The total number of ordered, rooted $d$-ary trees of cardinality $k$ is
\begin{equation}\label{treecount}
T(k)=\frac{1}{(d-1)k+1}\binom{dk}{k}.
\end{equation}
\end{lem}

In particular, note that $T(k)$ depends only upon the tree order $d$ and the sparsity $k$, and not upon the signal length $N$. It is for this reason that we are able to obtain quantitative bounds in the simplified proportional-growth asymptotic, i.e. in terms of $d$ and the variable $\rr:=\lim_{n\ra\infty}\frac{k}{n}$ only.

Before defining bounds, it will be useful to define the Shannon entropy in the usual way.

\begin{defn}[\textbf{Shannon entropy~\cite{lqphase}}]
Given $p\in(0,1)$, define the Shannon entropy with base $e$ logarithms as 
\begin{equation}\label{shannon_def}
H(p):=-p\ln(p)-(1-p)\ln(1-p).
\end{equation}
\end{defn}

We define the following bounds on tree-based RIP constants for Gaussian matrices.

\begin{defn}[\textbf{Tree-based RIP bounds}]\label{RIPboundsdef}
Define, for $\rr\in(0,1)$ and $\lambda>0$,
\begin{equation}\label{psimaxdef}
\psi_{max}(\lm,\rr)=\frac{1}{2}\left[(1+\rr)\ln\lm+1+\rr-\rr\ln\rr-\lm\right]
\end{equation}
and
\begin{equation}\label{psimindef}
\psi_{min}(\lm,\rr)=H(\rr)+\frac{1}{2}\left[(1-\rr)\ln\lm+1-\rr+\rr\ln\rr-\lm\right],
\end{equation}
where $H(\cdot)$ is defined in (\ref{shannon_def}). Define $\lm^{max}(\rr)$ and $\lm^{min}(\rr)$ as the unique solution to (\ref{lambdamaxdef}) and (\ref{lambdamindef}) respectively:
\begin{equation}\label{lambdamaxdef}
\psi_{max}\left(\lm^{max}(\rr),\rr\right)+d\rr\cdot H(d^{-1})=0\;\;\;\;\mbox{for}\;\;\;\;\lm^{max}(\rr)>1+\rr;
\end{equation}
\begin{equation}\label{lambdamindef}
\psi_{min}\left(\lm^{min}(\rr),\rr\right)+d\rr\cdot H(d^{-1})=0\;\;\;\;\mbox{for}\;\;\;\;\lm^{min}(\rr)<1-\rr,
\end{equation}
and define $\TU(\rr)=\lm^{max}(\rr)-1$ and $\TL(\rr)=1-\lm^{min}(\rr)$.
\end{defn}

That there exists a unique solution to (\ref{lambdamaxdef}) follows since $\psi_{max}[\lm,\rr]$ is positive for $\lm=1+\rr$, tends to $-\infty$ as $\lm\ra\infty$, and is strictly decreasing in $\lm$. Similarly, that there exists a unique solution to (\ref{lambdamindef}) follows since $\psi_{min}[\lm,\rr]$ is positive for $\lm=1-\rr$, tends to $-\infty$ as $\lm\ra\infty$, and is strictly decreasing in $\lm$.

Intuition behind the form of the bounds given in Definition~\ref{RIPboundsdef} is as follows. For a given $n\times k$ submatrix $A_{\Gamma}$, the asymptotic distributions of $\lambda^{max}$ and $\lambda^{min}$, the extreme eigenvalues of its corresponding Gram matrix $A_{\Gamma}^T A_{\Gamma}$, depend asymptotically upon $\rho$, and both decay exponentially away from $1$, with exponents given by $\gamma_{max}(\lambda^{max}(\rho),\rr)$ and $\gamma_{min}(\lambda^{min}(\rho),\rho)$ respectively. To bound the extreme eigenvalues of all possible such Gram matrices requires a union bound over the number of permissible support sets. For the standard notion of RIP analyzed in~\cite{lqphase}, all $\binom{N}{k}$ support sets must be considered, which leads to an exponent which depends upon $\rho$ and also $\delta:=\lim_{n\ra\infty}n/N$. In the tree-based setting, however, the number of permissible support sets is given by (\ref{treecount}), which has no dependence upon the ambient dimension $N$, and the resulting exponent  $d\rr\cdot H(d^{-1})$ depends only upon $\rho$ (for a given tree order $d$). The asymptotic bounds $\TU(\rho)$ and $\TL(\rho)$ are defined in such a way that they are satisfied in the asymptotic limit when the net exponents in (\ref{lambdamaxdef}) and (\ref{lambdamindef}) respectively are negative.

Counterparts of the bounds in Definition~\ref{RIPboundsdef} for the standard notion of asymmetric RIP constants were shown to hold asymptotically for Gaussian matrices in~\cite{lqphase}. Following their method of proof, we obtain an analogous result for tree-based RIP constants in the simplified proportional-growth asymptotic. 

\begin{lem}[\textbf{Validity of tree-based RIP bounds}]\label{RIPbounds}
Suppose Assumption~\ref{gaussianass} holds and let $\e>0$. In the simplified proportional-growth asymptotic,
\begin{equation}\label{TU_result}
\PP\left(TU_k\geq\TU(\rr)+\e\right)\ra 0,
\end{equation}
\begin{equation}\label{TL_result}
\PP\left(TL_k\le\TL(\rr)-\e\right)\ra 0,
\end{equation}
both exponentially in $n$.
\end{lem}

\textbf{Proof:} See Appendix~\ref{large_dev}.\\
\\
Closely following the approach in~\cite{greedy}, we show that a naive replacement of each $TL_{pk}$ and $TU_{qk}$ by the tree-based RIP bounds $\TL(p\rr)$ and $\TU(q\rr)$ is valid, provided the functions $\mu^{ITP_{\al}}_{RIP}$ and $\xi^{ITP_{\al}}_{RIP}$ satisfy certain properties given in Appendix~\ref{RIP_gauss}. We finally arrive at asymptotic recovery results of the following form for both variants of ITP and Gaussian matrices. 

\begin{thm}[\textbf{RIP-based recovery}]\label{ALG_RIPphase}
Consider Problem~\ref{treenoiseprob} and suppose Assumption~\ref{gaussianass} holds. Define $\rh^{ALG}_{RIP}$ as the unique solution to $\mu^{ALG}_{RIP}(\rr)=1$. Choose $\e\in(0,1)$ and suppose that 
\begin{equation}\label{CIHTrhocond}
\rr<(1-\e)\rh^{ALG}_{RIP}.
\end{equation}
Suppose $\hat{x}$ is the output of variant ALG of ITP at iteration $m$. Then
\begin{equation}\label{mdr_IHT}
\mu^{ALG}_{RIP}((1+\e)\rr)<1,
\end{equation}
and, in the simplified proportional-growth asymptotic\footnote{In other words, we consider instances of the Gaussian random variables $A$ for a sequence of triples
$(k,n,N)$ where $n\rightarrow\infty$, where $n$ is the number of measurements, $N$, the signal dimension and $k$, the sparsity of the underlying signal.},
\begin{equation}\label{CIHTdrerror}
\|\hat{x}-\xs\|\le\left(\mu^{ALG}_{RIP}((1+\e)\rr)\right)^m\|\xs\|+\frac{\xi^{ALG}_{RIP}((1+\e)\rr)}{1-\mu^{ALG}_{RIP}((1+\e)\rr)}\|e\|,
\end{equation}
for all $k$-tree sparse vectors $\xs$, with probability tending to $1$ exponentially in $n$.
\end{thm}

\textbf{Proof:} See Appendix~\ref{RIP_proofs}.\hfill$\Box$\\
\\
In the idealized case of zero measurement noise, we can deduce from Theorem~\ref{ALG_RIPphase} guaranteed convergence of ITP variants at a linear rate.

\begin{cor}[\textbf{RIP-based recovery: noiseless case}]\label{ALG_RIPphase_noiseless}
Consider Problem~\ref{treesparseprob} and suppose Assumption~\ref{gaussianass} holds. Choose $\e\in(0,1)$ and suppose that (\ref{CIHTrhocond}) holds, where $\rh^{ALG}_{RIP}$ and $\mu^{ALG}_{RIP}(\rr)$ are defined as in Theorem~\ref{ALG_RIPphase}. Then, in the simplified proportional-growth asymptotic, the iterates of variant ALG of ITP converge to $\xs$ at a linear rate, for all $k$-tree sparse vectors $\xs$, with probability tending to $1$ exponentially in $n$.
\end{cor}

\textbf{Proof:} See Appendix~\ref{RIP_proofs}.\hfill$\Box$\\
\\
Specifically, Theorem~\ref{ALG_RIPphase} and Corollary~\ref{ALG_RIPphase_noiseless} hold for ITP with stepsize $\alpha$ if $\mu^{ALG}_{RIP}(\rho):=\mu^{ITP_{\al}}_{RIP}(\rho)$ and $\xi^{ALG}:=\xi^{ITP_{\al}}_{RIP}(\rho)$, while Theorem~\ref{ALG_RIPphase} and Corollary~\ref{ALG_RIPphase_noiseless} hold for NITP with shrinkage parameter $\kappa$ if $\mu^{ALG}_{RIP}(\rho):=\mu^{NITP_{\kappa}}_{RIP}(\rho)$ and $\xi^{ALG}:=\xi^{NITP_{\kappa}}_{RIP}(\rho)$, defined as follows (compare with Definition~\ref{muxiITP_det}).

\begin{defn}[\textbf{Asymptotic convergence and stability factors}]\label{muxiALG_def}
Define, for $\rr\in(0,1/3)$,
\begin{equation}\label{CIHTmudrdef}
\mu^{ITP_{\al}}_{RIP}(\rr):=\sqrt{3}\max\{\al[1+\TU(3\rr)]-1,1-\al[1-\TL(3\rr)]\},
\end{equation}
\begin{equation}\label{CIHTxidrdef}
\xi^{ITP_{\al}}_{RIP}(\rr):=\al\sqrt{3[1+\TU(2\rr)]},
\end{equation}
\begin{equation}\label{NIHTmudrdef}
\mu^{NITP_{\kappa}}_{RIP}(\rr):=\sqrt{3}\max\left\{\frac{1+\TU(3\rr)}{1-\TL(\rr)}-1,1-\frac{1-\TL(3\rr)}{\kappa[1+\TU(2\rr)]}\right\},
\end{equation}
\begin{equation}\label{NIHTxidrdef}
\xi^{NITP_{\kappa}}_{RIP}(\rr):=\frac{\sqrt{3[1+\TU(2\rr)]}}{1-\TL(\rr)},
\end{equation}
where $\TU$ and $\TL$ are given in Definition~\ref{RIPboundsdef}.
\end{defn} 

In the case of ITP with constant stepsize, Theorem~\ref{ALG_RIPphase} and Corollary~\ref{ALG_RIPphase_noiseless} give a continuous range of oversampling thresholds for any $0<\al<2$. For $\al\geq 2$, the result gives $\rh^{IHT_{\al}}_{RIP}=0$ for all $\dd\in(0,1)$. It is clear that $\mu^{IHT_{\al}}_{RIP}(\rr)$ takes its minimum value when the two expressions inside the maximum in (\ref{CIHTmudrdef}) are equal, which implies that the optimal oversampling threshold is obtained when the stepsize is taken to be 
\begin{equation}\label{optalpha}
\hat{\al}:=2/[2+\TU(3\rr)-\TL(3\rr)].
\end{equation}
We will adopt the optimal stepsize choice $\hat{\al}$ in all our numerical computations of oversampling thresholds\footnote{Note that this optimal stepsize $\hat{\alpha}$ is closely related to the (constant) maximal stepsize in gradient methods 
for strongly convex optimization that ensures global linear rate of convergence (see \cite[Theorem 2.1.15]{yurii}). In particular, the objective (\ref{psidef}) restricted to 
a face of $\mathcal{T}_k$ is a strongly convex objective and ITP is taking a steepest descent step on this face, scaled by $\hat{\alpha}$. 
Then $\mu:=1-\mathcal{TL}(3\rho)$ can be regarded as a lower bound on the smallest eigenvalue of 
the reduced Hessian of the objective (\ref{psidef}) and $L:=1+\mathcal{TU}(3\rho)$ as an upper bound on the largest eigenvalue of the same
matrix. With this correspondence, the constant step sizes prescribed by (\ref{optalpha})  and the maximal one in \cite[Theorem 2.1.15]{yurii} coincide;
see \cite{NIPS2010_3984} for full details and similar analogies.}. 

\subsection{Prior bounds on tree-based RIP}\label{prior}

A result quantifying tree-based RIP for Gaussian matrices was proved in~\cite{union_subspaces} as a special case of a more general result on restricted isometry constants for subgaussian random matrices and signals drawn from a union of linear subspaces. A symmetric notion of tree-based RIP was considered, in which no distinction is made between the upper and lower tails. For a given measurement matrix, the symmetric tree-based RIP constant $TR_k$ is thus
\begin{equation}\label{TR_def}
TR_k:=\max(TL_k,TU_k).
\end{equation}

\begin{figure}[htp]
\centering
\includegraphics[width=0.6\textwidth]{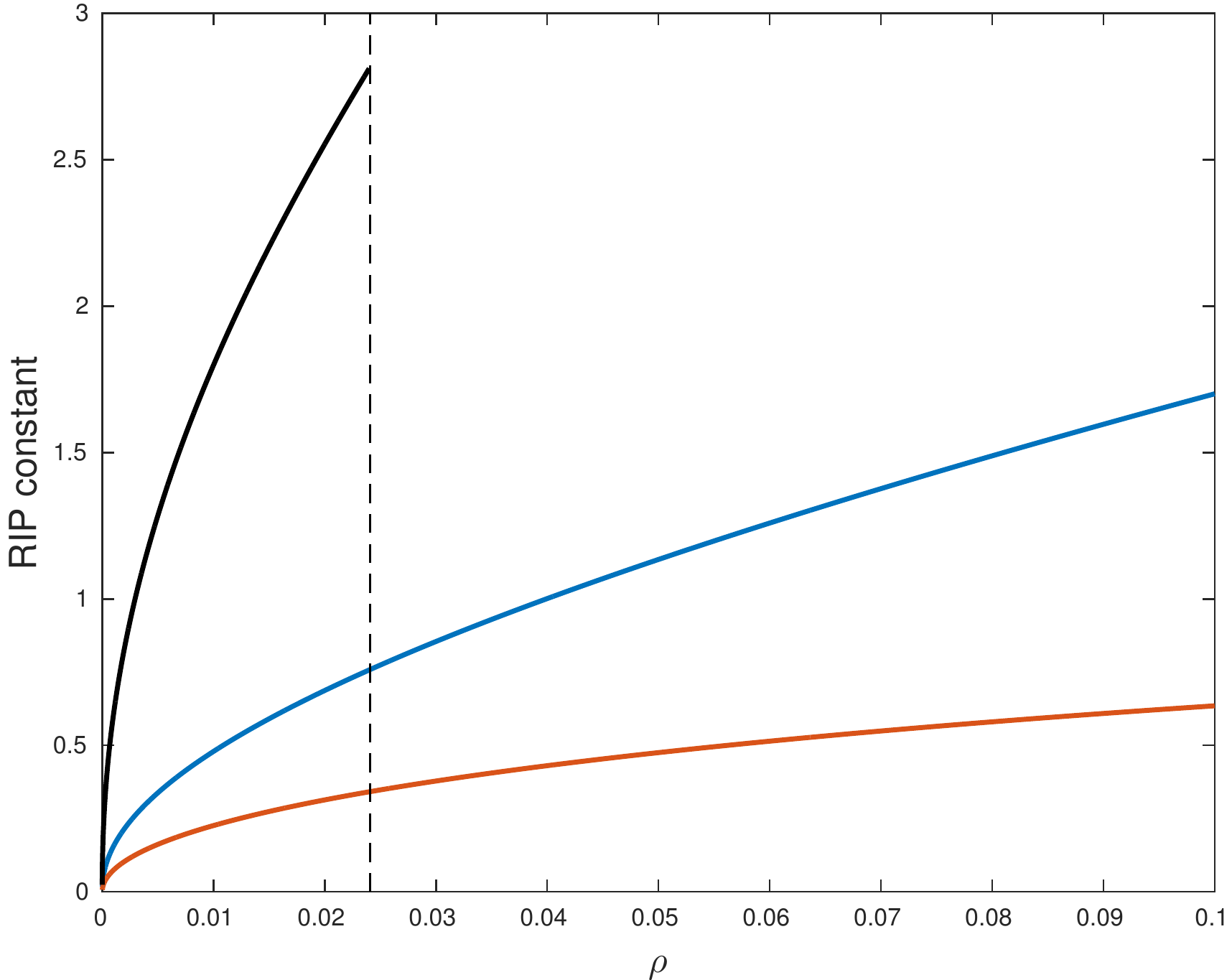}
\caption{A comparison of $\TL(\rho)$ (red), $\TU(\rho)$ (blue) and $\mathcal{TR}(\rho)$ (black) where defined, for $\rho\in(0,0.1)$.}
\label{RIP_compare_plot} 
\end{figure}

\begin{thm}[\textbf{\cite[Corollary 4.2]{union_subspaces}}]\label{union_thm}
Suppose Assumption~\ref{gaussianass} holds and choose $t>0$ and let the tree order be $d=2$. Then, with probability at least $1-e^{-t}$, $TR_k\le r$ provided
$$n\geq\left(\frac{r^2}{144}-\frac{r^3}{1296}\right)^{-1}\left[k\left(1+\ln\frac{72}{r}\right)-\ln\left(\frac{k+1}{2}\right)+t\right]$$
\end{thm}
We may deduce from this result a bound on (symmetrical) tree-based RIP, $\mathcal{TR}(\rho)$, within the simplified proportional-growth asymptotic. The resulting bound is plotted in Figure~\ref{RIP_compare_plot} alongside the bounds $\TU(\rho)$ and $\TL(\rho)$ of Definition~\ref{RIPboundsdef}.
\begin{defn}\label{union_tree_bound}
If $\rho\in(0,0.024)$\footnote{Elementary calculus shows that (\ref{union_tree_eqn}) only has a solution for $\rho$ below approximately $0.02407$ for $d=2$ (binary trees).}, define $\mathcal{TR}(\rho)$ to be the unique solution in $r>0$ to
\begin{equation}\label{union_tree_eqn}
r^2(9-r)=1296\rho\left[1+\ln\left(\frac{72}{r}\right)\right].
\end{equation}
\end{defn}

To make the quantification of recovery results for ITP algorithms explicit, one may combine the bound $\mathcal{TR}(\rho)$ with the symmetric versions of the RIP-based recovery results for each variant. For ITP, the condition $TR_{3k}<1/\sqrt{3}$ was proved in~\cite{HTP}. For NIHT, the condition $TR_{3k}<(11-\sqrt{3})/(11+21\sqrt{3})\approx 0.1956$ follows by combining (\ref{TR_def}) with Theorem~\ref{ALG_result} and Definition~\ref{muxiITP_det}, taking $\kappa:=1.1$. A quantitative comparison in the case of binary trees between the recovery conditions obtainable from this prior analysis and those presented in Sections~\ref{det_results} and~\ref{asympt_results} is given in Section~\ref{RIP_results}.

\section{Recovery results using a tree-based stable point analysis}\label{statement2}

Our second analysis, which broadly follows the approach used to analyze IHT in~\cite{stablepoint}, considers the \textit{stable points} of ITP, a concept which can be viewed as a generalization of the notion of a fixed point to accommodate variable stepsize schemes, see \cite[Section 3.1]{stablepoint}.

\begin{defn}[\textbf{Stable points of generic ITP}]\label{stable_ITP}
Given $\au>0$ and an index set $\Gm\in\Tk$, we say $\xb\in\RR^N$ is an $\au$-stable point of generic ITP on $\Gm$ if $\mbox{supp}(\xb)\subseteq\Gm$ and
\begin{equation}\label{stable1}
\left\{A^T(b-A\xb)\right\}_{\Gm}=0\;\;\;\;\mbox{and}
\end{equation}
\begin{equation}\label{stable2}
\|\xb_{\Gm\sm\Omega}\|\geq\au\|A_{\Omega\sm\Gm}^T(b-A\xb)\|\;\;\;\;\forall\;\Omega\in\Tk.
\end{equation}
\end{defn}

For brevity's sake, we will often drop the `of generic ITP' label, and at times we will also drop the reference to the support set $\Gm$. We will be interested in values of $\au$ that lower bound the stepsize $\al^m$ of generic ITP.

\subsection{Results for ITP}

First considering ITP with constant stepsize $\al$, our approach is two-stage: on the one hand, we give conditions guaranteeing convergence of ITP to \textit{some} stable point. Meanwhile, by analysing a necessary condition for the existence of a stable point on a given support (which we refer to as the \textit{stable point condition}), we give conditions guaranteeing that all stable points are `close' to the original signal. Thus, if both conditions are satisfied, we ensure recovery of the original signal.

We will require the following assumption for the deterministic results given in this section.

\begin{ass}\label{genposass}
The columns of $A$ are in $2k$-general position, namely any collection
of $2k$ of its columns are linearly independent.
\end{ass}

Assumption~\ref{genposass} is a typical (weak) assumption in compressed sensing, and which guarantees a unique solution to Problem~\ref{treesparseprob}. We denote by $A_{\Gm}^{\dag}$ the Moore-Penrose pseudoinverse $(A_{\Gm}^T A_{\Gm})^{-1}A_{\Gm}^T$, which is well-defined under Assumption~\ref{genposass}. We next state a necessary condition for a stable point on a given support $\Gm$ in terms of only $\xs$, $A$ and $e$ and their restrictions to certain support sets. 

We next give a deterministic condition guaranteeing convergence to some $\al$-stable point in terms of the tree-based RIP.

\begin{thm}[\textbf{ITP convergence}]\label{CITPconv}
Consider Problem~\ref{treenoiseprob}. Suppose that Assumption~\ref{genposass} holds, and suppose that the stepsize in ITP satisfies 
\begin{equation}\label{CITPconvcond}
\al<\frac{1}{1+TU_{2k}},
\end{equation} 
where $TU$ is defined in (\ref{treeRIPdef}). Then ITP with stepsize $\al$ converges to an $\al$-stable point $\xb$ of generic ITP.
\end{thm}

We next state the \textit{stable point condition}, that is, a necessary condition for the existence of a stable point on a given support. It will help to first define $\Lm\in\Tk$ to be the support of the original signal, namely
\begin{equation}\label{lambda_def_tree}
\Lm:=\mbox{supp}(\xs),
\end{equation}
so that $|\Lm|=k$.

\begin{thm}[\textbf{Stable point condition}]\label{stableprop}
Consider Problem~\ref{treenoiseprob}. Suppose Assumption~\ref{genposass} holds and suppose there exists an $\au$-stable point on some $\Gm$ such that $\Gm\neq\Lm$. Then
\begin{equation}\label{stablecond}
\left\|A_{\Gm}^{\dag}A_{\Lm\sm\Gm}\xs_{\Lm\sm\Gm}\right\|+\left\|A_{\Gm}^{\dag}e\right\|\geq\au\left\{\left\|A_{\Lm\sm\Gm}^T(I-A_{\Gm}A^{\dag}_{\Gm})A_{\Lm\sm\Gm}\xs_{\Lm\sm\Gm}\right\|-\left\|A_{\Lm\sm\Gm}^T(I-A_{\Gm}A^{\dag}_{\Gm})e\right\|\right\},
\end{equation}
where $\Lm$ is defined in (\ref{lambda_def_tree}). 
\end{thm}

\textbf{Proof:} See Appendix~\ref{stable_det}.\hfill$\Box$\\
\\
While it would be possible to analyse the stable point condition using the tree-based RIP, we take a different approach. The stable point condition is especially amenable to probabilistic analysis for Gaussian matrices under the average-case (but realistic) assumption that the original signal and measurement matrix are statistically independent.

\begin{ass}\label{independentass}
The original signal $\xs$ and the measurement matrix $A$ are statistically independent.
\end{ass}

The crucial independence assumption will allow us to obtain better quantitative results than could be achieved through the purely worst-case RIP-based analysis of Section~\ref{statement1}. However, it is worth noting that independence is the only average-case assumption we invoke: we assume nothing further about the coefficient values of $\xs$. In keeping with the spirit of average-case analysis, we also assume that the noise is Gaussian and independent of both $A$ and $\xs$, which we formalize as follows. 

\begin{ass}\label{noiseass}
The noise vector $e$ has i.i.d. Gaussian entries $e_i\sim
N(0,\sigma^2/n)$, independently of $A$ and $\xs$.
\end{ass}

Note that, under Assumption~\ref{noiseass}, $\EE\,\|e\|^2=\sg^2$, so that $\|e\|\approx\sg$.

Assumption~\ref{genposass} is satisfied with probability $1$ by a Gaussian matrix, see~\cite[Section 4.1]{stablepoint}, and so may now be replaced with Assumption~\ref{gaussianass}. 

Under Assumptions~\ref{gaussianass},~\ref{independentass} and~\ref{noiseass}, each of the terms in (\ref{stablecond}), viewed as a Rayleigh quotient over $\|x_{\Lm\sm\Gm}\|^2$, is distributed according to either the $\chi^2$ or the $\FF$ distribution. We write $\chi^2_s$ for the (univariate) $\chi^2$-distribution with $s\geq 1$ degrees of freedom. Furthermore, if $P\sim\frac{1}{s}\chi^2_s$ and $Q\sim\frac{1}{t}\chi^2_t$ are independent random variables, we say that $P/Q$ follows the $\FF$-distribution, and we write $P/Q\sim\mathcal{F}(s,t)$. The following lemma, which was proved in~\cite{stablepoint}, gives the precise distributions.

\begin{lem}[\textbf{Distribution results for the stable point condition~\cite[Lemma 4.4]{stablepoint}}]\label{dist}
Suppose Assumptions~\ref{gaussianass},~\ref{independentass} and~\ref{noiseass} hold, and let $\Gm$ be an index set of cardinality $k$, where $k<n$. Then
\begin{equation}\label{PQresult}
\frac{\|A_{\Gm}^{\dag}A_{\Lm\sm\Gm}x_{\Lm\sm\Gm}\|^2}{\|x_{\Lm\sm\Gm}\|^2}=F_{\Gm},\;\;\;\mbox{where}\;\;\;F_{\Gm}\sim\frac{k}{n-k+1}\FF(k,n-k+1);
\end{equation}
\begin{equation}\label{Rbound}
\frac{\|A_{\Lm\sm\Gm}^T(I-A_{\Gm}A^{\dag}_{\Gm})A_{\Lm\sm\Gm}x_{\Lm\sm\Gm}\|^2}{\|x_{\Lm\sm\Gm}\|^2}\geq{\left(\frac{n-k}{n}\right)}^2\cdot R_{\Gm}^2,\;\;\;\mbox{where}\;\;\;R_{\Gm}\sim\frac{1}{n-k}\chi^2_{n-k};
\end{equation}
\begin{equation}\label{lhsnoise}
\|A_{\Gm}^{\dag}e\|\le\sg\cdot\sqrt{G_{\Gm}},\;\;\;\mbox{where}\;\;\;G_{\Gm}\sim\frac{k}{n-k+1}\FF(k,n-k+1);
\end{equation}
\begin{equation}\label{rhsnoise}
\|A_{\Lm\sm\Gm}^T(I-A_{\Gm}A_{\Gm}^{\dag})e\|\le\sg\sqrt{\frac{k(n-k)}{n^2}\cdot(S_{\Gm})(T_{\Gm})},\;\;\;\mbox{where}\;\;\;S_{\Gm}\sim\frac{1}{n-k}\chi^2_{n-k},\;\;\;T_{\Gm}\sim\frac{1}{k}\chi^2_k.
\end{equation}
\end{lem}

Recalling the stable point condition, we wish to show that all stable points are `close' to the original signal, which can be achieved by bounding each of the constituent terms over all permissible support sets. We can make an analogy with the tree-based RIP, where upper bounds on tree-based RIP constants are obtained in the simplified porportional-growth asymptotic by union bounding the tail probabilities of extreme singular values of submatrices of $A$ corresponding to permissible support sets. Similarly, large deviation bounds over $|\Tk|$ instances of $\chi^2$ and $\FF$ distributed random variables can be derived in the same asymptotic framework. One can view the resulting bounds as a kind of `independent RIP', where the assumption of independence between the measurement matrix and the original signal allows the tightening of bounds on Rayleigh quotients. Such an analysis is only possible if matrix-vector independence can be assumed, which is the case for the stable point condition (\ref{stablecond}). We define three tail bound functions.

\begin{defn}[\textbf{$\chi^2$ tail bounds}]\label{Idef_tree}
Let $\rr\in(0,1)$ and $\lm\in(0,1]$. Let $\TIU(\rr,\lm)$ be the unique solution to
\begin{equation}\label{udef_tree}
\nu-\ln(1+\nu)=\frac{2d\rr\cdot H(d^{-1})}{\lm}\;\;\;\;\mbox{for}\;\;\;\;\nu>0,
\end{equation}
and let $\TIL(\rr,\lm)$ be the unique solution to
\begin{equation}\label{ldef_tree}
-\nu-\ln(1-\nu)=\frac{2d\rr\cdot H(d^{-1})}{\lm}\;\;\;\;\mbox{for}\;\;\;\;\nu\in(0,1),
\end{equation}
where $H(\cdot)$ is defined in (\ref{shannon_def}).
\end{defn}

That $\TIU$ is well-defined follows since the left-hand side of (\ref{udef_tree}) is zero at $\nu=0$, tends to infinity as $\nu\ra\infty$, and is strictly increasing on $\nu>0$. Similarly, $\TIL$ is well-defined since the left-hand side of (\ref{ldef_tree}) is zero at $\nu=0$, tends to infinity as $\nu\ra 1$, and is strictly increasing on $\nu\in(0,1)$. 

\begin{defn}[\textbf{$\FF$ tail bound}]\label{IFdef_tree}
Let $\rr\in(0,1/2]$. Let $\TIF(\rr)$ be the unique solution in $f$ to
\begin{equation}\label{Fdef_tree}
\ln(1+f)-\rr\ln f=2d\rr\cdot H(d^{-1})+H(\rr)\;\;\;\;\mbox{for}\;\;\;\;f>\frac{\rr}{1-\rr},
\end{equation}
where $H(\cdot)$ is defined in (\ref{shannon_def}).
\end{defn}

That $\TIF$ is well-defined follows since the left-hand side of (\ref{Fdef_tree}) is equal to $H(\rr)$ at $f=\rr/(1-\rr)$, tends to infinity as $f\ra\infty$, and is strictly increasing on $f>\rr/(1-\rr)$. 

The bounds given in Definitions~\ref{Idef_tree} and ~\ref{IFdef_tree} are related to those given in the context of standard sparsity in~\cite[Definitions 4.4 and 4.5]{stablepoint}, and their intuition is as follows. The expressions on the left-hand sides of (\ref{udef_tree}), (\ref{ldef_tree}) and (\ref{Fdef_tree}) capture the rate of exponential decay of the $\chi^2$ and $F$ distributions, and these expressions are identical to the corresponding expressions in~\cite{stablepoint}. The difference lies in the expressions on the right-hand side, which capture the effect of the union bound over all permissible support sets. As was observed for the bounds on tree-based RIP in Section~\ref{statement1}, the number of permissible support sets in the tree-based setting is given by (\ref{treecount}), which has no dependence upon the ambient dimension $N$, which is why the expressions on the right-hand sides of (\ref{udef_tree}), (\ref{ldef_tree}) and (\ref{Fdef_tree}) depend only upon $\rho$ (for a given tree order $d$).

\begin{lem}[\textbf{Tree-based large deviations result for $\chi^2$}]\label{chisq_tree}
Let $l\in\{1,\ldots,n\}$ and let the random variables $X_l^i\sim\displaystyle\frac{1}{l}\chi^2_l$ for all $i\in S_n$, where $|S_n|=T(k)$, and let $\e>0$. In the simplified proportional growth asymptotic, let $l/n\ra\lm\in(0,1]$. Then
\begin{equation}\label{chisqresult1_tree}
\PP\left\{\cup_{i\in S_n}[X_l^i\geq 1+\TIU(\rr,\lm)+\e]\right\}\rightarrow
0
\end{equation}
and
\begin{equation}\label{chisqresult2_tree}
\PP\left\{\cup_{i\in S_n}[X_l^i\le 1-\TIL(\rr,\lm)-\e]\right\}\rightarrow
0,
\end{equation}
exponentially in $n$, where $\TIU(\rr,\lm)$ and $\TIL(\rr,\lm)$ are defined in (\ref{udef_tree}) and (\ref{ldef_tree}) respectively.
\end{lem}

\begin{lem}[\textbf{Tree-based large deviations results for $F$}]\label{Fdist_tree}
Let the random variables $X_n^i\sim\frac{k}{n-k+1}\;\mathcal{F}(k,n-k+1)$ for all $i\in S_n$, where $|S_n|=T(k)$, and let $\e>0$. In the simplified proportional growth asymptotic,
\begin{equation}\label{Fresult_tree}
\PP\left\{\cup_{i\in S_n}[X_n^i\geq \TIF(\rr)+\e]\right\}\rightarrow 0,
\end{equation}
exponentially in $n$, where $\TIF(\rr)$ is defined in (\ref{Fdef_tree}).
\end{lem}

We define oversampling thresholds for ITP algorithms in terms of the above tail bounds.

\begin{defn}[\textbf{Stable point recovery oversampling threshold for ITP}]\label{rhITP1}
Define $\rh^{ITP}_{SP}$ to be the unique solution to
\begin{equation}\label{rhoITPdef}
\frac{\sqrt{\TIF(\rr)}}{(1-\rr)\left[1-\TIL(\rr,1-\rr)\right]}=\frac{1}{1+\TU(2\rr)}\;\;\;\;\mbox{for}\;\;\;\;\rr\in(0,1/2],
\end{equation}
where $\TIF$ is defined in (\ref{Fdef_tree}), $\TIL$ is defined in (\ref{ldef_tree}) and $\TU$ is defined in Definition~\ref{RIPboundsdef}.
\end{defn}

The oversampling threshold (\ref{rhoITPdef}) is a counterpart of the phase transitions given in~\cite[Section 5.2]{thesis} for IHT algorithms, with the only changes being the switch to tree-based tail bounds and the disappearance of the $\dd$ variable. A proof that (\ref{rhoITPdef}) admits a unique solution proceeds analogously to the one given for the counterpart phase transitions in~\cite[Section 5.2]{thesis}. Next, we define a function $\xi^{ITP_{\al}}_{SP}(\rr)$ which will represent a stability factor in our results, bounding the approximation error of the output of ITP as a multiple of the noise level $\sg$. 

\begin{defn}[\textbf{Stability factor for ITP}]\label{massive_def_tree}
Consider Problem~\ref{treenoiseprob}. Given $\rr\in(0,1/2]$ and $\au>0$, provided
\begin{equation}\label{ITPstablecond}
\rr<\rh^{ITP}_{SP},
\end{equation} 
define
\begin{equation}\label{adefn_tree}
a(\rr):=\frac{\sqrt{\TIF(\rr)}+\al\sqrt{\rr(1-\rr)[1+\TIU(\rr,1-\rr)][1+\TIU(\rr,\rr)]}}{\al(1-\rr)[1-\TIL(\rr,1-\rr)]-\sqrt{\TIF(\rr)}},
\end{equation}
and
\begin{equation}\label{xidef_tree}
\xi^{ITP_{\al}}_{SP}(\rr):=\sqrt{\TIF(\rr)\left[1+a(\rr)\right]^2+\left[a(\rr)\right]^2},
\end{equation}
where $\TIF$ is defined in (\ref{Fdef_tree}), and where $\TIU$ and $\TIL$ are defined in (\ref{udef_tree}) and (\ref{ldef_tree}) respectively.
\end{defn}

Note that (\ref{ITPstablecond}) ensures that the denominator in (\ref{adefn_tree}) is strictly positive and that $a(\rr)$ is therefore well-defined. We proceed to our recovery result for constant stepsize ITP.

\begin{thm}[\textbf{Stable point recovery for ITP}]\label{recov1noise_tree}
Consider Problem~\ref{treenoiseprob} and suppose Assumptions~\ref{gaussianass},~\ref{independentass} and~\ref{noiseass} hold. If (\ref{ITPstablecond}) holds and the stepsize $\al$ satisfies
\begin{equation}\label{alphabound1_tree}
\frac{\sqrt{\TIF(\rr)}}{(1-\rr)\left[1-\TIL(\rr,1-\rr)\right]}<\al<\frac{1}{1+\TU(2\rr)},
\end{equation}
then, in the simplified proportional-growth asymptotic\footnote{In other words, we consider instances of $k$-tree sparse vectors $\xs$ and Gaussian random variables $A$ and $e$ for a sequence of triples
$(k,n,N)$ where $n\rightarrow\infty$, where $n$ is the number of measurements, $N$, the signal dimension and $k$, the sparsity of the underlying signal.}, ITP with stepsize $\al$ converges to $\xb$ such that
\begin{equation}\label{error3}
\|\xb-\xs\|\le\xi^{ITP_{\al}}_{SP}(\rr)\cdot\sg,
\end{equation}
with probability tending to $1$ exponentially in $n$.
\end{thm}

\textbf{Proof:} See Appendix~\ref{stable_proofs}.\hfill$\Box$\\
\\
In the special case of Problem~\ref{treesparseprob}, the same oversampling threshold guarantees exact recovery of the underlying signal $\xs$.

\begin{cor}[\textbf{Stable point recovery for ITP: noiseless case}]\label{recov1noiseless_tree}
Consider Problem~\ref{treesparseprob}. Suppose Assumptions~\ref{gaussianass} and~\ref{independentass} hold, suppose that (\ref{ITPstablecond}) holds, and suppose that $\al$ satisfies (\ref{alphabound1_tree}). Then, in the simplified proportional-growth asymptotic, ITP with stepsize $\al$ converges to $\xs$ with probability tending to $1$ exponentially in $n$.
\end{cor}

\textbf{Proof:} See Appendix~\ref{stable_proofs}.\hfill$\Box$

\subsubsection{Results for NITP}

We now turn our attention to NITP, and define the following oversampling threshold and stability factor in this case.

\begin{defn}[\textbf{Stable point recovery oversampling threshold for ITP}]\label{rhITP2}
Define $\rh^{NITP_{\kappa}}_{SP}$ to be the unique solution to
\begin{equation}\label{rhoNITPdef}
\frac{\sqrt{\TIF(\rr)}}{(1-\rr)\left[1-\TIL(\rr,1-\rr)\right]}=\frac{1}{\kappa[1+\TU(2\rr)]}\;\;\;\;\mbox{for}\;\;\;\;\rr\in(0,1/2],
\end{equation}
where $\TIF$ is defined in (\ref{Fdef_tree}), $\TIL$ is defined in (\ref{ldef_tree}) and $\TU$ is defined in Definition~\ref{RIPboundsdef}.
\end{defn}

A proof that (\ref{rhoNITPdef}) admits a unique solution proceeds analogously to the one given for the counterpart phase transitions in~\cite[Section 5.2]{thesis}. We define the following stability factor for NITP.

\begin{defn}[\textbf{Stability factor for NITP}]\label{massive_def2_tree}
Consider Problem~\ref{treesparseprob}. Given $\rr\in(0,1/2]$, provided
\begin{equation}\label{NITPstablecond}
\rr<\rh^{NITP_{\kappa}}_{SP},
\end{equation}
define
\begin{equation}\label{adefn2_tree}
a(\rr):=\frac{\sqrt{\TIF(\rr)}+\{\kappa[1+\TU(2\rr)]\}^{-1}\sqrt{\rr(1-\rr)[1+\TIU(\rr,1-\rr)][1+\TIU(\rr,\rr)]}}{(1-\rr)\{\kappa[1+\TU(2\rr)]\}^{-1}[1-\TIL(\rr,1-\rr)]-\sqrt{\TIF(\rr)}},
\end{equation}
and
\begin{equation}\label{xidef2_tree}
\xi^{NITP_{\kappa}}_{SP}(\rr):=\sqrt{\TIF(\rr)\left[1+a(\rr)\right]^2+\left[a(\rr)\right]^2},
\end{equation}
where $\TIF$ is defined in (\ref{Fdef_tree}), where $\TIU$ and $\TIL$ are defined in (\ref{udef_tree}) and (\ref{ldef_tree}) respectively, and where $\TU$ is defined in Definition~\ref{RIPbounds}.
\end{defn}

\begin{thm}[\textbf{Stable point recovery for NITP}]\label{recov2noise_tree}
Consider Problem~\ref{treenoiseprob}, suppose Assumptions~\ref{gaussianass},~\ref{independentass} and~\ref{noiseass} hold, and suppose (\ref{NITPstablecond}) holds. Then, in the simplified proportional-growth asymptotic, NITP with shrinkage parameter $\kappa$ converges to $\xb$ such that
\begin{equation}\label{error4}
\|\xb-\xs\|\le\xi^{NITP_{\kappa}}_{SP}(\rr)\cdot\sg,
\end{equation}
with probability tending to $1$ exponentially in $n$.
\end{thm}

\textbf{Proof:} See Appendix~\ref{stable_proofs}.\hfill$\Box$\\
\\
In the case of Problem~\ref{treesparseprob}, Theorem~\ref{recov2noise_tree} also simplifies to an exact recovery result.

\begin{cor}[\textbf{Stable point recovery for NITP: noiseless case}]\label{recov2noiseless_tree}
Consider Problem~\ref{treesparseprob}. Suppose Assumptions~\ref{gaussianass} and~\ref{independentass} hold and suppose that (\ref{NITPstablecond}) holds. Then, in the simplified proportional-growth asymptotic, NITP with shrinkage parameter $\kappa$ converges to $\xs$ with probability tending to $1$ exponentially in $n$.
\end{cor}

\textbf{Proof:} See Appendix~\ref{stable_proofs}.\hfill$\Box$

\section{Discussion of recovery results}\label{discussion}

\subsection{Tree-based RIP recovery results}\label{RIP_results}

\begin{figure}[h]
\centering
\subfigure[]{\includegraphics[width=3.2in,height=2.2in]{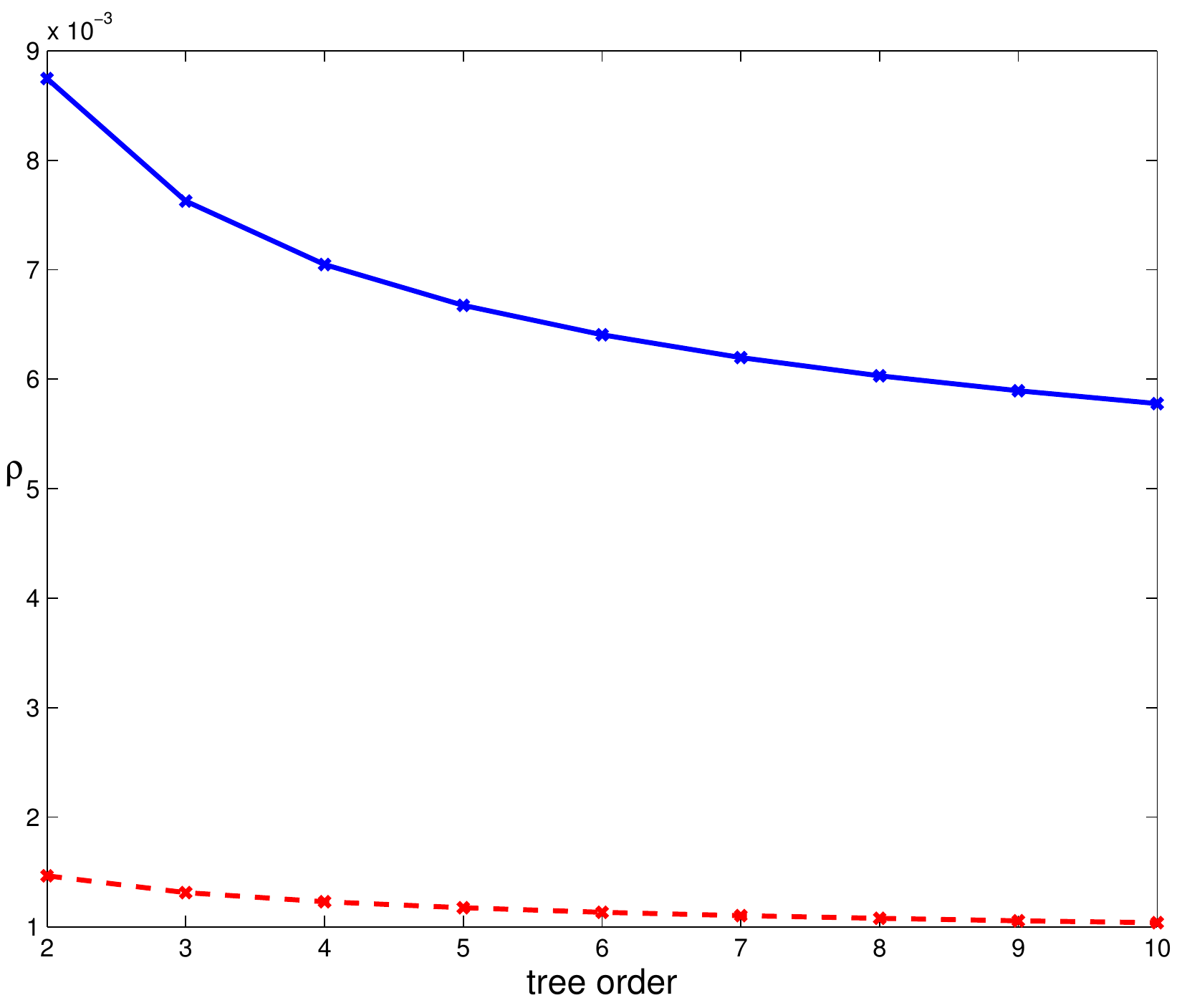}}
\subfigure[]{\includegraphics[width=3.2in,height=2.2in]{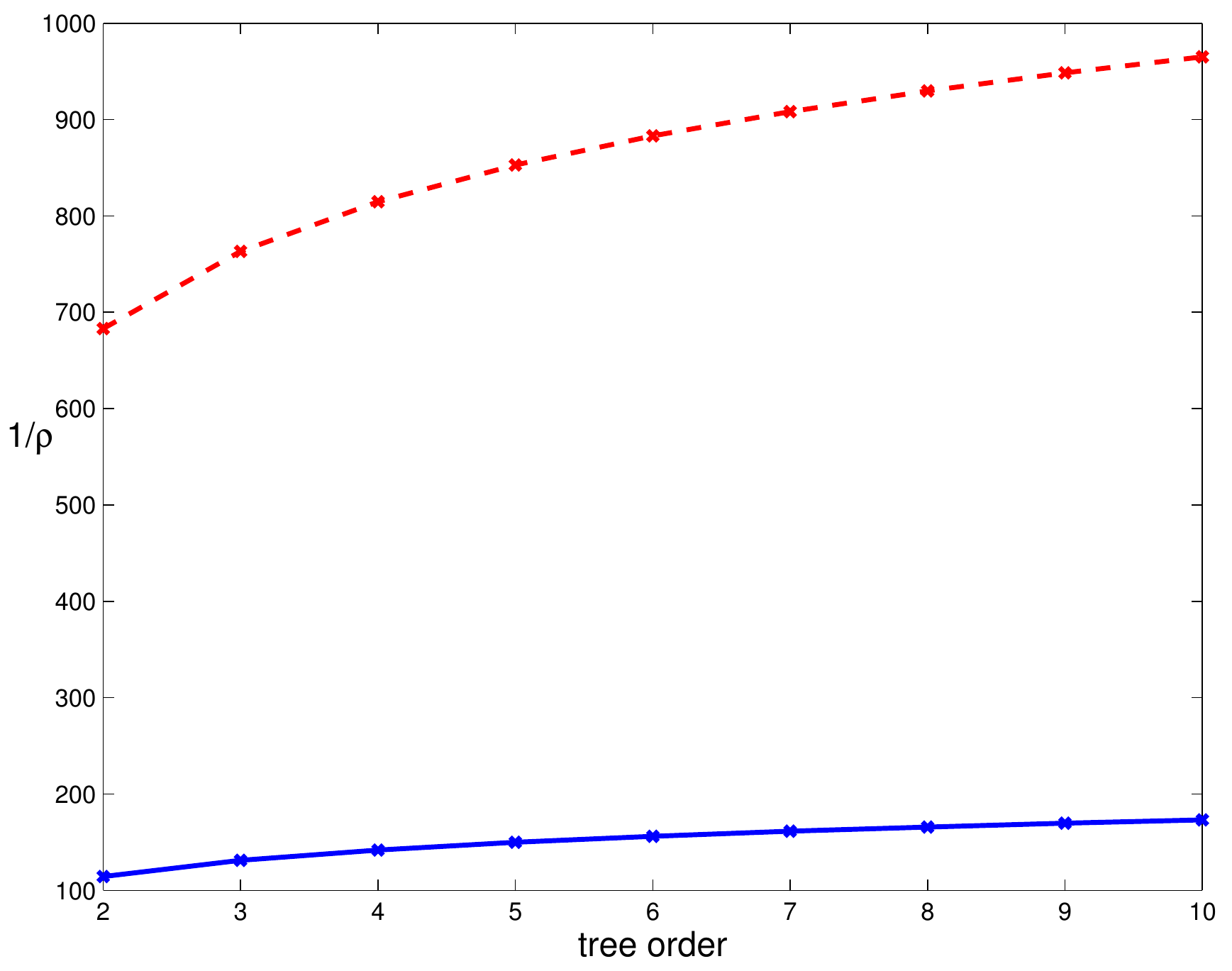}}
\caption{(a) Critical $\rr$-values for different tree orders from tree-based RIP analysis: ITP -- unbroken; NITP -- dashed. (b) Corresponding oversampling factors (reciprocals of $\hat{\rr}$).}
\label{oversampleplots}
\end{figure}

{\bf Noiseless case.}\quad The oversampling thresholds for ITP and NITP given by Definition~\ref{muxiALG_def} and Corollary~\ref{ALG_RIPphase_noiseless} are displayed in Figure~\ref{oversampleplots}(a) for different tree orders $d$. For binary trees, for example, we have $\hat{\rr}^{ITP_{\ha}}_{RIP}\approx 0.00875$ for ITP and $\hat{\rr}^{{NITP}_{\kappa}}_{RIP}\approx 0.00146$ for NITP (taking $\kappa=1.1$ for the shrinkage parameter in NITP). In both cases, exact recovery in the noiseless case is asymptotically guaranteed provided the limiting value of the ratio $\rr$ is less than the given threshold. We see a measured deterioration in the results for higher tree orders: the corresponding thresholds for quad-trees ($d=4$) -- which arise in image analysis using 2D wavelets -- are $0.00705$ and $0.00123$ for ITP and NITP respectively. Figure~\ref{oversampleplots}(b) shows the inverse of the oversampling ratio, which indicates the number of measurements required by the analysis as a multiple of the sparsity. We find, for binary trees, that $n\geq 115k$ measurements guarantees recovery by ITP, while $n\geq 683k$ measurements guarantees recovery by NITP. Provided the oversampling thresholds are respected, convergence to the original signal is guaranteed at a linear rate. The quantities $\mu^{ITP_{\ha}}_{RIP}(\rr)$ and $\mu^{NITP_{1.1}}_{RIP}(\rr)$ represent guaranteed bounds on the convergence rate for each variant.

\begin{figure}[h]
\centering
\subfigure[]{\includegraphics[width=3.2in,height=2.2in]{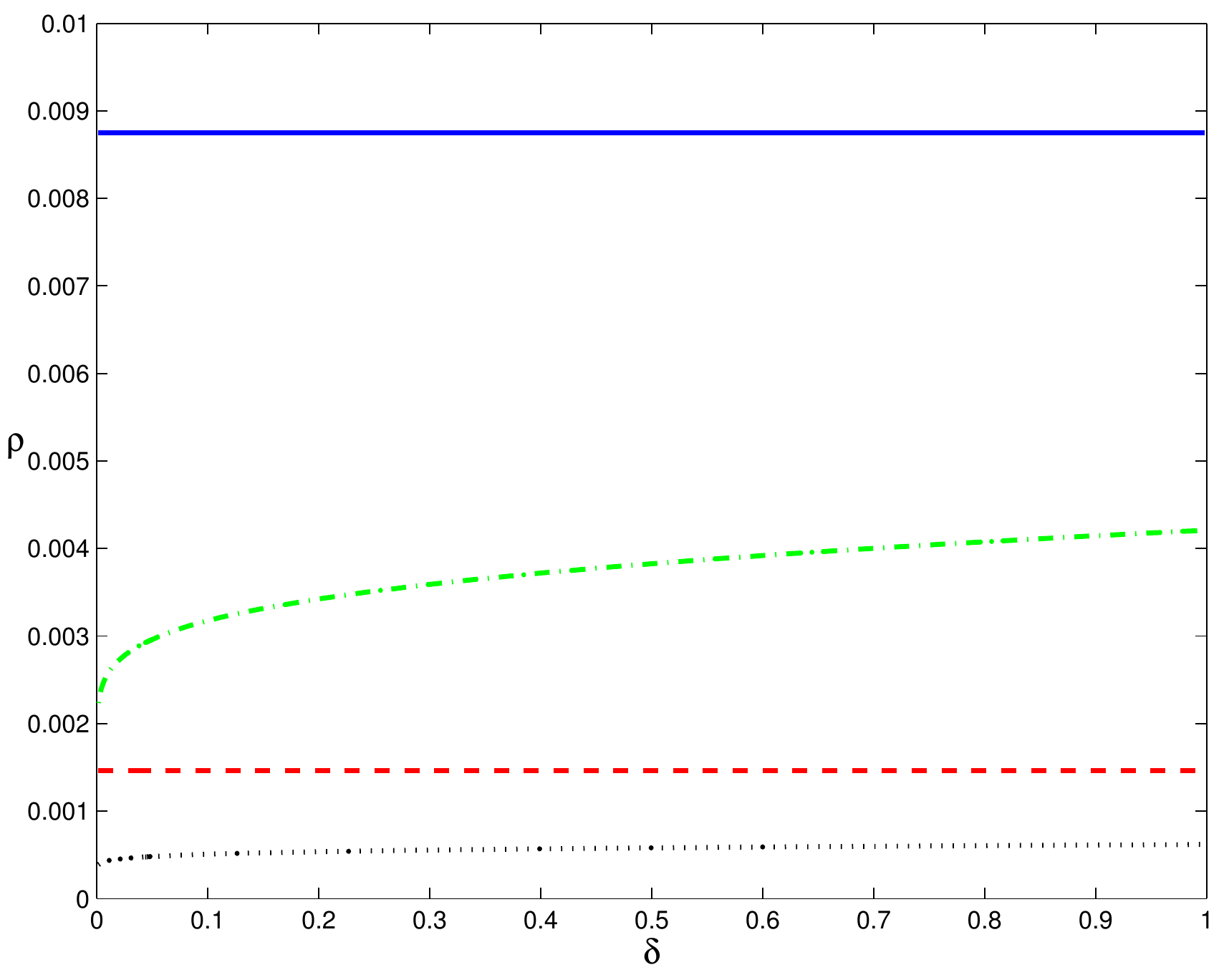}}
\subfigure[]{\includegraphics[width=3.2in,height=2.2in]{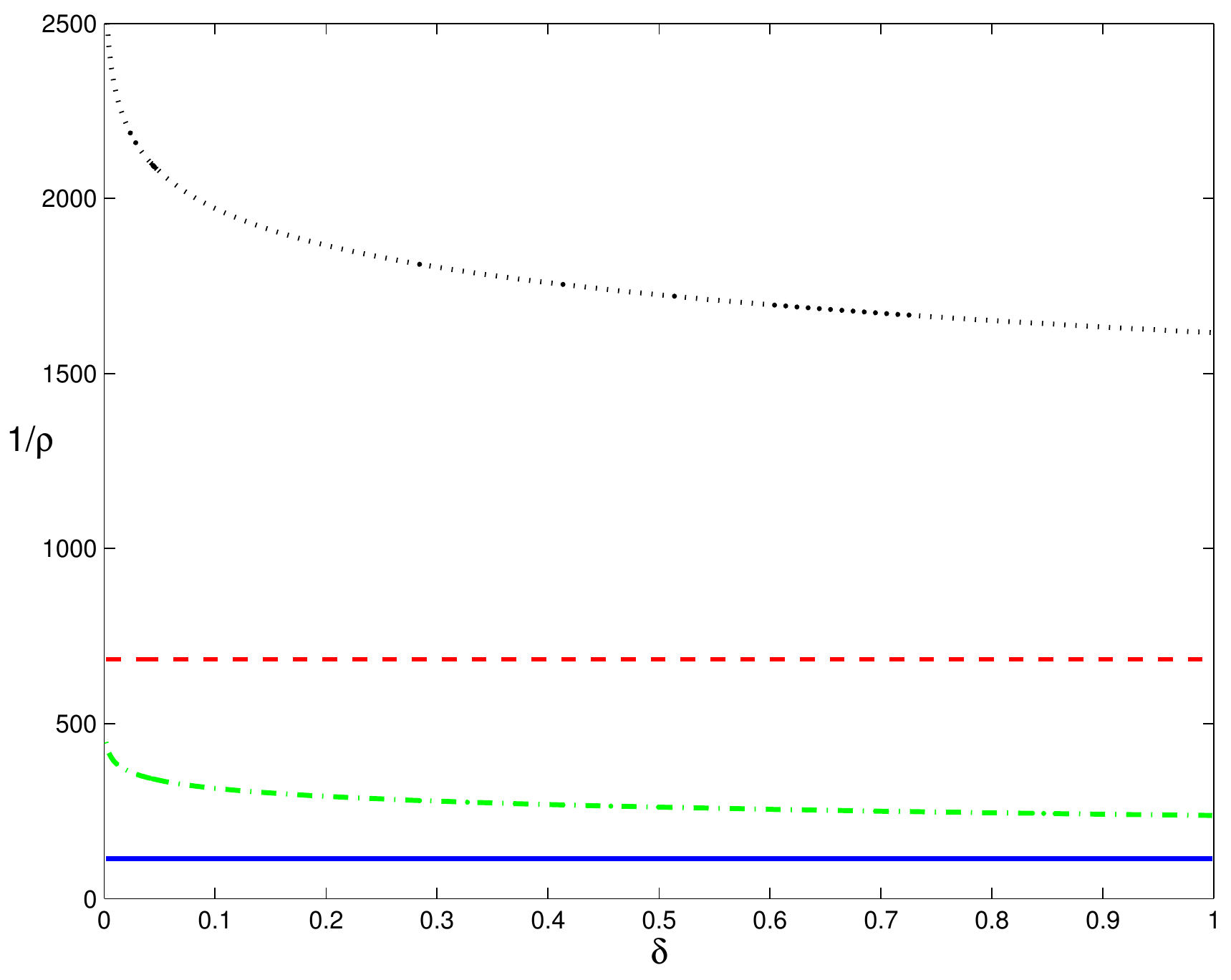}}
\caption{(a) Phase transitions from RIP analysis in the $(\dd,\rr)$ framework for binary trees (ITP -- unbroken; NITP -- dashed) and non-tree-based (IHT -- dash-dot; NIHT -- dotted). (b) Corresponding inverses of the phase transition.}
\label{phaseplots}
\end{figure}

While in the present paper we have dispensed with the undersampling ratio $\dd=n/N$, we may also frame our results in the $(\dd,\rr)$ asymptotic in order to make a comparison with analogous results derived in the non-tree-based setting for IHT based upon the standard notion of RIP~\cite{thesis}. Since there is no dependence upon $\dd$ in our case, the phase transitions we obtain are simply horizontal lines in the $(\dd,\rr)$-plane. Exact recovery phase transitions for binary trees are displayed in Figure~\ref{phaseplots} alongside the phase transitions derived in~\cite{stablepoint}: recovery is guaranteed asymptotically beneath the respective curves. We observe that the switch to the tree-based setting leads to significantly improved results, especially for small $\dd$.

{\bf Comparison with prior work}\quad Analogous oversampling ratios can be explicitly obtained in the case of binary trees using the prior analysis in~\cite{union_subspaces}. We observe that the oversampling thresholds given by Definition~\ref{muxiALG_def} and Corollary~\ref{ALG_RIPphase_noiseless} represent a scale factor improvement of around $100$ over those obtainable using the analysis in~\cite{union_subspaces}. The precise thresholds for binary trees are given in Table~\ref{bin_thresh} for comparison, along with the scale factor improvement. For the prior analysis, the optimal stepsize for ITP is $\al:=1$, and the parameter $\kappa$ is again taken to be $1.1$. The dramatic improvement in oversampling thresholds is due to the tightening of the tree-based RIP bounds in Definition~\ref{RIPboundsdef} over those in Definition~\ref{union_tree_bound}. This tightening is achieved through an asymmetric treatment of the tree-based RIP accompanied by a tighter large deviations analysis based on the PDF of the extreme singular values of the submatrices of Gaussian matrices, as opposed to the more generic sphere-covering argument relied upon in~\cite{union_subspaces}. 

\begin{table}
\begin{center}
\begin{tabular}{|c|c|c|c|c|c|}
\hline
&\multicolumn{2}{|c|}{Current paper}&\multicolumn{2}{|c|}{Analysis in \cite{union_subspaces}}&Factor\\
&$\rho$&$\rho^{-1}$&$\rho$&$\rho^{-1}$&improvement\\
\hline
ITP&$8.75\times 10^{-3}$&$115$&$1.24\times 10^{-4}$&$8068$&$70$\\
NITP&$1.46\times 10^{-3}$&$683$&$1.25\times 10^{-5}$&$79705$&$116$\\
\hline
\end{tabular}
\end{center}
\caption{Comparison of oversampling thresholds obtained from the current analysis and the prior analysis in~\cite{union_subspaces}, in the case of binary trees.}
\label{bin_thresh}
\end{table}

\begin{figure}[h]
\centering
\subfigure[]{\includegraphics[width=3.2in,height=2.2in]{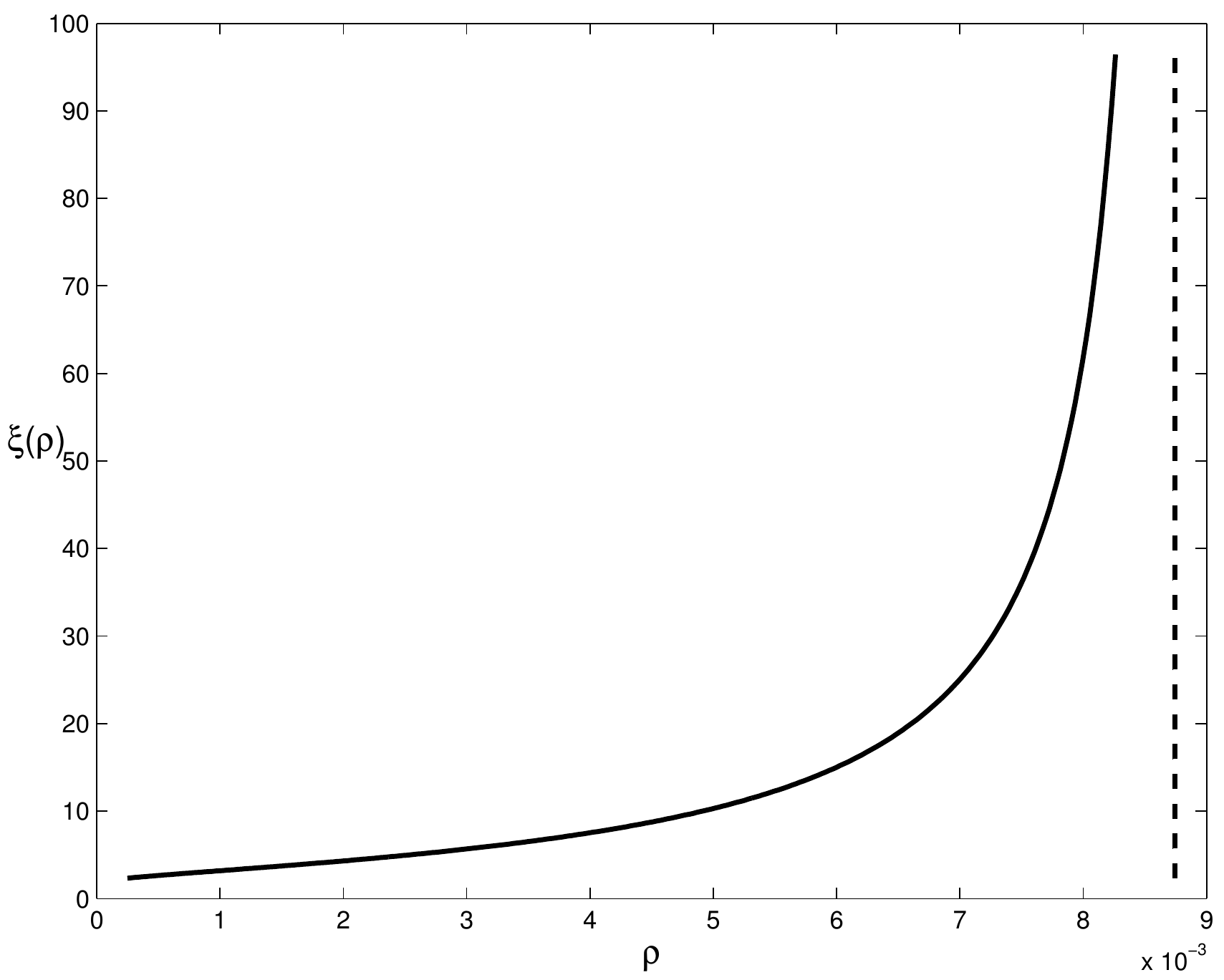}}
\subfigure[]{\includegraphics[width=3.2in,height=2.2in]{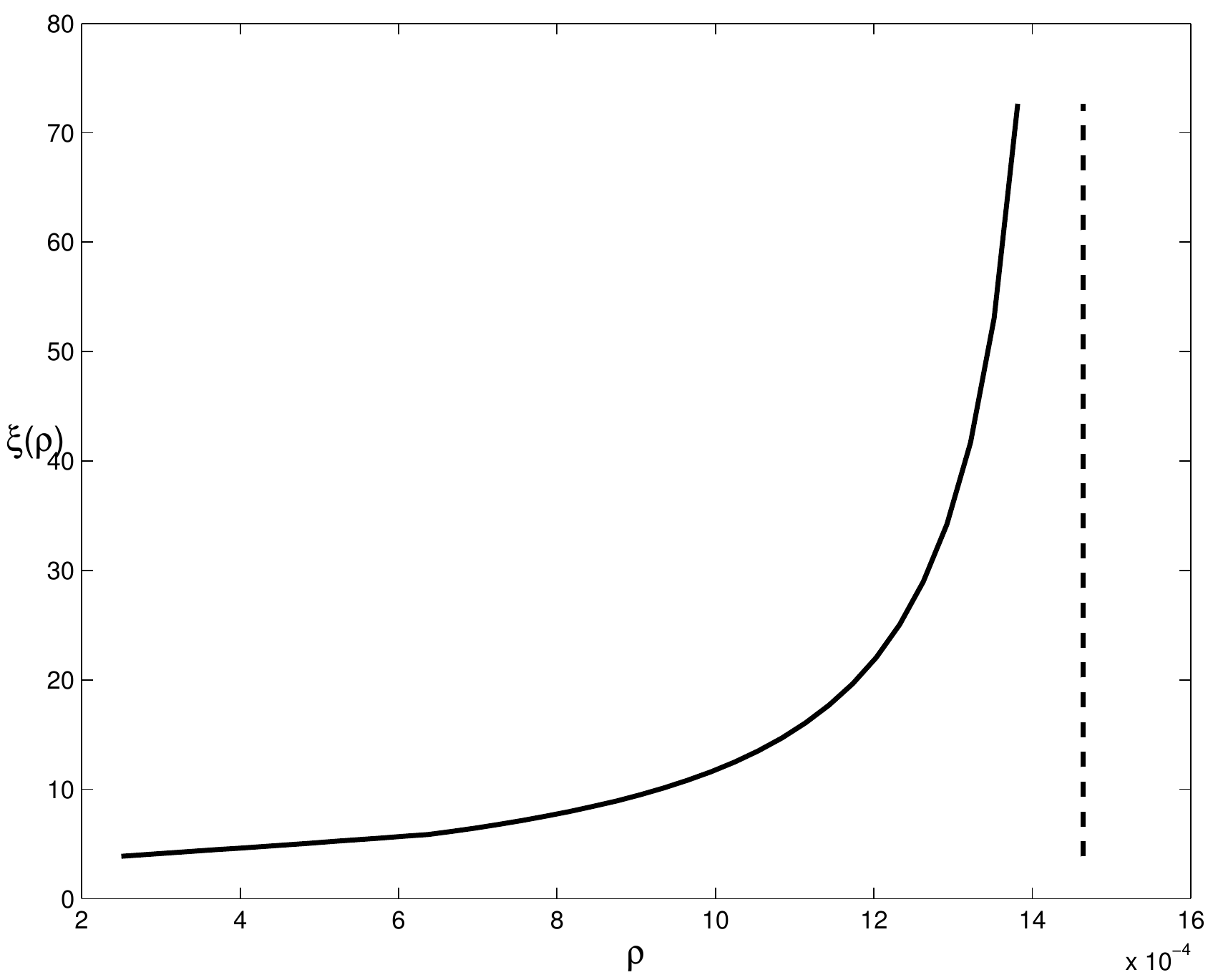}}
\caption{Plot of the stability factor $\xi(\rr)/[1-\mu(\rr)]$ from tree-based RIP analysis for binary trees: (a) ITP; (b) NITP.}
\label{stabilityplots}
\end{figure}

{\bf Extension to noise.}\quad In the case where measurements are contaminated by noise, exact recovery of the original signal is an unrealistic aim. However, provided the limiting value of the ratio $\rr$ falls below the respective oversampling threshold, Theorem~\ref{ALG_RIPphase} gives bounds on the limiting approximation error. More precisely, the results state that the limiting approximation error of the iterates of ITP/NITP is asymptotically bounded by some known stability factor multiplied by the noise level $\sg$. However, neither result necessarily implies convergence of the algorithm in the case of noise. The  Figure~\ref{stabilityplots} plots the noise stability factor $\xi(\rr)/[1-\mu(\rr)]$ for binary trees, for each of the two stepsize schemes considered ($\kappa=1.1$ for NITP). In keeping with~\cite{lqphase},~\cite{greedy} and~\cite{stablepoint}, we observe that the stability factor tends to infinity as the transition point is reached, \textit{i.e.}  $\xi(\rr)/[1-\mu(\rr)]\ra\infty$ as $\rr\ra\hat{\rr}$. For both ITP and NITP, given any value of $\rr$ for which the stability factors derived in this paper are defined, they are always lower than the corresponding stability factors derived from analysis of IHT based upon the standard RIP~\cite{HTP}; see \cite[Section 2.4]{thesis} for a comparison.

\subsection{Recovery results from the tree-based stable point analysis}\label{stable_results}

\begin{figure}[h]
\centering
\subfigure[]{\includegraphics[width=3.2in,height=2.2in]{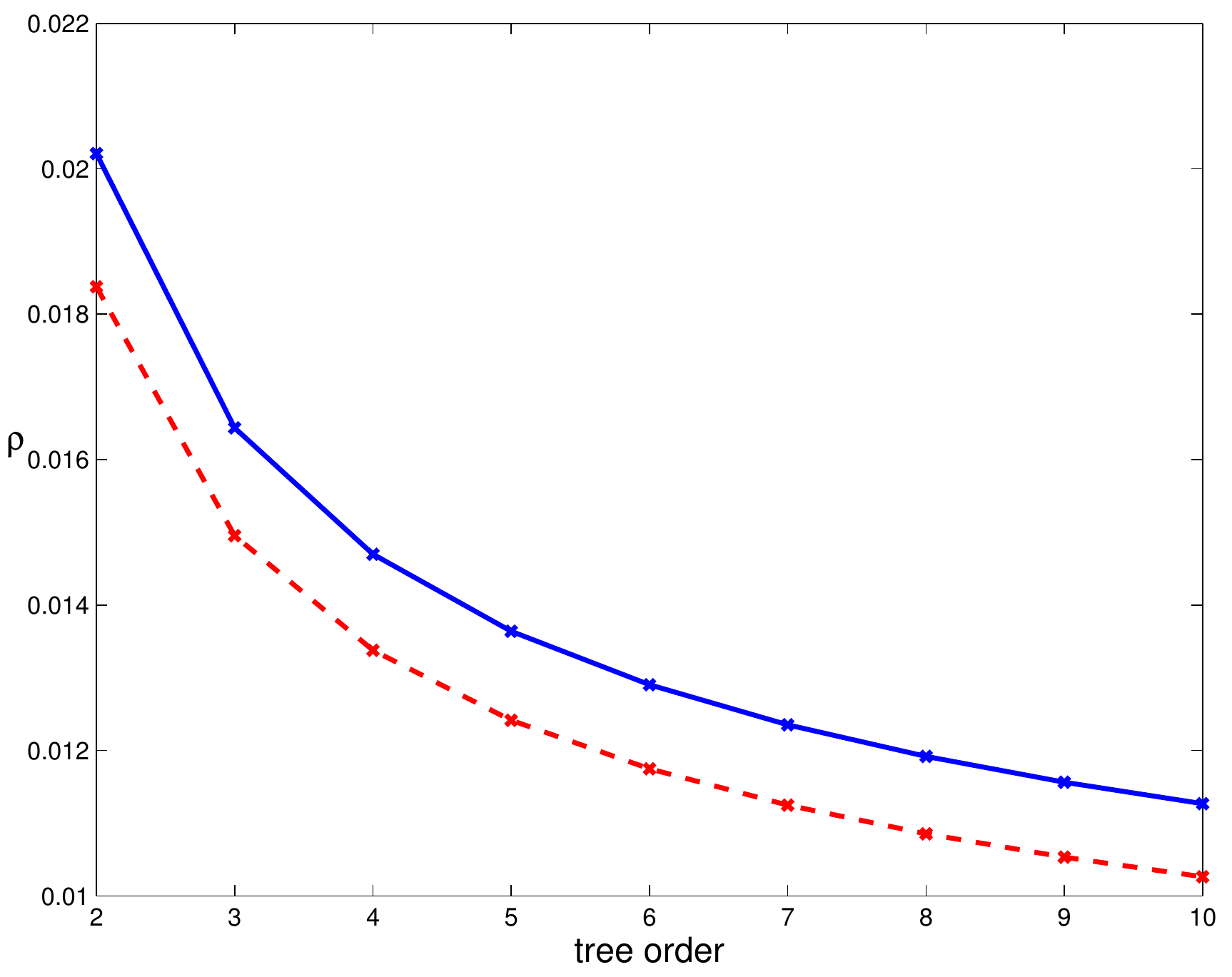}}
\subfigure[]{\includegraphics[width=3.2in,height=2.2in]{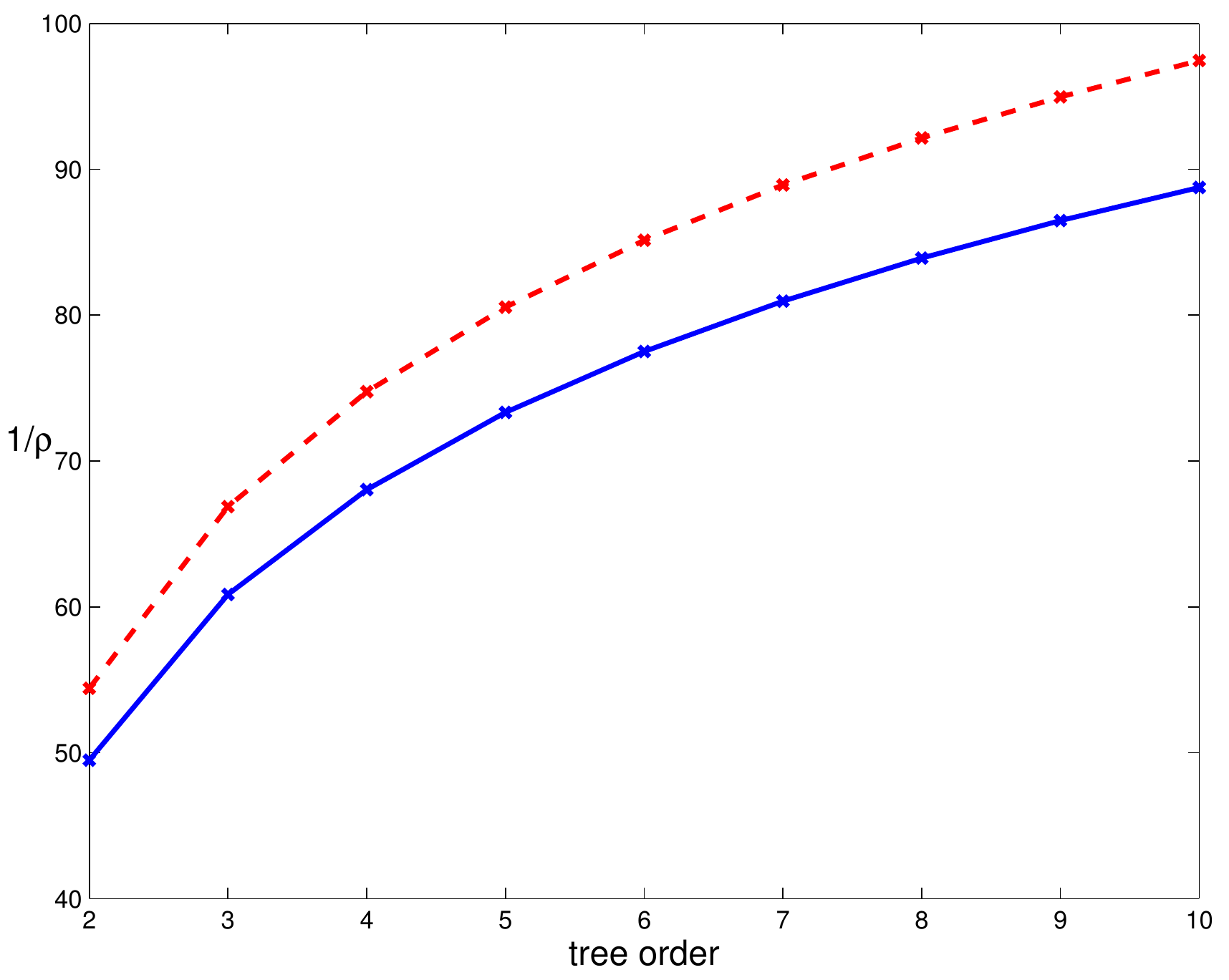}}
\caption{(a) Critical $\rr$-values for different tree orders from stable point analysis: ITP -- unbroken; NITP -- dashed. (b) Corresponding oversampling factors (reciprocals of $\hat{\rr}$).}
\label{oversampleplots2}
\end{figure}

\begin{figure}[h]
\centering
\subfigure[]{\includegraphics[width=3.2in,height=2.2in]{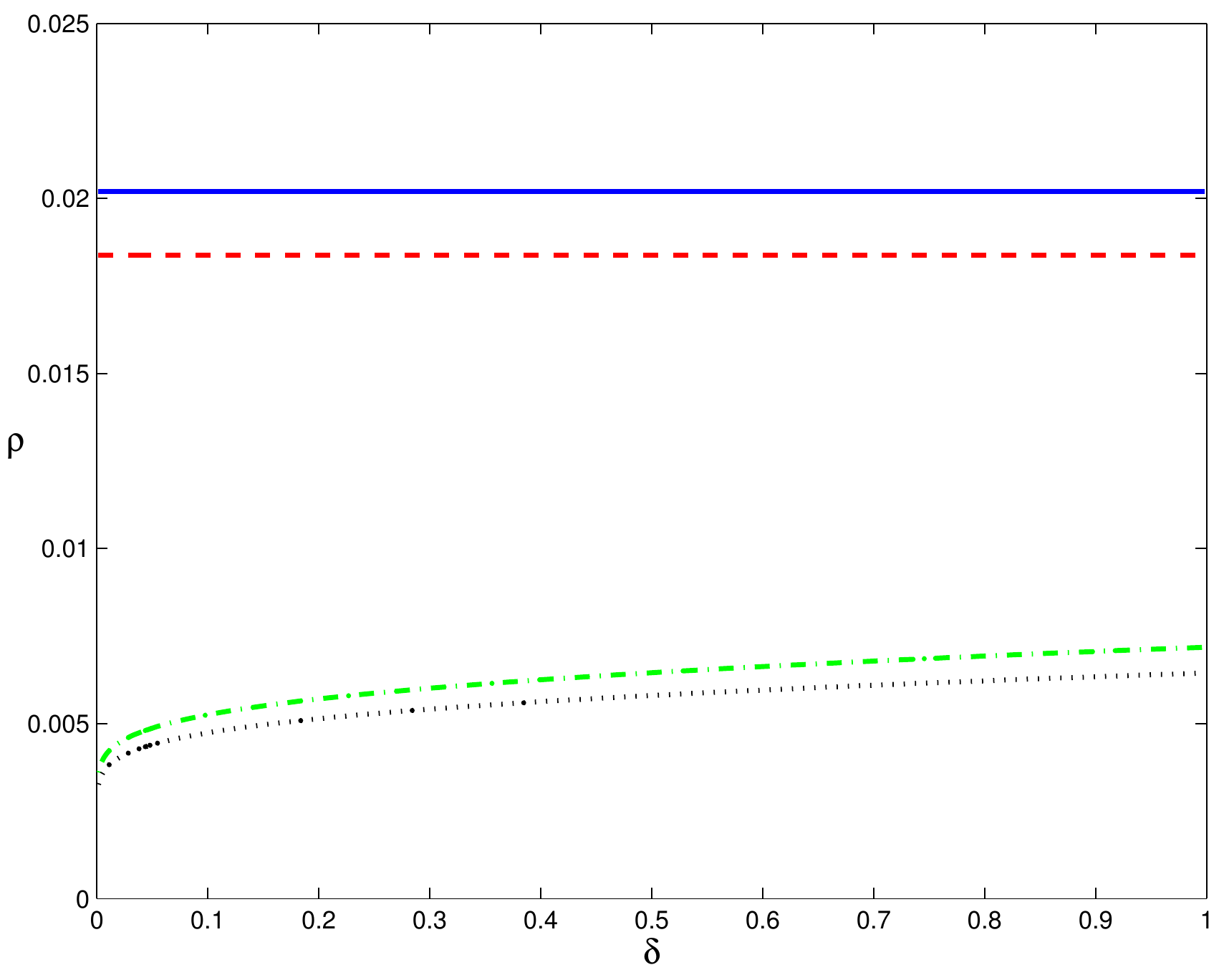}}
\subfigure[]{\includegraphics[width=3.2in,height=2.2in]{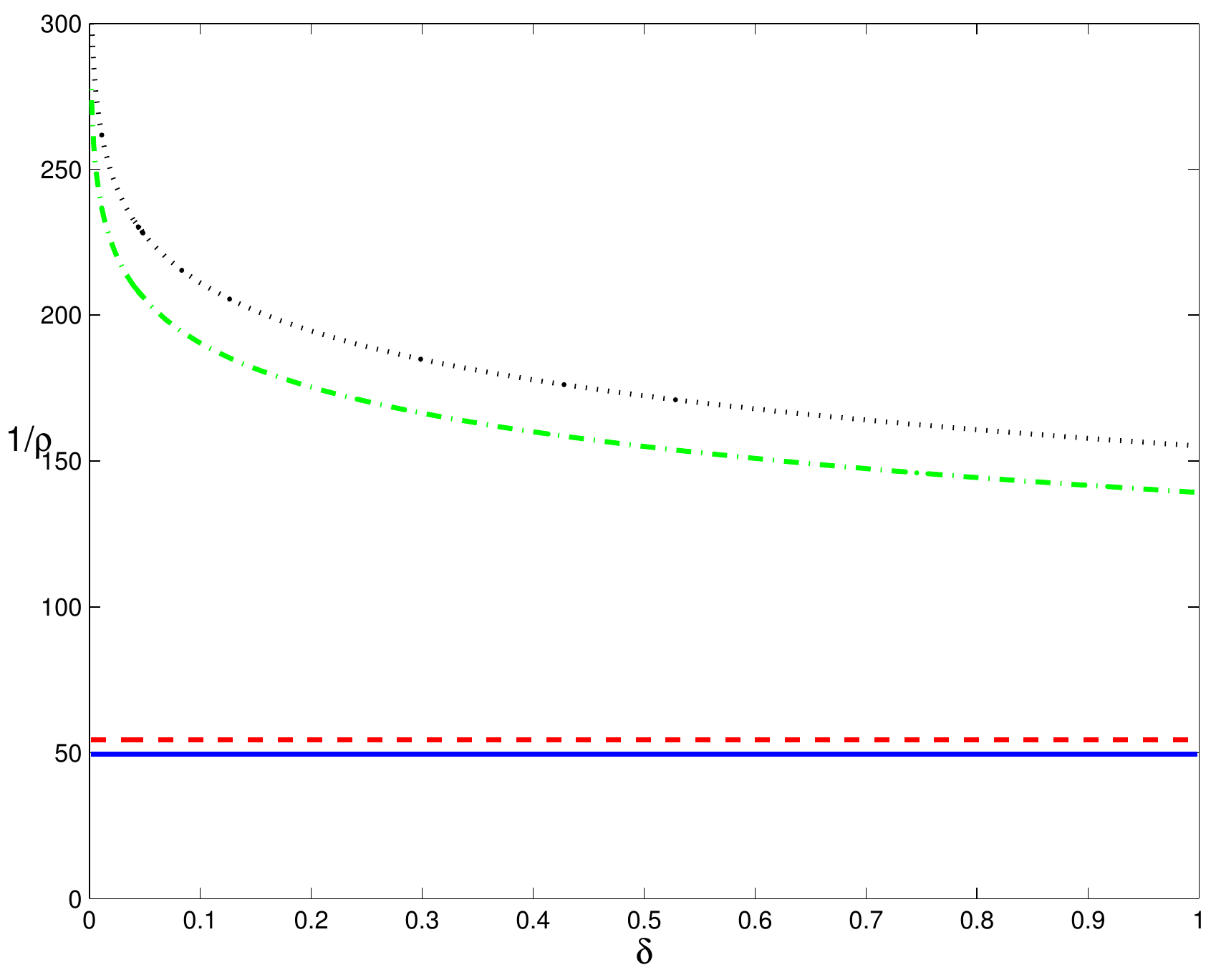}}
\caption{(a) Phase transitions from stable point analysis in the $(\dd,\rr)$ framework for binary trees (ITP -- unbroken; NITP -- dashed) and non-tree-based (IHT -- dash-dot; NIHT -- dotted). (b) Corresponding inverses of the phase transition.}
\label{phaseplots2}
\end{figure}

{\bf Noiseless case.}\quad The oversampling thresholds for ITP and NITP defined in Corollaries~\ref{recov1noiseless_tree} and~\ref{recov2noiseless_tree} are displayed in Figure~\ref{oversampleplots2}(a) for different tree orders $d$. For binary trees, we have $\hat{\rr}^{ITP_{\al}}_{RIP}\approx 0.0202$ for ITP and $\hat{\rr}^{{NITP}_{1.1}}_{RIP}\approx 0.0184$ for NITP, and the corresponding thresholds for quad-trees ($d=4$) are $0.0147$ and $0.0134$ respectively. Figure~\ref{oversampleplots2}(b) shows the inverse of the oversampling ratio: we find, for binary trees, that $n\geq 50k$ measurements guarantees recovery by ITP, while $n\geq 55k$ measurements guarantees recovery by NITP. The same exact recovery thresholds for binary trees are presented in the form of phase transitions in the $(\dd,\rr)$ asymptotic in Figure~\ref{phaseplots2}, alongside the phase transitions for IHT/NIHT derived in~\cite{stablepoint}. Again, we observe improved results by switching to the tree-based setting, especially for small $\dd$. 

Comparing the oversampling thresholds derived from the stable point analysis (Figure~\ref{oversampleplots2}) with those derived from tree-based RIP analysis (Figure~\ref{oversampleplots}), we observe a significant quantitative improvement for both algorithm variants, by over a factor of $10$ for NITP in fact for all tree orders under consideration. We have obtained improved oversampling thresholds by exploiting average-case assumptions, and we should point out the difference between the results in Sections~\ref{RIP_results} and~\ref{stable_results}. The tree-based RIP results are worst-case in nature: given a sequence of randomly generated Gaussian matrices, it is asymptotically guaranteed that ITP/NITP will in fact recover an accurate approximation to \textit{any} $k$-tree sparse signal vector. On the other hand, the results derived from our stable point analysis have a more average-case flavour: given a sequence of randomly generated Gaussian measurement matrices \textit{along with} a sequence of signal and noise vectors which are both independent of the measurement matrix, recovery is asymptotically guaranteed in this sense. It is not surprising that our average-case framework leads to an improvement over tree-based RIP since the assumption of independence between signal and measurement matrix rules out the practically unlikely case in which one chooses the very worst possible signal for a given measurement matrix. For a comparison of phase transitions derived from both stable point and RIP analysis in the context of IHT and simple sparsity, we refer the reader to~\cite[Section 6]{stablepoint}.

\begin{figure}[h]
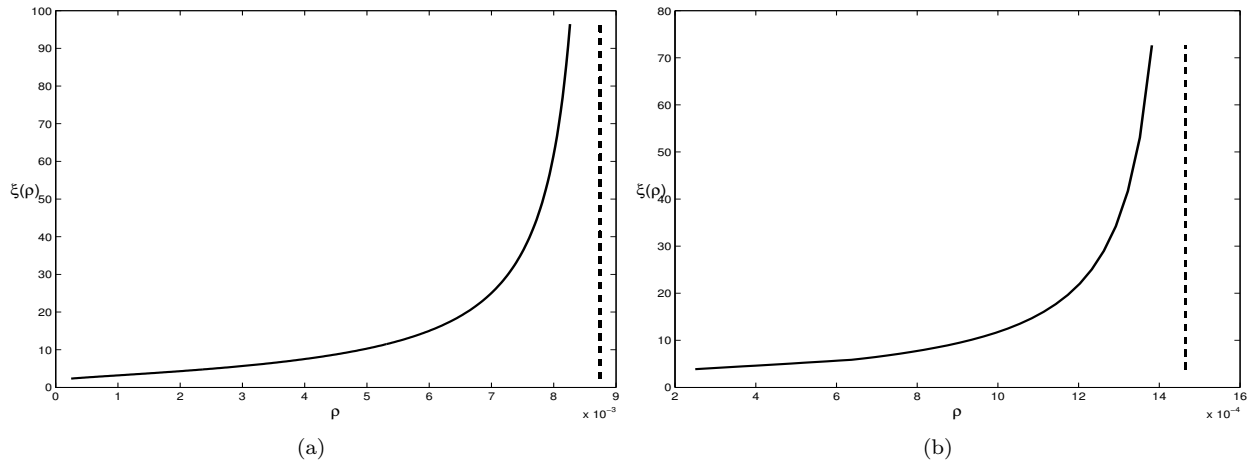

\centering
\subfigure[]{\includegraphics[width=3.2in,height=2.2in]{treenoise1_RIP.pdf}}
\subfigure[]{\includegraphics[width=3.2in,height=2.2in]{treenoise2_RIP.pdf}}
\caption{Plot of the stability factor $\xi(\rr)$ from stable point analysis for binary trees: (a) ITP; (b) NITP.}
\label{stabilityplots2}
\end{figure}

{\bf Extension to noise.}\quad Below the same oversampling thresholds, Theorems~\ref{recov1noise_tree} and~\ref{recov2noise_tree} go further than the tree-based RIP analysis in proving convergence of ITP/NITP to a limit point --- whose approximation error is asymptotically bounded by some known stability factor multiplied by the noise level $\sg$. Figure~\ref{stabilityplots2} plots the noise stability factor $\xi(\rr)$ for binary trees, for each of the two stepsize schemes considered ($\kappa=1.1$ for NITP). For both ITP and NITP, given any value of $\rr$ for which the stability factors derived in this paper are defined, they are always lower than the corresponding stability factors derived from analysis of IHT based upon the standard RIP~\cite{HTP}; see \cite[Section 2.4]{thesis} for a comparison. 

Comparing Figure~\ref{stabilityplots2} with Figure~\ref{stabilityplots}, we also observe a significant quantitative improvement in the stability factors for both algorithm variants compared with those achieved by means of tree-based RIP, in the case of binary trees. It should be pointed out that we have obtained improved stability results by imposing additional restrictions upon the noise, namely that the noise is Gaussian distributed and independent of the signal and measurement matrix. This assumption is in keeping with our aim of exploiting average-case assumptions. Our analysis could, however, be altered to deal with the case of non-independent noise by making more use of the RIP, though this would lead to larger stability constants.

\subsection{Extension to tree compressible signals}

While we have assumed so far in this paper that signals are exactly $k$-tree sparse, it is more realistic to expect that signals are \textit{tree compressible}, meaning that they are well approximated by a $k$-tree sparse vector. An important consideration for any compressed sensing recovery analysis is, therefore, whether it can be extended to the tree compressible case. From the point of view of worst-case analysis, a difference emerges in this respect between standard and tree-based compressed sensing. In the case of standard compressed sensing, the extension to compressible signals can be achieved using the RIP, which can be used to bound the amplification factor of the signal tail~\cite{cosamp}. However, it was argued in~\cite{modelbased} that the RIP is not sufficient to control this amplification factor for more general structured sparsity models (including the tree-based model). This deficit was partially addressed by the introduction of the Restricted Amplification Property (RAmP), and the extension to model-compressible signals was established provided the sparsity model has a certain `nested' property~\cite{modelbased}, which unfortunately is not the case for the rooted tree model.

On the other hand, the stable point approach in which we consider independent Gaussian noise is much more amenable to the analysis of the tree-compressible case. In~\cite[Chapter 7]{thesis}, the main results of the present paper are extended to the tree-compressible case. More precisely, the assumption that $\xs$ is $k$-tree sparse is relaxed, and $\xs_k$ is defined to be the closest $k$-tree sparse approximation to $\xs$, namely $\xs_k:=\mP_k(\xs)$. Defining $\Lm^k$ to be the support of this optimal tree-sparse approximation, that is $\Lm^k:=\mbox{supp}(\xs_k)$, a measure of unrecoverable energy, $\Sg$, is defined to be
$$\Sg:=\sg+\|\xs_{\Lm^k}\|,$$
which represents the combined inaccuracy due to both measurement noise and signal model violation. It is shown in~\cite[Theorems 7.23 and 7.29]{thesis} that, beneath the same oversampling thresholds given in Theorems~\ref{recov1noise_tree} and~\ref{recov2noise_tree} of the present paper, the approximation error of the output of ITP/NITP amplifies the unrecoverable energy by no more than some (different) stability factor. See~\cite[Chapter 7]{thesis} for an explicit quantification of the stability factor in this case.

The observation that controlling stability to noise in tree-based compressed sensing is alleviated by switching to average-case assumptions is not new, see for example~\cite{best_kterm,indyk_price,approx_tol}.

\section{Concluding remarks and future directions}

We have introduced a simplified proportional-growth asymptotic framework, and used it to quantify recovery guarantees for ITP algorithms. Recovery guarantees in terms of tree-based RIP have also been obtained for tree-based CoSaMP~\cite{modelbased}, while recovery guarantees based on the standard notion of RIP for other greedy algorithms including Conjugate Gradient Iterative Hard Thresholding~\cite{cgiht}, Subspace Pursuit~\cite{sp} and Orthogonal Matching Pursuit~\cite{foucart_rauhut} could also be translated into the tree-based framework. Our asymptotic framework could equally well be used to quantify the existing RIP-based recovery guarantees for these algorithms and for Gaussian matrices.

Our focus in this paper has been on the tree-based model, but other variants of the model-based compressed sensing paradigm are possible, including block sparsity~\cite{modelbased}. Many of the arguments we have presented for the tree-based model would apply equally to other union-of-subspaces model. We leave the extension of our analysis to other models as future work.

The calculation of an exact tree projection is computationally burdensome, and the more recently proposed approximate ITP algorithm~\cite{approx_alg} is attractive in practice. An interesting direction for future work is to extend our results by obtaining quantified oversampling thresholds for approximate ITP algorithms. 

\appendix

\section{Proofs for the tree-based RIP analysis}\label{RIP_proofs}

We begin with a brief roadmap of the proofs found in the following appendices. The current appendix gives proofs for the tree-based RIP analysis, and Appendix~\ref{stable_proofs} gives proofs for the stable point analysis. In both cases, we first obtain recovery conditions for deterministic matrices (in Sections~\ref{RIP_det} and~\ref{stable_det} respectively). We then perform a probabilistic analysis of these conditions for Gaussian matrices in the simplified proportional-growth asymptotic (in Sections~\ref{RIP_gauss} and~\ref{stable_gauss} respectively). In both cases, the analysis for Gaussian matrices relies on large deviations results for certain quantities related to Gaussian matrices (including bounds on tree-based RIP constants). These large deviations results, which extend to the tree-based setting those given originally in~\cite{lqphase,stablepoint}, are stated and proved in Appendix~\ref{large_dev}.

\subsection{Deterministic recovery conditions}\label{RIP_det}

The following lemma gives some further consequences of the tree-based RIP.
 
\begin{lem}[\textbf{Consequences of the tree-based RIP}]\label{RIP_lemma} Given some positive integer $s$, suppose that $A\in\RR^{n\times N}$ has lower and upper tree-based RIP constants $TL_s$ and $TU_s$ respectively, as defined in (\ref{treeRIPdef}). Let $\Omega\in\mathcal{T}_s$, and let $\Omega=\Omega_1\cup\Omega_2$ where $|\Omega_1|=s_1$, $|\Omega_2|=s_2$ and $s=s_1+s_2$. Then
\begin{equation}\label{transpose_inequality}
\|A^T_{\Omega}y\|\le\sqrt{1+TU_s}\|y\|\;\;\;\;\mbox{for all}\;y\in\RR^n;
\end{equation}
\begin{equation}\label{gramm_inequality}
(1-TL_s)\|x\|\le\|A^T_{\Omega}A_{\Omega}x\|\le(1+TU_s)\|x\|\;\;\;\;\mbox{for all}\;x\in\RR^s;
\end{equation}
\begin{equation}\label{inverse_inequality}
\frac{1}{1+TU_s}\|x\|\le\|(A^T_{\Omega}A_{\Omega})^{-1}x\|\le\frac{1}{1-TL_s}\|x\|\;\;\;\;\mbox{for all}\;x\in\RR^s;
\end{equation}
\begin{equation}\label{pseudo_inequality}
\|A^{\dag}_{\Omega}y\|\le\frac{1}{\sqrt{1-TL_s}}\|y\|\;\;\;\;\mbox{for all}\;y\in\RR^n,\;\;\mbox{provided $A^{\dag}_{\Omega}$ is well-defined};
\end{equation}
\begin{equation}\label{distinct_inequality}
\|A^T_{\Omega_1}A_{\Omega_2}z\|\le\frac{1}{2}(TL_s+TU_s)\|z\|\;\;\;\;\mbox{for all}\;z\in\RR^{s_2};
\end{equation}
\begin{equation}\label{max_inequality}
\|(I-\omega A^T_{\Omega}A_{\Omega})x\|\le\max\{\omega(1+TU_s)-1,1-\omega(1-TL_s)\}\|x\|\;\;\;\;\mbox{for all}\;x\in\RR^s\;\mbox{and all}\;\omega>0.
\end{equation}
\end{lem}

\textbf{Proof:} All the above results were proved for the standard notion of RIP in~\cite[Lemma 15]{greedy_tech}. The results extend trivially by restricting all support sets to rooted trees.\hfill$\Box$\\
\\
Next, by largely following the analysis in~\cite{HTP}, we use the tree-based RIP to obtain a result for generic ITP with bounded stepsize. 

\begin{lem}[\textbf{Iteration invariant for bounded stepsize}]\label{generic_lemma}
Consider Problem~\ref{treenoiseprob}. Let the stepsizes of generic ITP satisfy 
\begin{equation}\label{step_limits}
\au\le\al^m\le\ao
\end{equation}
for all $m\geq 0$. Then
\begin{equation}\label{iter_invariant}
\|x^{m+1}-\xs\|\le\sqrt{3}\max\{\ao(1+TU_{3k})-1,1-\au(1-TL_{3k})\}\|x^m-\xs\|+\ao\sqrt{3(1+TU_{2k})}\|e\|.
\end{equation}
\end{lem}
\textbf{Proof:} Let us write $y^m:=x^m+\al^m A^T(b-Ax^m)$, which can be rearranged to give
\begin{equation}\label{am_formula}
y^m=x^m+\al^m A^T(A\xs+e-Ax^m)=\xs+(I-\al^mA^T A)(x^m-\xs)+\al^m A^T e.
\end{equation}
Let $\Gm^m=\mbox{supp}(x^m)$, let $\Gm^{m+1}=\mbox{supp}(x^{m+1})$ and let us further define
\begin{equation}\label{omega_def}
\Omega=\Lm\cup\Gm^m\cup\Gm^{m+1},
\end{equation} 
where $\Lm$ is defined in (\ref{lambda_def_tree}). By (\ref{Pk_proj}), we have
$$\|y^m_{\Lm}\|^2\le\|y^m_{\Gm^{m+1}}\|^2,$$
which cancels to give
\begin{equation}\label{am_proj}
\|y^m_{\Lm\sm\Gm^{m+1}}\|^2\le\|y^m_{\Gm^{m+1}\sm\Lm}\|^2.
\end{equation}
Substituting (\ref{am_formula}) into (\ref{am_proj}) gives
$$\begin{array}{l}
\left\|\left\{\xs+(I-\al^mA^T A)(x^m-\xs)+\al^m A^T e\right\}_{\Lm\sm\Gm^{m+1}}\right\|\\
\;\;\;\;\;\;\;\;\le\left\|\left\{\xs+(I-\al^mA^T A)(x^m-\xs)+\al^m A^T e\right\}_{\Gm^{m+1}\sm\Lm}\right\|,\end{array}$$
and the triangle inequality, along with $\xs_{\Gm^{m+1}\sm\Lm}=0$, implies
\begin{eqnarray}\label{am_ineq}
\left\|\xs_{\Lm\sm\Gm^{m+1}}\right\|-\left\|\left\{(I-\al^mA^T A)(x^m-\xs)+\al^m A^T e\right\}_{\Lm\sm\Gm^{m+1}}\right\|&&\nonumber\\
\;\;\;\;\;\;\;\;\le\left\|\left\{(I-\al^mA^T A)(x^m-\xs)+\al^m A^T e\right\}_{\Gm^{m+1}\sm\Lm}\right\|.&&
\end{eqnarray}
The sets $\Lm\sm\Gm^{m+1}$ and $\Gm^{m+1}\sm\Lm$ are disjoint, and we may therefore apply the Cauchy-Schwarz inequality, namely $(a+b)^2\le\sqrt{2}(a^2+b^2)$, to (\ref{am_ineq}), yielding 
$$\left\|\xs_{\Lm\sm\Gm^{m+1}}\right\|\le\sqrt{2}\left\|\left\{(I-\al^mA^T A)(x^m-\xs)+\al^m A^T e\right\}_{\Lm\cup\Gm^{m+1}}\right\|,$$
from which a further application of the triangle inequality and (\ref{omega_def}) leads us to deduce
\begin{equation}\label{unrecovered}
\left\|\xs_{\Lm\sm\Gm^{m+1}}\right\|\le\sqrt{2}\left\{\left\|(I-\al^m A^T_{\Omega}A_{\Omega})(x^m-\xs)_{\Omega}\right\|+\al^m\left\|A^T_{\Lm\cup\Gm^{m+1}}e\right\|\right\}.
\end{equation}
Meanwhile, splitting on $\Gm^{m+1}$ and $\Lm\sm\Gm^{m+1}$, and using the definition of $\Gm^{m+1}=\mbox{supp}(x^{m+1})$,
$$\begin{array}{rcl}
\|x^{m+1}-\xs\|^2&=&\left\|(x^{m+1}-\xs)_{\Gm^{m+1}}\right\|^2+\left\|(x^{m+1}-\xs)_{\Lm\sm\Gm^{m+1}}\right\|^2\\
&=&\left\|\left\{(I-\al^mA^T A)(x^m-\xs)+\al^m A^T e\right\}_{\Gm^{m+1}}\right\|^2+\left\|\xs_{\Lm\sm\Gm^{m+1}}\right\|^2,
\end{array}$$
where the second inequality follows from (\ref{am_formula}). We then apply the triangle inequality and (\ref{omega_def}) to deduce
\begin{eqnarray}\label{overall}
\|x^{m+1}-\xs\|^2&\le&\left\{\left\|\left\{(I-\al^mA^T A)(x^m-\xs)\right\}_{\Gm^{m+1}}\right\|+\left\|\left\{\al^m A^T e\right\}_{\Gm^{m+1}}\right\|\right\}^2+\left\|\xs_{\Lm\sm\Gm^{m+1}}\right\|^2\nonumber\\
&\le&\left\{\left\|(I-\al^mA^T_{\Omega} A_{\Omega})(x^m-\xs)_{\Omega}\right\|+\al^m\left\|A^T_{\Lm\cup\Gm^{m+1}} e\right\|\right\}^2+\left\|\xs_{\Lm\sm\Gm^{m+1}}\right\|^2.
\end{eqnarray}
Substituting (\ref{unrecovered}) into (\ref{overall}) then gives
\begin{equation}\label{omega_invariant}
\|x^{m+1}-\xs\|^2\le 3\left\{\left\|(I-\al^mA^T_{\Omega} A_{\Omega})(x^m-\xs)_{\Omega}\right\|+\al^m\left\|A^T_{\Lm\cup\Gm^{m+1}} e\right\|\right\}^2.
\end{equation}
Since $|\Omega|\le 3k$ and $|\Lm\cup\Gm^{m+1}|\le 2k$, the result now follows by applying (\ref{transpose_inequality}), (\ref{max_inequality}) and (\ref{step_limits}) to (\ref{omega_invariant}), and taking square roots.\hfill$\Box$\\
$ $

Both the ITP and NITP stepsize schemes have bounded stepsizes: trivially in the case of ITP, and bounds for NITP are given next.

\begin{lem}[\textbf{NITP stepsize bounds}]\label{NITP_stepsize}
Let $\al^m$ be chosen according to Algorithm~\ref{normalisedITP}. Then
\begin{equation}\label{NITP_bound}
\frac{1}{\kappa(1+TU_{2k})}\le\al^m\le\frac{1}{1-TL_k}.
\end{equation}
\end{lem}

\textbf{Proof of Lemma~\ref{NITP_stepsize}:} If (\ref{exactchoice_tree}) is accepted, then $\al^m\le 1/(1-TL_k)$ by (\ref{treeRIPeqn}). On the other hand, if (\ref{exactchoice_tree}) is rejected, the backtracking phase can only reduce the stepsize further, which proves the upper bound in (\ref{NITP_bound}). To prove the lower bound, we also distinguish two cases. If (\ref{exactchoice_tree}) is accepted, then $\al^m\geq 1/(1+TU_k)$ by (\ref{treeRIPeqn}). Since $\kappa>1$, and since $TU_{2k}\geq TU_k$ by the nonincreasing property of tree-based RIP constants, the lower bound in (\ref{NITP_bound}) holds in this case. On the other hand, if (\ref{exactchoice_tree}) is rejected, the penultimate stepsize calculated in the backtracking phase must also have been rejected. Writing $\tilde{\al}^m$ for the penultimate stepsize, since $\tilde{\al}^m$ was rejected, we have
\begin{equation}\label{penultimate_tree}
\tilde{\al}^m\geq(1-c)\frac{\|\tilde{x}^{m+1}-x^m\|^2}{\|A(\tilde{x}^{m+1}-x^m)\|^2}\geq\frac{1-c}{1+TU_{2k}},
\end{equation}
where the last step follows from (\ref{treeRIPeqn}). But $\al^m=\tilde{\al}^m/[\kappa(1-c)]$, which combines with (\ref{penultimate_tree}) to give the lower bound in (\ref{NITP_bound}) in this case also.\hfill$\Box$

We may therefore deduce the following results.

\begin{thm}[\textbf{Iteration invariant for ITP}]\label{CIHT_thm}
Consider Problem~\ref{treenoiseprob}. Then the iterates of ITP with stepsize $\al$ satisfy
\begin{equation}\label{CIHT_invariant}
\|x^{m+1}-\xs\|\le\mu^{ITP_{\al}}\|x^m-\xs\|+\xi^{ITP_{\al}}\|e\|,
\end{equation}
where  $\mu^{ITP_{\al}}$ and $\xi^{ITP_{\al}}$ are defined in (\ref{CIHTmudef}) and (\ref{CIHTxidef}) respectively.
\end{thm}
\textbf{Proof:} For ITP with stepsize $\al$, we have $\au=\al$ and $\ao=\al$, and the result follows by applying Lemma~\ref{generic_lemma}.\hfill$\Box$

\begin{thm}[\textbf{Iteration invariant for NITP}]\label{NIHT_thm}
Consider Problem~\ref{treenoiseprob} and suppose Assumption~\ref{genposass} holds. Then the iterates of NITP with shrinkage parameter $\kappa$ satisfy
\begin{equation}\label{NIHT_invariant}
\|x^{m+1}-\xs\|\le\mu^{NITP_{\kappa}}\|x^m-\xs\|+\xi^{NITP_{\kappa}}\|e\|,
\end{equation}
where $\mu^{ITP_{\al}}$ and $\xi^{ITP_{\al}}$ are defined in (\ref{NIHTmudef}) and (\ref{NIHTxidef}) respectively.
\end{thm}
\textbf{Proof:} For a given $\kappa>1$, the stepsize bounds (\ref{NITP_stepsize}) apply to NITP, and the result follows by applying Lemma~\ref{generic_lemma} with $\au:=1/(1-L_k)$ and $\ao:=1/[\kappa(1+U_{2k})]$.\hfill$\Box$

In order to prove recovery results, we will need the following lemma.

\begin{lem}\label{induction} Suppose there exist $\mu\in[0,1)$ and $\xi>0$ such that the sequence of iterates $\{x^m\}$ satisfies, for each $m\geq 0$,
\begin{equation}\label{gen_invariant}
\|x^{m+1}-\xs\|\le\mu\|x^m-\xs\|+\xi\|e\|.
\end{equation}
Then, for all $m\geq 0$,
\begin{equation}\label{gen_error}
\|\xb-\xs\|\le\mu^m\|\xs\|+\frac{\xi}{1-\mu}\|e\|.
\end{equation}
\end{lem}
\textbf{Proof:} We first prove by induction that, for all $m\geq 0$,
\begin{equation}\label{induction_result}
\|x^{m}-\xs\|\le\mu^m \|\xs\|+\xi\left(\frac{1-\mu^m}{1-\mu}\right)\|e\|.
\end{equation}
Supposing (\ref{induction_result}) holds for some $m\geq 0$, then we may apply (\ref{gen_invariant}) to (\ref{induction_result}) to deduce
$$\begin{array}{rcl}
\|x^{m+1}-\xs\|&\le&\mu\left[\mu^m \|\xs\|+\xi\left(\frac{1-\mu^m}{1-\mu}\right)\|e\|\right]+\xi\|e\|\\
&=&\mu^{m+1} \|\xs\|+\left[\xi\left(1+\frac{\mu-\mu^{m+1}}{1-\mu}\right)\right]\|e\|\\
&=&\mu^{m+1} \|\xs\|+\xi\left(\frac{1-\mu^{m+1}}{1-\mu}\right)\|e\|,
\end{array}$$
and so (\ref{induction_result}) also holds for $m+1$. Since $x^0=0$, the result holds trivially for $m=0$, and therefore for all $m\geq 0$ by induction. Since $\mu^m\in(0,1)$ for all $m\geq 0$, (\ref{gen_error}) now follows.\hfill$\Box$\\

Provided $\mu<1$, the $\mu^m\|\xs\|$ term in (\ref{gen_error}) tends to zero, and the expression $\xi/(1-\mu)$ may be viewed as a \textit{stability factor}, giving a limiting bound on the approximation error as a multiple of the noise level $\|e\|$. We now proceed to the proof of the recovery results for arbitrary matrices. 

\textbf{Proof of Theorem~\ref{ALG_result}}: The result for ITP follows by combining Theorem~\ref{CIHT_thm} and Lemma~\ref{induction}. The result for NITP follows by combining Theorem~\ref{NIHT_thm} and Lemma~\ref{induction}.\hfill$\Box$

Though Theorems~\ref{ALG_result} gives a limiting bound on the approximation error, it does not necessarily imply convergence of the algorithm. In the simplified noiseless case however, both results can be used to deduce convergence to $\xs$ at a linear rate. 

\subsection{Analysis for Gaussian matrices}\label{RIP_gauss}

The next two lemmas are needed to enable a translation of Theorem~\ref{ALG_result} for arbitrary matrices into the asymptotic framework for Gaussian matrices. 

\begin{lem}\label{lemma12} For some $\tau<1$, define the set
$\mathcal{Z}:=(0,\tau)^p\times(0,\infty)^q$ and let
$F:\mathcal{Z}\rightarrow \RR$ be continuously differentiable on
$\mathcal{Z}$. Let $A\in\RR^{n\times N}$ be a Gaussian matrix with tree-based RIP constants $TL_k,\dots,TL_{pk}$ and $TU_k,\dots,TU_{qk}$, and let
$\TL(\rr),\dots,\TL(p\rr)$ and $\TU(\rr),\dots,\TU(q\rr)$ be defined as in Definition~\ref{RIPboundsdef}.
Define $\mathbf{1}$ to be the vector of all ones, and
\begin{align}
z(k,n,N)&:=[TL_k,\dots,TL_{pk},TU_k,\dots,TU_{qk}]\nonumber,\\
z(\rr)&:=[\TL(\rr),\dots,\TL(p\rr),\TU(\rr),\dots,\TU(q\rr)].\nonumber
\end{align}
Suppose, for all $t\in \mathcal{Z}$, $\left(\nabla
    F[t]\right)_i \ge 0$ for all $i=1,\dots,p+q$ and there exists
  $j\in\{1,\dots,p\}$ such that $\left(\nabla F[t]\right)_j > 0$.
  Then, for any $\e\in(0,1)$, in the simplified proportional-growth asymptotic, 
\begin{equation}\label{lemma12_eqn}
\PP\left(F[z(k,n,N)]<F[z((1+\e)\rr)]\right)\ra 1\;\;\;\;\hbox{as}\;\;n\ra\infty,
\end{equation}
exponentially in $n$ on the draw of $A$. Also, $F[z(\rr)]$ is strictly increasing in $\rr$. 
\end{lem}
\textbf{Proof:} A proof was given in~\cite[Lemma 12]{greedy} for the case where $\TL(\cdot)$ and $\TU(\cdot)$ are replaced by $\LL(\dd,\cdot)$ and $\UU(\dd,\cdot)$, bounds on the standard notion of RIP constants given in~\cite{lqphase}. Note first that a function that depends only upon $\rr$ is a trivial special case of a function that depends upon both $\dd$ and $\rr$. Only two assumptions are made in the proof concerning the bounds: first that they are indeed upper bounds, and second that $\LL(\dd,\rr)$ is strictly increasing in $\rr$ and $\UU(\dd,\rr)$ is nondecreasing in $\rr$. The first condition holds in our case by Lemma~\ref{RIPbounds}, and it is straightforward to show that the second property also holds in our case. More precisely, $\TL(\rr)$ and $\TU(\rr)$ are both strictly increasing on $\rr\in(0,1)$. It follows that the argument in~\cite[Lemma 12]{greedy} extends.\hfill$\Box$

\begin{lem}\label{lemma13}
For some $\tau<1$, define the set $\mathcal{Z}:=(0,\tau)^p\times(0,\infty)^q$ and let $F,G,H:\mathcal{Z}\rightarrow \RR$ satisfy the conditions of Lemma~\ref{lemma12}. Suppose that
\begin{equation}\label{muxisplit}
\mu(k,n,N)=\max\big\{F[z(k,n,N)],G[z(k,n,N)]\big\},\;\;\;\;\xi(k,n,N)=H[z(k,n,N)],
\end{equation}
and
\begin{equation}\label{muxidrsplit}
\mu(\rr)=\max\big\{F[z(\rr)],G[z(\rr)]\big\},\;\;\;\;\xi(\rr)=H[z(\rr)].
\end{equation}
Then $\mu(\rr)$ and $\xi(\rr)$ are both strictly increasing in $\rr$ and, for any $\e\in(0,1)$, in the proportional-growth asymptotic, 
\begin{equation}\label{gen_muprob}
\PP\big\{\mu(k,n,N)\geq\mu((1+\e)\rr)\big\}\ra 0,
\end{equation}
and
\begin{equation}\label{gen_xiprob}
\PP\big\{\xi(k,n,N)\geq\xi((1+\e)\rr)\big\}\ra 0,
\end{equation}
both exponentially in $n$. Furthermore, define $\rh$ as the unique solution to $\mu(\rr)=1$, and suppose that 
\begin{equation}\label{gen_rhocond}
\rr<(1-\e)\rh.
\end{equation}
Then
\begin{equation}\label{mdr_below1}
\mu((1+\e)\rr)<1,
\end{equation}
and, in the simplified proportional-growth asymptotic,
\begin{equation}\label{gen_mubelowone2}
\PP\big\{\mu(k,n,N)\geq 1\big\}\ra 0,
\end{equation}
exponentially in $n$. 
\end{lem}

\textbf{Proof:} By assumption, we may apply Lemma~\ref{lemma12} to each of $F(z)$, $G(z)$ and $H(z)$, deducing from (\ref{lemma12_eqn}) that
\begin{eqnarray}
&&\PP\left(F[z(k,n,N)]<F[(1+\e)\rr)]\right)\ra 1\;\;\;\;\hbox{as}\;\;n\ra\infty,\\\label{Fprob}
&&\PP\left(G[z(k,n,N)]<G[z((1+\e)\rr)]\right)\ra 1\;\;\;\;\hbox{as}\;\;n\ra\infty,\\\label{Gprob}
&&\PP\left(H[z(k,n,N)]<H[z((1+\e)\rr)]\right)\ra 1\;\;\;\;\hbox{as}\;\;n\ra\infty,\label{Hprob}
\end{eqnarray}
exponentially in $n$, and that $F[z(\rr)]$, $G[z(\rr)]$ and $H[z(\rr)]$ are each strictly increasing in $\rr$, from which it immediately follows that both $\mu(\rr)$ and $\xi(\rr)$ are also strictly increasing in $\rr$. Combining (\ref{muxisplit}), (\ref{muxidrsplit}), (\ref{Fprob}) and (\ref{Gprob}), we have
\begin{eqnarray}
&&\PP\big\{\mu(k,n,N)\geq\mu((1+\e)\rr)\big\}\nonumber\\
&=&\PP\Big\{\max\big\{F[z(k,n,N)],G[z(k,n,N)]\big\}\geq\max\big\{F[z((1+\e)\rr)],G[z((1+\e)\rr)]\big\}\Big\}\nonumber\\
&\le&\PP\big\{F[z(k,n,N)]\geq F[z((1+\e)\rr)]\big\}+\PP\big\{G[z(k,n,N)]\geq G[z((1+\e)\rr)]\big\}\nonumber\\
&&\ra 0\;\;\mbox{as}\;\;n\ra\infty,\label{muprob}
\end{eqnarray}
and therefore (\ref{gen_muprob}) holds. Meanwhile, combining (\ref{muxisplit}), (\ref{muxidrsplit}) and (\ref{Hprob}) immediately yields (\ref{gen_xiprob}). Now suppose (\ref{gen_rhocond}) holds. Since $1-\e<(1+\e)^{-1}$ for any $\e\in(0,1)$, (\ref{gen_rhocond}) implies that
\begin{equation}\label{epsinvert}
(1+\e)\rr<\rh,
\end{equation}
Since $\mu(\rr)$ is strictly increasing in $\rr$, it follows from (\ref{epsinvert}) and the definition of $\rh$ that
$$\mu((1+\e)\rr)<\mu(\rh)=1,$$
which proves (\ref{mdr_below1}), and from which it also follows that
$$\PP\big\{\mu(k,n,N)\geq 1\big\}\le\PP\big\{\mu(k,n,N)\geq\mu((1+\e)\rr)\big\},$$
to which we may apply (\ref{muprob}) to deduce (\ref{gen_mubelowone2}).\hfill$\Box$\\
\\
We now proceed to the proofs of the main results.\\
\\
\textbf{Proof of Theorem~\ref{ALG_RIPphase} for ITP:} Select $\e\in(0,1)$, fix $\tau<1$ and let
$$z(k,n,N):=[TL_{3k},TU_{2k},TU_{3k}]\;\;\;\;\mbox{and}\;\;\;\;z(\rr):=[\TL(3\rr),\TU(2\rr),\TU(3\rr)].$$
Define $\mZ:=(0,\tau)\times(0,\infty)^2$, and define the functions $F_{\al}(z),G_{\al}(z),H_{\al}(z):\mZ\ra\RR$ as
\begin{eqnarray}
F_{\al}(z)&=&F_{\al}(z_1,z_2,z_3):=\sqrt{3}[\al(1+z_3)-1],\\
G_{\al}(z)&=&G_{\al}(z_1,z_2,z_3):=\sqrt{3}[1-\al(1-z_1)],\\
H_{\al}(z)&=&H_{\al}(z_1,z_2,z_3):=\al\sqrt{3(1+z_2)},
\end{eqnarray}
noting that
$$\mu^{ITP_{\al}}=\max\big\{F_{\al}[z(k,n,N)],G_{\al}[z(k,n,N)]\big\},\;\;\;\;\xi^{ITP_{\al}}=H_{\al}[z(k,n,N)],$$
where $\mu^{ITP_{\al}}$ and $\xi^{ITP_{\al}}$ are defined in (\ref{CIHTmudef}) and (\ref{CIHTxidef}) respectively, and
$$\mu^{ITP_{\al}}(\rr)=\max\big\{F_{\al}[z(\rr)],G_{\al}[z(\rr)]\big\},\;\;\;\;\xi^{ITP_{\al}}(\rr)=H_{\al}[z(\rr)],$$
where $\mu^{ITP_{\al}}(\rr)$ and $\xi^{ITP_{\al}}(\rr)$ are defined in (\ref{CIHTmudrdef}) and (\ref{CIHTxidrdef}) respectively. Now $F_{\al}(z)$, $G_{\al}(z)$ and $H_{\al}(z)$ are continuously differentiable and nondecreasing in $(z_1,z_2,z_3)\in\mZ$, and strictly increasing in $z_3$, $z_1$ and $z_2$ respectively due to $\al>0$, and therefore each satisfies the conditions of Lemma~\ref{lemma12}. We may therefore apply Lemma~\ref{lemma13}, deducing
\begin{equation}\label{CIHT_muprob}
\PP\big\{\mu^{ITP_{\al}}\geq\mu^{ITP_{\al}}_{RIP}((1+\e)\rr)\big\}\ra 0,
\end{equation}
and
\begin{equation}\label{CIHT_xiprob}
\PP\big\{\xi^{ITP_{\al}}\geq\xi^{ITP_{\al}}_{RIP}((1+\e)\rr)\big\}\ra 0,
\end{equation}
exponentially in $n$, and furthermore that $\mu^{ITP_{\al}}_{RIP}(\rr)$ and $\xi^{ITP_{\al}}_{RIP}(\rr)$ are both strictly increasing in $\rr$, from which it follows that $\rh^{ITP_{\al}}_{RIP}$ is unique. Since (\ref{CIHTrhocond}) holds, we may also use Lemma~\ref{lemma13} to deduce (\ref{mdr_IHT}), and furthermore that
\begin{equation}\label{CIHT_mubelowone2}
\PP\big\{\mu^{ITP_{\al}}\geq 1\big\}\ra 0,
\end{equation}
exponentially in $n$, and we may apply Theorem~\ref{ALG_result} to deduce (\ref{ALGerror}) with probability tending to $1$ exponentially in $n$. Since $\mu^{ITP_{\al}}_{RIP}(\rr)$ and $\xi^{ITP_{\al}}_{RIP}(\rr)$ are strictly increasing in $\rr$, (\ref{CIHTdrerror}) now follows from (\ref{ALGerror}), (\ref{CIHT_muprob}) and (\ref{CIHT_xiprob}).\hfill$\Box$\\

\textbf{Proof of Corollary~\ref{ALG_RIPphase_noiseless} for ITP:} Since we consider Problem~\ref{treesparseprob}, we have $e:=0$. Provided (\ref{CIHTrhocond}) holds, we can apply Theorem~\ref{ALG_RIPphase} with $e:=0$, deducing that, for any $m\geq 0$,
$$\|x^m-\xs\|\le\left(\mu^{ITP_{\al}}_{RIP}(\dd,(1+\e)\rr)\right)^m\|\xs\|,$$
where
$$\mu^{ITP_{\al}}_{RIP}((1+\e)\rr)<1,$$
and so we have convergence to $\xs$ with convergence rate $\mu^{ITP_{\al}}_{RIP}((1+\e)\rr)$.\hfill$\Box$\\

\textbf{Proof of Theorem~\ref{ALG_RIPphase} for NITP:} Select $\e\in(0,1)$, fix $\tau<1$ and let
$$z(k,n,N):=[TL_k,TL_{3k},TU_{2k},TU_{3k}]\;\;\;\;\mbox{and}\;\;\;\;z(\rr):=[\TL(\rr),\TL(3\rr),\TU(2\rr),\TU(3\rr)].$$
Define $\mZ:=(0,\tau)^2\times(0,\infty)^2$, and define the functions $F_{\al}(z),G_{\al}(z),H_{\al}(z):\mZ\ra\RR$ as
\begin{eqnarray}
F_{\kappa}(z)&=&F_{\kappa}(z_1,z_2,z_3,z_4):=\sqrt{3}\left[\frac{1+z_4}{1-z_1}-1\right],\nonumber\\
G_{\kappa}(z)&=&G_{\kappa}(z_1,z_2,z_3,z_4):=\sqrt{3}\left[1-\frac{1-z_2}{\kappa(1+z_3)}\right],\nonumber\\
H_{\kappa}(z)&=&H_{\kappa}(z_1,z_2,z_3,z_4):=\frac{\sqrt{3(1+z_3)}}{1-z_1},\nonumber
\end{eqnarray}
noting that
$$\mu^{NITP_{\kappa}}=\max\big\{F_{\kappa}[z(k,n,N)],G_{\kappa}[z(k,n,N)]\big\},\;\;\;\;\xi^{NITP_{\kappa}}=H_{\kappa}[z(k,n,N)],$$
where $\mu^{NITP_{\kappa}}$ and $\xi^{NITP_{\kappa}}$ are defined in (\ref{NIHTmudef}) and (\ref{NIHTxidef}) respectively, and
$$\mu^{NITP_{\kappa}}_{RIP}(\rr)=\max\big\{F_{\kappa}[z(\dd,\rr)],G_{\kappa}[z(\rr)]\big\},\;\;\;\;\xi^{NITP_{\kappa}}_{RIP}(\rr)=H_{\kappa}[z(\rr)],$$
where $\mu^{NITP_{\kappa}}_{RIP}(\rr)$ and $\xi^{NITP_{\kappa}}_{RIP}(\rr)$ are defined in (\ref{NIHTmudrdef}) and (\ref{NIHTxidrdef}) respectively. Now $F_{\kappa}(z)$, $G_{\kappa}(z)$ and $H_{\kappa}(z)$ are continuously differentiable and nondecreasing in $(z_1,z_2,z_3,z_4)$, and strictly increasing componentwise in $(z_1,z_4)$, $(z_2,z_3)$ and $(z_1,z_3)$ respectively, and therefore each satisfies the conditions of Lemma~\ref{lemma12}. We may therefore apply Lemma~\ref{lemma13}, deducing
\begin{equation}\label{NIHT_muprob}
\PP\big\{\mu^{NITP_{\kappa}}\geq\mu^{NITP_{\kappa}}_{RIP}((1+\e)\rr)\big\}\ra 0
\end{equation}
and
\begin{equation}\label{NIHT_xiprob}
\PP\big\{\xi^{NITP_{\kappa}}\geq\xi^{NITP_{\kappa}}_{RIP}((1+\e)\rr)\big\}\ra 0,
\end{equation}
exponentially in $n$, and furthermore that $\mu^{NITP_{\kappa}}_{RIP}(\rr)$ and $\xi^{NITP_{\kappa}}_{RIP}(\rr)$ are both strictly increasing in $\rr$, from which it follows that $\rh^{NITP_{\kappa}}_{RIP}$ is unique. Since (\ref{CIHTrhocond}) holds, we may also use Lemma~\ref{lemma13} to deduce (\ref{mdr_IHT}), and furthermore that
\begin{equation}\label{NIHT_mubelowone2}
\PP\big\{\mu^{NITP_{\kappa}}\geq 1\big\}\ra 0,
\end{equation}
exponentially in $n$, and we may apply Theorem~\ref{ALG_result} to deduce (\ref{ALGerror}) with probability tending to $1$ exponentially in $n$. Since $\mu^{NITP_{\kappa}}_{RIP}(\rr)$ and $\xi^{NIHT_{\kappa}}_{RIP}(\rr)$ are strictly increasing in $\rr$, (\ref{CIHTdrerror}) now follows from (\ref{ALGerror}), (\ref{NIHT_muprob}) and (\ref{NIHT_xiprob}).\hfill$\Box$\\
\\
\textbf{Proof of Corollary~\ref{ALG_RIPphase_noiseless} for NITP:} Since we consider Problem~\ref{treesparseprob}, we have $e:=0$. Provided (\ref{CIHTrhocond}) holds, we can apply Theorem~\ref{ALG_RIPphase} with $e:=0$, deducing that, for any $m\geq 0$,
$$\|x^m-\xs\|\le\left(\mu^{NITP_{\kappa}}_{RIP}((1+\e)\rr)\right)^m\|\xs\|,$$
where
$$\mu^{NIHT_{\kappa}}_{RIP}((1+\e)\rr)<1,$$
and so we have convergence to $\xs$ with convergence rate $\mu^{NITP_{\kappa}}_{RIP}((1+\e)\rr)$.\hfill$\Box$

\section{Proofs for the tree-based stable point analysis}\label{stable_proofs}

\subsection{Analysis for deterministic matrices}\label{stable_det}

We will follow the approach first introduced by the present authors in~\cite{stablepoint}, central to which is the concept of an $\au$-\textit{stable point}, defined in Definition~\ref{stable_ITP}.  

We will analyse the stable points of generic ITP, and our final goal is to prove quantitative conditions that guarantee that all stable points of the algorithm are `close' to the original signal $x$, in the context of Gaussian matrices. Provided we also have guaranteed convergence to some stable point, we may then conclude that ITP outputs a good approximation to $x$. For this reason, the results derived in this section come in two parts: a necessary condition for there to be a stable point on some support $\Gm$, and conditions guaranteeing convergence to some stable point for our two stepsize schemes.

\subsubsection{A necessary condition for the existence of a stable point}\label{stableanal}

Any $\au$-stable point of generic ITP may also be characterized as a minimum-norm solution on some $k$-subspace. 

\begin{lem}\label{pseudo}
Suppose Assumption~\ref{genposass} holds and suppose $\xb$ is an $\au$-stable point of generic ITP on $\Gm$ for some $\au>0$. Then $\xb_{\Gm}=A_{\Gm}^{\dag}b$.
\end{lem}

\textbf{Proof}: It follows from (\ref{stable1}) that
 $A^T_{\Gm}(b-A_{\Gm}\xb_{\Gm})=0$ where $\mbox{supp}(\xb)\subseteq\Gm$ and $|\Gm|=k$. Under Assumption~\ref{genposass}, the pseudoinverse $A_{\Gm}^{\dag}$ is well-defined and we may rearrange to give $\xb_{\Gm}=A_{\Gm}^{\dag}b$.\hfill$\Box$\\

While this lemma tells us that any stable point is necessarily a minimum-norm solution on some $k$-subspace, the converse may not hold. We next prove Theorem~\ref{stableprop}, which gives a necessary condition for a stable point on a given support.\\
\\
\textbf{Proof of Theorem~\ref{stableprop}:} Supposing that $\xb$ is an $\au$-stable point on $\Gm$, choosing $\Omega:=\Lm$ in (\ref{stable2}) yields
$$\|\xb_{\Gm\sm\Lm}\|^2\geq\au^2\|A^T_{\Lm\sm\Gm}(b-A\xb)\|^2.$$
We may now follow the argument of~\cite[Theorem 3.2]{stablepoint} to deduce (\ref{stablecond}).\hfill$\Box$\\

\subsubsection{Conditions guaranteeing convergence}\label{conv}

In addition to the result of the previous section, in order to show recovery of $x^{\ast}$, we must also show that ITP converges to an $\au$-stable point. In this section we derive convergence conditions for generic ITP used in conjunction with the two stepsize schemes introduced in Section~\ref{stepsize}. A sufficient condition for convergence of generic ITP is given next.

\begin{lem}[\textbf{Sufficient condition for convergence}]\label{convlem_ITP}
Consider Problem~\ref{treenoiseprob}. Suppose Assumption~\ref{genposass} holds, and suppose the iterates of generic ITP satisfy
\begin{equation}\label{convsuff}
\|x^{m+1}-x^m\|^2\le c\left[\Psi(x^m)-\Psi(x^{m+1})\right]\;\;\;\;\mbox{for all}\;m\geq 0,
\end{equation}
for some $c>0$ which does not depend upon $m$, where $\Psi(\cdot)$ is defined in (\ref{psidef}). Assume that there exist $\ao\geq \au>0$ such that
\begin{equation}\label{lbalpham}
\ao\geq \alpha^m\geq \au\;\;\;\;\mbox{for all}\;m\geq 0.
\end{equation}
Then  $x^m\ra\xb$ as $m\ra\infty$, where $\xb$ is an $\au$-stable point of generic ITP.
\end{lem}

\textbf{Proof:} We may follow the proof of \cite[Lemma 3.5]{stablepoint} to deduce that $x^m\rightarrow\xb$ , where $\xb_{\Gm}=A_{\Gm}^{\dag}b$ and $\xb_{\Gm^C}=0$, for some $\Gm$ such that $|\Gm|=k$. The proof still holds since all that is assumed about the hard threshold projection $\mH_k(\cdot)$ is that it preserves the value of selected coefficients, a property which is also shared by the tree projection $\mathcal{P}_k(\cdot)$ by (\ref{treeproj_def2}). Since $\Gm=\Gm^m$ for some $m\geq 0$, it follows that, in the case of ITP, $\Gm\in\Tk$. Therefore (\ref{stable1}) holds for $\xb$.\\
It remains to establish that $\xb$ satisfies (\ref{stable2}). Defining
\begin{equation}\label{gamma1}
\Gm_1=\{i\in\Gm\;:\;\xb_i\neq 0\},
\end{equation}
it follows that $\Gm_1\subseteq\Gm^m$ for all $m$ sufficiently large. It follows from (\ref{treeproj_def2}) that, for any $\Omega\in\Tk$,
$$\|x^{m+1}_{\Gm^{m+1}}\|^2\geq\|\{x^m-\al^m g^m\}_{\Omega}\|^2,\;\;\mbox{for all}\;\;m\geq 0.$$
and therefore, for all $m$ sufficiently large,
$$\|x^{m+1}_{\Gm_1}\|^2+\|x^{m+1}_{\Gm^{m+1}\sm\Gm_1}\|^2\geq\|x^{m+1}_{\Omega\cap\Gm_1}\|^2+\|\{x^m-\al^m g^m\}_{\Omega\sm\Gm_1}\|^2,$$
which cancels to
\begin{equation}\label{optimal_choice}
\|x^{m+1}_{\Gm_1\sm\Omega}\|^2+\|x^{m+1}_{\Gm^{m+1}\sm\Gm_1}\|^2\geq\|\{x^m+\al^m g^m\}_{\Omega\sm\Gm_1}\|^2.
\end{equation}
Furthermore, it follows from (\ref{gamma1}) that 
\begin{equation}\label{xcomp}
\|x^{m+1}_{\Gm^{m+1}\sm\Gm_1}\|^2\rightarrow 0.
\end{equation}
By (\ref{lbalpham}), there exists a convergent subsequence of stepsizes,
\begin{equation}\label{subsequence}
\al^{m_r}\rightarrow\tilde{\al}\geq\au\;\;\mbox{as}\;\;r\ra\infty
\end{equation}
Passing to the limit in (\ref{optimal_choice}) on the subsequence $m_r$ for which (\ref{subsequence}) holds, we deduce that $\displaystyle\|\xb_{\Gm_1\sm\Omega}\|\geq\au\displaystyle\|\{A^T(b-A\xb)\}_{\Omega\sm\Gm_1}\|$, from which it follows trivially that
\begin{equation}\label{convresult}
\|\xb_{\Gm\sm\Omega}\|\geq\au\|\{A^T(b-A\xb)\}_{\Omega\sm\Gm}\|.
\end{equation}
Since (\ref{convresult}) holds for any $\Omega\in\Tk$, $\xb$ satisfies (\ref{stable2}), and the result is proved.\hfill$\Box$\\
\\
\textbf{Proof of Theorem~\ref{CITPconv}:} We may follow the proof of \cite[Theorem 3.6]{stablepoint}, replacing $U_{2k}$ with $TU_{2k}$, to deduce that (\ref{convsuff}) holds with  $c:=2\al/[1-\al(1+TU_{2k})]$. Due to (\ref{CITPconvcond}), (\ref{lbalpham}) trivially holds with $\ao=\au=\al$. Thus  Lemma~\ref{convlem_ITP} applies, and the ITP iterates $x^m$ converge to an $\al$-stable point.\hfill$\Box$\\

We next obtain a convergence result for NITP. In this case, there is no explicit requirement for a tree-based RIP condition to be satisfied; however, the tree-based RIP this time appears in the choice of $\au$.

\begin{thm}[\textbf{NITP convergence}]\label{NITPconv}
Suppose Assumption~\ref{genposass} holds. Then NITP with shrinkage parameter $\kappa$ converges to a $[\kappa(1+TU_{2k})]^{-1}$-stable point $\xb$ of generic ITP.
\end{thm}

\textbf{Proof:} By replacing $L_{2k}$ with $TL_{2k}$, the proof given for \cite[Theorem 3.7]{stablepoint} holds.\hfill$\Box$

\subsection{Analysis for Gaussian matrices}\label{stable_gauss}

In this section, we build upon the results for arbitrary matrices in Section~\ref{stable_det} and obtain quantitative oversampling thresholds for ITP algorithms of the form $\rr<\rh$ in the case of Gaussian measurement matrices.

\subsubsection{Proof for ITP}\label{ITP_proof}
 
The present analysis broadly follows the same lines as that in~\cite{stablepoint}, but differs in two respects. First, we switch to using the tree-based tail bounds defined in Section~\ref{large_dev}. Second, since there is now no dependence upon $\dd$, we can prove results in the simplified proportional-growth asymptotic (Definition~\ref{propdimdef2}). The changes are nontrivial, and therefore we present full proofs of the new results. We begin by defining a support set partition.

\begin{defn}[\textbf{Support set partition for ITP}]\label{zeta_def_tree}
Consider Problem~\ref{treenoiseprob} and suppose $\rr\in(0,1/2]$ and $\al>0$. Given $\zeta>0$, let us write
\begin{equation}\label{astardef_tree}
\astart:=a(\rr)+\zeta,
\end{equation}
let us write $\{\Gm_i:i\in\Tk\}$ for the set of all possible support sets which form a rooted tree of cardinality $k$, and let us disjointly partition $\Tk:=\Ta\cup\Tb$ such that
\begin{equation}\label{thetadefn_tree}
\Ta:=\left\{i\in\Tk\;\;:\;\;\|\xs_{\Lm\sm\Gm_i}\|>\sg\cdot\astart\right\}\;\;\;\;\mbox{and}\;\;\;\;\Tb:=\left\{i\in\Tk\;\;:\;\;\|\xs_{\Lm\sm\Gm_i}\|\le\sg\cdot\astart\right\},
\end{equation}
where $\Lm$ is defined in (\ref{lambda_def_tree}).
\end{defn}

The partition in (\ref{thetadefn_tree}) has been defined in such a way that, provided (\ref{ITPstablecond}) holds, an analysis of the stable point condition (\ref{stablecond}) shows that ITP must necessarily converge to some $\al$-stable point on $\Gm_i$ such that $i\in\Tb$, and this is proved in Lemma~\ref{theta1lem}. On the other hand, it is also possible to use the large deviations results of Section~\ref{large_dev} to bound the error in approximating $\xs$ by any $\al$-stable point on $\Gm_i$ such that $i\in\Tb$, which is achieved by Lemma~\ref{theta2lem}. It follows that, for any $\al>0$, all $\al$-stable points have bounded approximation error. Combining these two results, we have convergence to some $\al$-stable point with guaranteed approximation error, provided the conditions in each lemma hold; combining the conditions leads to the oversampling threshold defined in (\ref{rhoITPdef}).

We first show that, asymptotically, there are no $\al$-stable points on any $\Gm_i$ such that $i\in\Ta$, and we write $NSP_{\al}$ for this event.

\begin{lem}\label{theta1lem}
Consider Problem~\ref{treenoiseprob} and choose $\zeta>0$. Suppose Assumptions~\ref{gaussianass},~\ref{independentass} and \ref{noiseass} hold, suppose (\ref{ITPstablecond}) holds, and suppose that $\al$ is chosen to satisfy 
\begin{equation}\label{alphabound1}
\al<\frac{1}{1+\TU(2\rr)}.
\end{equation}
Then, in the proportional-growth asymptotic, ITP converges to an $\al$-stable point supported on some $\Gm_i$ such that $i\in\Theta_2$, with probability tending to $1$ exponentially in $n$.
\end{lem}

\textbf{Proof}: Given (\ref{alphabound1}), we may apply Lemma \ref{RIPbounds} with $\e$ sufficiently small to deduce $\al(1+TU_{2k})<1$, with probability tending to $1$ exponentially in $n$. Under Assumption~\ref{gaussianass}, we may apply Lemma~\ref{CITPconv} and deduce convergence of ITP to an $\al$-stable point. We now show that this stable point must be supported on $\Gm_i$ such that $i\in\Theta_2$. For any $\Gm_i$ such that $i\in\Ta$, we have $\Gm_i\neq\Lm$, and we may therefore combine Theorem~\ref{stableprop} with Lemma~\ref{dist} to deduce that a necessary condition for there to be an $\al$-stable point on $\Gm_i$ is
\begin{equation}\label{stablecond2}
\begin{array}{ll}
\|\xs_{\Lm\sm\Gm_i}\|\cdot\sqrt{F_{\Gm_i}}+\|\xs_{\Gm_i\sm\Lm}\|+\sg\cdot\sqrt{G_{\Gm_i}}\\
\;\;\;\;\;\;\;\geq\au\left[\left(\frac{n-k}{n}\right)\|\xs_{\Lm\sm\Gm_i}\|\cdot R_{\Gm_i}-\sg\sqrt{\frac{k(n-k)}{n^2}(S_{\Gm_i})(T_{\Gm_i})}\right],
\end{array}
\end{equation}
where
$$\begin{array}{c}F_{\Gm_i}\sim\displaystyle\frac{k}{n-k+1}F(k,n-k+1);\;\;\;\;G_{\Gm_i}\sim\displaystyle\frac{k}{n-k+1}F(k,n-k+1);\\
R_{\Gm_i}\sim\displaystyle\frac{1}{n-k}\chi^2_{n-k};\;\;\;\;S_{\Gm_i}\sim\displaystyle\frac{1}{n-k}\chi^2_{n-k};\;\;\;\;T_{\Gm_i}\sim\displaystyle\frac{1}{k}\chi^2_{k}.\end{array}$$
We also have, by (\ref{thetadefn_tree}),
\begin{equation}\label{Sgbound}
\sg\le\frac{\|\xs_{\Lm\sm\Gm_i}\|}{\astar}
\end{equation}
for any $\Gm_i$ such that $i\in\Ta$. Since $\Gm_i\neq\Lm$, $\|\xs_{\Lm\sm\Gm}\|>0$, and substitution of (\ref{Sgbound}) into (\ref{stablecond2}), rearrangement and division by $\|\xs_{\Lm\sm\Gm_i}\|$ yields
$$\astar\left[\al\left(\frac{n-k}{n}\right)\cdot R_{\Gm_i}-\sqrt{F_{\Gm_i}}\right]\le\sqrt{G_{\Gm_i}}+\al\sqrt{\frac{k(n-k)}{n^2}\cdot S_{\Gm_i}\cdot T_{\Gm_i}}.$$
Consequently,
\begin{eqnarray}\label{necrhon}
\PP(\bN)&=&\PP\left\{\cup_{i\in\Theta_1}(\exists\;\mbox{an $\al$-stable point
  supported on}\;\Gm_i)\right\}\nonumber\\
&=&\PP\left\{\bigcup_{i\in\Theta_1}\left[\astar\left[\al\left(1-\rr_n\right)\cdot R_{\Gm_i}-\sqrt{F_{\Gm_i}}\right]\le 1+\sqrt{G_{\Gm_i}}+\al\sqrt{\rr_n(1-\rr_n)(S_{\Gm_i})(T_{\Gm_i})}\right]\right\},\\
&&
\end{eqnarray}
where we write $\rr_n$ for the sequence of values of the ratio $k/n$. For brevity's sake, let us define
\begin{equation}\label{fdef}
\Phi[\rr,F,G,R,S,T]:=1+\sqrt{G}+\au\sqrt{\rr(1-\rr)(S)(T)}-\astar\cdot\left[\au(1-\rr)\cdot R-\sqrt{F}\right],
\end{equation}
so that (\ref{necrhon}) may equivalently be written as
\begin{equation}\label{necsplit0}
\PP(\bN)=\PP\left\{\cup_{i\in\Theta_1}\left(\Phi[\rr_n,\astar,F_{\Gm_i},G_{\Gm_i},R_{\Gm_i},S_{\Gm_i},T_{\Gm_i}]\geq
  0\right)\right\}.
\end{equation}
Given some $\e>0$, we now define
\begin{equation}\label{astdef}
F^{\ast}=G^{\ast}:=\TIF(\rr)+\e;\;\;\;\;\;R^{\ast}:=1-\TIL(\rr,1-\rr)-\e;\;\;\;\;\;S^{\ast}:=1+\TIU(\rr,1-\rr)+\e;\;\;\;\;\;T^{\ast}:=1+\TIU(\rr,\rr)+\e,
\end{equation}
Using (\ref{astdef}), we deduce from (\ref{necsplit0}) that
\begin{eqnarray}
&&\PP(\bN)\nonumber\\
&\le&\PP\left\{\cup_{i\in\Theta_1}\left(\Phi[\rr_n,F_{\Gm_i},G_{\Gm_i},R_{\Gm_i},S_{\Gm_i},T_{\Gm_i}]\geq \Phi[\rr_n,F^{\ast},G^{\ast},R^{\ast},S^{\ast},T^{\ast}]\right)\right\}\label{necsplitnoise1}\\
&+&\PP\left\{\Phi[\rr_n,F^{\ast},G^{\ast},R^{\ast},S^{\ast},T^{\ast}]\geq \Phi[\rr,F^{\ast},G^{\ast},R^{\ast},S^{\ast},T^{\ast}]+\e\right\}\label{necsplitnoise2}\\
&+&\PP\left\{\Phi[\rr,F^{\ast},G^{\ast},R^{\ast},S^{\ast},T^{\ast}]+\e\geq
  0\right\}\label{necsplitnoise3},
\end{eqnarray}
since the event in the right-hand side of (\ref{necsplit0}) lies in the union of the three
events in (\ref{necsplitnoise1}), (\ref{necsplitnoise2}) and
(\ref{necsplitnoise3}). Now (\ref{necsplitnoise3}) is a deterministic event, and $\astar$ has been defined in such a way that, for any $\zeta>0$, provided $\e$ is taken sufficiently small, the event has probability
$0$. This follows from (\ref{ITPstablecond}), (\ref{adefn_tree}), \ref{astardef_tree}, and by the continuity of $\Phi$. The event (\ref{necsplitnoise2}) is also deterministic, and by continuity and since $\rr_n\rightarrow\rr$, it follows that there exists some $\tilde{n}$ such that
$$\PP\left\{\Phi[\rr_n,F^{\ast},G^{\ast},R^{\ast},S^{\ast},T^{\ast}]\geq \Phi[\rr,F^{\ast},G^{\ast},R^{\ast},S^{\ast},T^{\ast}]+\e\right\}=0\;\;\;\;\mbox{for all}\;\;n\geq\tilde{n}.$$
Taking limits as $n\rightarrow\infty$, the terms (\ref{necsplitnoise2}) and
(\ref{necsplitnoise3}) are zero, leaving only (\ref{necsplitnoise1}), and we have
\begin{eqnarray}
&&\lim_{n\rightarrow\infty}\PP(\bN)\nonumber\\
&\le&\lim_{n\rightarrow\infty}\PP\left\{\cup_{i\in\Theta_1}\left(\Phi[\rr_n,\astar,F_{\Gm_i},G_{\Gm_i},R_{\Gm_i},S_{\Gm_i},T_{\Gm_i}]\geq \Phi[\rr_n,\astar,F^{\ast},G^{\ast},R^{\ast},S^{\ast},T^{\ast}]\right)\right\}\nonumber\\
&\le&\lim_{n\rightarrow\infty}\PP\left\{\cup_{i\in\Theta_1}(F_{\Gm_i}\geq F^{\ast})\right\}+\lim_{n\rightarrow\infty}\PP\left\{\cup_{i\in\Theta_1}(G_{\Gm_i}\geq G^{\ast})\right\}+\lim_{n\rightarrow\infty}\PP\left\{\cup_{i\in\Theta_1}(R_{\Gm_i}\le R^{\ast})\right\}\nonumber\\
&+&\lim_{n\rightarrow\infty}\PP\left\{\cup_{i\in\Theta_1}(S_{\Gm_i}\geq S^{\ast})\right\}+\lim_{n\rightarrow\infty}\PP\left\{\cup_{i\in\Theta_1}(T_{\Gm_i}\geq T^{\ast})\right\},\label{neclemnoise}
\end{eqnarray}
where the last line follows from the monotonicity of $\Phi$ with respect to $F$, $G$, $R$, $S$ and $T$. Since
  $\Theta_1\subseteq\Tk$, we may apply
  Lemmas~\ref{chisq_tree} and~\ref{Fdist_tree} to (\ref{neclemnoise}), and since $\Theta_1$ and $\Theta_2$ partition $\Tk$, the result follows.\hfill$\Box$\\

Next we show that all $\al$-stable points supported on some $\Gm_i\in\Theta_2$ have bounded approximation error. 

\begin{lem}\label{theta2lem}
Suppose Assumptions~\ref{gaussianass},~\ref{independentass} and~\ref{noiseass} hold, suppose that (\ref{ITPstablecond}) holds, and suppose that $\al$ is chosen to satisfy (\ref{alphabound1}). There exists $\zeta$ sufficiently small such that, in the proportional-growth asymptotic, any $\al$-stable point $\xb$ of ITP on $\Gm_i$ such that $i\in\Theta_2$ satisfies 
\begin{equation}\label{error}
\|\xb-\xs\|\le\xi(\rr)\cdot\sg,
\end{equation}
where $\xi(\rr)$ is defined in (\ref{xidef_tree}).
\end{lem}

\textbf{Proof}: Suppose $\xb$ is a minimum-norm solution on $\Gm$, so that $\xb_{\Gm}=A_{\Gm}^{\dag}b$ and $\xb_{\Gm^C}=0$. Then, using $A_{\Gm}^{\dag}A_{\Gm}=I$, we have
\begin{eqnarray}
(\xb-\xs)_{\Gm}&=&A_{\Gm}^{\dag}(A_{\Gm}\xs_{\Gm}+A_{\Gm^C}\xs_{\Gm^C}+e)-\xs_{\Gm}\nonumber\\
&=&\xs_{\Gm}+A_{\Gm}^{\dag}(A_{\Lm\sm\Gm}\xs_{\Lm\sm\Gm}+A_{(\Lm\cup\Gm)^C}\xs_{(\Lm\cup\Gm)^C}+e)-\xs_{\Gm}\nonumber\\
&=&A_{\Gm}^{\dag}(A_{\Lm\sm\Gm}\xs_{\Lm\sm\Gm}+e)+\xs_{\Gm}-\xs_{\Gm}\nonumber\\
&=&A_{\Gm}^{\dag}(A_{\Lm\sm\Gm}\xs_{\Lm\sm\Gm}+e),\label{errorgam}
\end{eqnarray}
while
\begin{equation}\label{errorgamC}
(\xb-\xs)_{\Gm^C}=-\xs_{\Gm^C}.
\end{equation}
Combining (\ref{errorgam}) and (\ref{errorgamC}) using the triangle inequality, we may bound
\begin{eqnarray}
\|\xb-\xs\|^2&\le&\|(\xb-\xs)_{\Gm}\|^2+\|(\xb-\xs)_{\Gm^C}\|^2\nonumber\\
&=&\|A_{\Gm}^{\dag}(A_{\Lm\sm\Gm}\xs_{\Lm\sm\Gm}+e)\|^2+\|\xs_{\Gm^C}\|^2\nonumber\\
&\le&\left[\|A_{\Gm}^{\dag}A_{\Lm\sm\Gm}\xs_{\Lm\sm\Gm}\|+\|A_{\Gm}^{\dag}e\|\right]^2+\|\xs_{\Lm\sm\Gm}\|^2+\|\xs_{(\Lm\cup\Gm)^C}\|^2.\label{errorbound1}
\end{eqnarray}
We may deduce, by (\ref{PQresult}) of Lemma~\ref{dist},
\begin{equation}\label{errorbound2}
\|A_{\Gm}^{\dag}A_{\Lm\sm\Gm}\xs_{\Lm\sm\Gm}\|^2=\|\xs_{\Lm\sm\Gm}\|^2\cdot P_{\Gm},\;\;\;\mbox{where}\;\;\;P_{\Gm}\sim\frac{k}{n-k+1}F(k,n-k+1),
\end{equation}
and by (\ref{lhsnoise}) of Lemma~\ref{dist},
\begin{equation}\label{errorbound3}
\|A_{\Gm}^{\dag}e\|^2=\sg^2\cdot Q_{\Gm},\;\;\;\mbox{where}\;\;\;Q_{\Gm}\sim\frac{k}{n-k+1}F(k,n-k+1).
\end{equation}
Substituting (\ref{errorbound2}) and (\ref{errorbound3})
into (\ref{errorbound1}), we have
\begin{equation}\label{PQ_bound}
\|\xb-\xs\|^2\le\left[\|\xs_{\Lm\sm\Gm}\|\cdot\sqrt{P_{\Gm}}+\sg\cdot\sqrt{Q_{\Gm}}\right]^2+\|\xs_{\Lm\sm\Gm}\|^2,
\end{equation}
and we may use (\ref{thetadefn_tree}) to further deduce
\begin{eqnarray}
\|\xb-\xs\|^2&\le&\sg^2\left[\astar\cdot\sqrt{P_{\Gm}}+\sqrt{Q_{\Gm}}\right]^2+\left[\astar\right]^2\cdot\sg^2\nonumber\\
&=&\sg^2\left\{\left[\astar\cdot\sqrt{P_{\Gm}}+\sqrt{Q_{\Gm}}\right]^2+\left[\astar\right]^2\right\}.\label{errorbound}
\end{eqnarray}
For the sake of brevity, let us define
\begin{equation}\label{psidef2}
\Psi(P,Q):=\sqrt{\left(\astar\cdot\sqrt{P}+\sqrt{Q}\right)^2+[\astar]^2},
\end{equation}
so that (\ref{errorbound}) may equivalently be written as
\begin{equation}\label{errorbound_brevity}
\|\xb-\xs\|\le\sg\cdot\Psi\left[P_{\Gm},Q_{\Gm}\right].
\end{equation}
Given $\zeta>0$, let us define
\begin{equation}\label{PQdef}
P^{\ast}=Q^{\ast}:=\TIF(\rr)+\zeta.
\end{equation} 
Now we use (\ref{errorbound_brevity}) to perform a union bound over all $\Gm_i$ such that $i\in\Theta_2$, writing $\xb_i$ for the minimum-norm solution supported on $\Gm_i$, giving
\begin{eqnarray}
&&\PP\left\{\exists\;\mbox{some}\;\Gm_i\;\mbox{such
    that}\;i\in\Theta_2\;\mbox{and}\;\|\xb_i-\xs\|\geq\sg\cdot\Psi[P^{\ast},Q^{\ast}]\right\}\nonumber\\
&=&\PP\left\{\bigcup_{i\in\Theta_2}\left(\|\xb_i-\xs\|\geq\sg\cdot\Psi[P^{\ast},Q^{\ast}]\right)\right\}\label{2necsplit0}\\
&\le&\PP\left\{\bigcup_{i\in\Theta_2}\left(\|\xb_i-\xs\|\geq\sg\cdot\Psi[P_{\Gm_i},Q_{\Gm_i}]\right)\right\}\label{2necsplit1}\\
&+&\PP\left\{\bigcup_{i\in\Theta_2}\left(\sg\cdot\Psi[P_{\Gm_i},Q_{\Gm_i}]\geq\sg\cdot\Psi[P^{\ast},Q^{\ast}]\right)\right\},\label{2necsplit2}
\end{eqnarray}
since the event in (\ref{2necsplit0}) lies in the union of the two
events in (\ref{2necsplit1}) and (\ref{2necsplit2}). It is an immediate consequence of
(\ref{errorbound}) that the event in (\ref{2necsplit1}) has probability
$0$. Taking limits of (\ref{2necsplit2}) as $n\rightarrow\infty$, we have
\begin{eqnarray}
&&\lim_{n\rightarrow\infty}\PP\left\{\exists\;\mbox{some}\;\Gm_i\;\mbox{such
    that}\;i\in\Theta_2\;\mbox{and}\;\|\xb_i-\xs\|\geq\sg\cdot\Psi[P^{\ast},Q^{\ast}]\right\}\nonumber\\
&\le&\lim_{n\rightarrow\infty}\PP\left\{\bigcup_{i\in\Theta_2}\left(\sg\cdot\Psi[P_{\Gm_i},Q_{\Gm_i}]\geq\sg\cdot\Psi[P^{\ast},Q^{\ast}]\right)\right\}\nonumber\\
&\le&\lim_{n\rightarrow\infty}\PP\left\{\cup_{i\in\Theta_2}(P_{\Gm_i}\geq P^{\ast})\right\}+\lim_{n\rightarrow\infty}\PP\left\{\cup_{i\in\Theta_2}(Q_{\Gm_i}\geq Q^{\ast})\right\},\label{2neclem}
\end{eqnarray}
where we used the monotonicity of $\Psi$ with respect to $P$ and $Q$ in the last line. Since
  $\Theta_2\subseteq\Tk$, and using (\ref{errorbound2}) and (\ref{errorbound3}), we may apply
  Lemma~\ref{Fdist_tree} to (\ref{2neclem}), yielding that each of the limits in the right-hand side of (\ref{2neclem}) converges to zero exponentially in $n$, and so finally
$$\lim_{n\rightarrow\infty}\PP\left\{\exists\;\mbox{some}\;\Gm_i\;\mbox{such
    that}\;i\in\Tb\;\mbox{and}\;\|\xb_i-\xs\|>\sg\cdot\Psi\left[\astar,P^{\ast},Q^{\ast}\right]\right\}=0,$$
exponentially in $n$. Since, by Lemma~\ref{pseudo}, any stable point is necessarily a minimum-norm solution, and recalling the definition of $\astar$ in (\ref{astardef_tree}), $\Psi(P,Q)$ in (\ref{psidef2}), and the definitions of $P^{\ast}$ and $Q^{\ast}$ in (\ref{PQdef}), we have 
$$\lim_{n\rightarrow\infty}\PP\left\{\exists\;\mbox{some $\al$-stable point $\xb_i$ on}\;\Gm_i\;\mbox{such
    that}\;i\in\Tb\;\mbox{and}\;\|\xb_i-\xs\|\geq\sg\cdot\Psi[P^{\ast},Q^{\ast}]\right\}=0,$$
with convergence exponential in $n$. Finally, by continuity, 
$$\|\xb_i-x\|>\sg\sqrt{\TIF(\rr)[1+a(\rr)]^2+[a(\rr)]^2}$$
$$\Longrightarrow\|\xb_i-\xs\|\geq\sg\sqrt{\TIF(\rr)[1+a(\rr)+\zeta]^2+[a(\rr)+\zeta]^2},$$
for some $\zeta$ suitably small, and the result now follows from the definition of $\xi^{ITP_{\al}}_{SP}(\rr)$ in (\ref{xidef_tree}).\hfill$\Box$\\
\\
It is now straightforward to prove the two main results for ITP.\\
\\
\textbf{Proof of Theorem~\ref{recov1noise_tree}}: By Lemma~\ref{theta1lem}, we have convergence to an $\al$-stable point supported on some $\Gm_i$ such that $i\in\Theta_2$, to which we can apply Lemma~\ref{theta2lem} deducing (\ref{error}) with probability tending to $1$ exponentially in $n$.\hfill$\Box$\\
\\
\textbf{Proof of Corollary~\ref{recov1noiseless_tree}}: The result follows by setting $\sg:=0$ in Theorem~\ref{recov1noise_tree}.\hfill$\Box$

\subsubsection{Proof for NITP}

In the case of NITP, it is possible to prove convergence to an $\au(\rr;\e)$-stable point, where
\begin{equation}\label{au_eps}
\au(\rr;\e):=\{\kappa[1+\TU(2\rr)+\e]\}^{-1},
\end{equation}
for some $\e>0$. 

The proof of Theorem~\ref{recov2noise_tree} for NITP takes broadly the same approach as for the corresponding result for ITP in Section~\ref{ITP_proof}. However, in order to finally eliminate the dependence upon $\e$ in $\au(\rr;\e)$, the results corresponding to Lemmas~\ref{theta1lem} and~\ref{convlem_ITP} for ITP need to be combined together. This is accomplished by Lemma~\ref{theta1lemNIHT}, which establishes that, provided (\ref{NITPstablecond}) holds and $\e$ is taken sufficiently small, NITP converges to an $\au(\rr;\e)$-stable point on some $\Gm_i$ such that $i\in\Tb$ (the NITP support set partition is given in (\ref{thetadefn2}) below). Lemma~\ref{theta2lemNIHT} corresponds to Lemma~\ref{theta2lem} for ITP, giving bounds on the approximation error of an $\au(\rr;\e)$-stable point on some $\Gm_i$ such that $i\in\Tb$, for any $\e>0$. Combining the two lemmas leads us to conclude that NITP converges to some limit point with bounded approximation error. We write $NSP_{\au}$ for the event that there is no $\au(\rr;\e)$-stable point on any $\Gm_i$ such that $i\in\Ta$.

We next introduce the support set partition definition relevant for NITP.

\begin{defn}[\textbf{Support set partition for NITP}]\label{zeta_def2}
Suppose $\rr\in(0,1/2]$. Given $\zeta>0$, let us write
\begin{equation}\label{astardef2}
\astar:=a(\rr)+\zeta,
\end{equation}
where $a(\rr)$ is defined in (\ref{adefn2_tree}), let us write $\{\Gm_i:i\in S_n\}$ for the set of all possible support sets of cardinality $k$, and let us disjointly partition $S_n:=\Ta\cup\Tb$ such that
\begin{equation}\label{thetadefn2}
\Ta:=\left\{i\in S_n\;\;:\;\;\|\xs_{\Lm\sm\Gm_i}\|>\Sg\cdot\astar\right\};\;\;\;\;\Tb:=\left\{i\in S_n\;\;:\;\;\|\xs_{\Lm\sm\Gm_i}\|\le\Sg\cdot\astar\right\}.
\end{equation}
\end{defn}

\begin{lem}\label{theta1lemNIHT}
Choose $\zeta>0$. Suppose Assumptions~\ref{gaussianass}, \ref{independentass} and \ref{noiseass} hold, and suppose that (\ref{NITPstablecond}) holds.
Then there exists $\e$ such that, in the proportional-growth asymptotic, NITP converges to an $\au(\rr;\e)$-stable point on some $\Gm_i$ such that $i\in\Tb$, with probability tending to $1$ exponentially in $n$.
\end{lem}

\textbf{Proof:} Under Assumption~\ref{gaussianass}, we have by Theorem~\ref{NITPconv} convergence of NITP to a $[\kappa(1+TU_{2k})]^{-1}$-stable point. By Definition~\ref{stable_ITP}, for any $\al_1<\al_2$, the set of $\al_1$-stable points includes the set of $\al_2$-stable points, and this observation combines with Lemma~\ref{RIPbounds} to imply convergence to an $\au(\rr;\e)$-stable point, where $\au(\rr;\e)$ is defined in (\ref{au_eps}), with probability tending to $1$ exponentially in $n$. We now rehearse the argument of Lemma~\ref{theta1lem} to show that, provided $\e$ is taken sufficiently small, this stable point must be on $\Gm_i$ such that $i\in\Tb$. For any $\Gm_i$ such that $i\in\Ta$, we have $\Gm_i\neq\Lm$, and we may therefore use Theorem~\ref{stableprop} and Lemma~\ref{dist} with $\Gm:=\Gm_i$ to deduce that, given some $\e>0$, a necessary condition for there to be an $\au(\rr;\e)$-stable point on $\Gm_i$ is
\begin{equation}\label{stablecond2_NIHT}
\begin{array}{l}
\|\xs_{\Lm\sm\Gm_i}\|\cdot\sqrt{F_{\Gm_i}}+\sg\cdot\sqrt{G_{\Gm_i}}\\
\;\;\;\;\;\;\;\;\geq\au(\rr;\e)\left[\left(\frac{n-k}{n}\right)\|\xs_{\Lm\sm\Gm_i}\|\cdot R_{\Gm_i}-\sg\cdot\sqrt{\frac{k(n-k)}{n^2}\cdot S_{\Gm_i}\cdot T_{\Gm_i}}\right],\end{array}
\end{equation}
where
$$\begin{array}{c}F_{\Gm_i}\sim\displaystyle\frac{k}{n-k+1}\mathcal{F}(k,n-k+1);\;\;\;\;G_{\Gm_i}\sim\displaystyle\frac{k}{n-k+1}\mathcal{F}(k,n-k+1);\\
R_{\Gm_i}\sim\displaystyle\frac{1}{n-k}\chi^2_{n-k};\;\;\;\;S_{\Gm_i}\sim\displaystyle\frac{1}{n-k}\chi^2_{n-k};\;\;\;\;T_{\Gm_i}\sim\displaystyle\frac{1}{k}\chi^2_{k}.\end{array}$$
We also have, by (\ref{thetadefn2}),
\begin{equation}\label{Sgbound2}
\sg\le\frac{\|\xs_{\Lm\sm\Gm_i}\|}{\astar}
\end{equation}
for any $\Gm_i$ such that $i\in\Ta$. Since $\Gm_i\neq\Lm$, $\|\xs_{\Lm\sm\Gm}\|>0$, and substitution of (\ref{Sgbound2}) into (\ref{stablecond2_NIHT}), rearrangement and division by $\|\xs_{\Lm\sm\Gm_i}\|$ yields
$$\astar\left[\au(\rr;\e)\left(\frac{n-k}{n}\right)\cdot R_{\Gm_i}-\sqrt{F_{\Gm_i}}\right]\le\sqrt{G_{\Gm_i}}+\au(\rr;\e)\sqrt{\frac{k(n-k)}{n^2}\cdot S_{\Gm_i}\cdot T_{\Gm_i}},$$
and consequently
\begin{eqnarray}\label{necsplit0_NIHT}
\PP(\bNN)&=&\PP\left\{\cup_{i\in\Ta}(\exists\;\mbox{an $\au(\rr;\e)$-stable point
  supported on}\;\Gm_i)\right\}\nonumber\\
&\le&\PP\left\{\cup_{i\in\Ta}\left(\Phi[\rr_n,F_{\Gm_i},G_{\Gm_i},R_{\Gm_i},S_{\Gm_i},T_{\Gm_i}]\geq
  0\right)\right\},
\end{eqnarray}
where we write $\rr_n$ for the sequence of values of the ratio $k/n$, and where
\begin{equation}\label{fdef2}
\Phi[\rr,F,G,R,S,T]:=\sqrt{G}+\au(\rr;\e)\sqrt{\rr(1-\rr)(S)(T)}-\astar\cdot\left[\au(\rr;\e)(1-\rr)\cdot R-\sqrt{F}\right].
\end{equation}
We now define
\begin{equation}\label{astdef2}
\begin{array}{c}
F^{\ast}=G^{\ast}:=\TIF(\rr)+\e;\;\;\;\;\;R^{\ast}:= 1-\TIL(\rr,1-\rr)-\e;\\
S^{\ast}:= 1+\TIU(\rr,1-\rr)+\e;\;\;\;\;\;T^{\ast}:= 1+\TIU(\rr,\rr)+\e.\end{array}
\end{equation}
Using (\ref{astdef2}), we deduce from (\ref{necsplit0_NIHT}) that
\begin{eqnarray}
&&\PP(\bNN)\nonumber\\
&\le&\PP\left\{\cup_{i\in\Ta}\left(\Phi[\rr_n,F_{\Gm_i},G_{\Gm_i},R_{\Gm_i},S_{\Gm_i},T_{\Gm_i}]\geq \Phi[\rr_n,F^{\ast},G^{\ast},R^{\ast},S^{\ast},T^{\ast}]\right)\right\}\label{necsplitnoise1_NIHT}\\
&+&\PP\left\{\Phi[\rr_n,F^{\ast},G^{\ast},R^{\ast},S^{\ast},T^{\ast}]\geq \Phi[\rr,F^{\ast},G^{\ast},R^{\ast},S^{\ast},T^{\ast}]+\e\right\}\label{necsplitnoise2_NIHT}\\
&+&\PP\left\{\Phi[\rr,F^{\ast},G^{\ast},R^{\ast},S^{\ast},T^{\ast}]+\e\geq
0\right\}\label{necsplitnoise3_NIHT},
\end{eqnarray}
since the event in (\ref{necsplit0_NIHT}) lies in the union of the three
events in (\ref{necsplitnoise1_NIHT}), (\ref{necsplitnoise2_NIHT}) and
(\ref{necsplitnoise3_NIHT}). Now (\ref{necsplitnoise3_NIHT}) is a deterministic event, and $\astar$ has been defined in such a way that, for any $\zeta>0$, provided $\e$ is taken sufficiently small, the event has probability
$0$. This follows from (\ref{NITPstablecond}), (\ref{adefn2_tree}), (\ref{astardef2}), and by the continuity of $\Phi$. The event (\ref{necsplitnoise2_NIHT}) is also deterministic, and by continuity and since $\rr_n\rightarrow\rr$, it follows that there exists some $\tilde{n}$ such that
$$\PP\left\{\Phi[\rr_n,F^{\ast},G^{\ast},R^{\ast},S^{\ast},T^{\ast}]\geq \Phi[\rr,F^{\ast},G^{\ast},R^{\ast},S^{\ast},T^{\ast}]+\e\right\}=0\;\;\;\;\mbox{for all}\;\;n\geq\tilde{n}.$$
Taking limits as $n\rightarrow\infty$, the terms (\ref{necsplitnoise2_NIHT}) and
(\ref{necsplitnoise3_NIHT}) are zero, leaving only (\ref{necsplitnoise1_NIHT}), and we have
\begin{eqnarray}
&&\lim_{n\rightarrow\infty}\PP(\bNN)\nonumber\\
&\le&\lim_{n\rightarrow\infty}\PP\left\{\cup_{i\in\Ta}\left(\Phi[\rr_n,F_{\Gm_i},G_{\Gm_i},R_{\Gm_i},S_{\Gm_i},T_{\Gm_i}]\geq \Phi[\rr_n,F^{\ast},G^{\ast},R^{\ast},S^{\ast},T^{\ast}]\right)\right\}\nonumber\\
&\le&\lim_{n\rightarrow\infty}\PP\left\{\cup_{i\in\Ta}(F_{\Gm_i}\geq F^{\ast})\right\}+\lim_{n\rightarrow\infty}\PP\left\{\cup_{i\in\Ta}(G_{\Gm_i}\geq G^{\ast})\right\}+\lim_{n\rightarrow\infty}\PP\left\{\cup_{i\in\Ta}(R_{\Gm_i}\le R^{\ast})\right\}\nonumber\\
&+&\lim_{n\rightarrow\infty}\PP\left\{\cup_{i\in\Ta}(S_{\Gm_i}\geq S^{\ast})\right\}+\lim_{n\rightarrow\infty}\PP\left\{\cup_{i\in\Ta}(T_{\Gm_i}\geq T^{\ast})\right\},\label{neclemnoise_NIHT}
\end{eqnarray}
where the last line follows from the monotonicity of $\Phi$ with respect to $F$, $G$, $R$, $S$ and $T$. Since
  $\Ta\subseteq S_n$, we may apply
  Lemmas~\ref{chisq_tree} and~\ref{Fdist_tree} to (\ref{neclemnoise_NIHT}), and we deduce $\PP(\bNN)\ra 0$ as $n\ra\infty$, exponentially in $n$, as required.\hfill$\Box$\\
\\
\begin{lem}\label{theta2lemNIHT}
Suppose Assumptions~\ref{gaussianass}, \ref{independentass} and \ref{noiseass} hold, and suppose that (\ref{NITPstablecond}) holds. Given any $\e>0$, there exists $\zeta$ sufficiently small such that, in the proportional-growth asymptotic, any $\au(\rr;\e)$-stable point on $\Gm_i$ such that $i\in\Tb$ satisfies (\ref{error4}), with probability tending to $1$ exponentially in $n$.
\end{lem}

\textbf{Proof:} Suppose $\xb$ is a minimum-norm solution on $\Gm$, so that $\xb_{\Gm}=A_{\Gm}^{\dag}b$. Then we may follow the argument of Lemma~\ref{theta2lem} to deduce (\ref{PQ_bound}), where
\begin{equation}\label{dist_def_NIHT}
P_{\Gm}\sim\frac{k}{n-k+1}\mathcal{F}(k,n-k+1);\;\;\;\;Q_{\Gm}\sim\frac{k}{n-k+1}\mathcal{F}(k,n-k+1).
\end{equation}
Combining (\ref{PQ_bound}) with (\ref{thetadefn2}), we may further deduce
\begin{eqnarray}
\|\xb-\xs\|^2&\le&\sg^2\left[\astar\cdot\sqrt{P_{\Gm}}+\sqrt{Q_{\Gm}}\right]^2+\left[\astar\right]^2\cdot\sg^2\nonumber\\
&=&\sg^2\left\{\left[\astar\cdot\sqrt{P_{\Gm}}+\sqrt{Q_{\Gm}}\right]^2+\left[\astar\right]^2\right\}.\label{errorbound_NIHT}
\end{eqnarray}
For the sake of brevity, let us define
\begin{equation}\label{psi_def2}
\Psi[P,Q]:=\sqrt{\left(\astar\cdot\sqrt{P}+\sqrt{Q}\right)^2+\astar^2},
\end{equation}
so that (\ref{errorbound_NIHT}) may equivalently be written as
\begin{equation}\label{errorbound_brevity2}
\|\xb-\xs\|\le\sg\cdot\Psi\left[P_{\Gm},Q_{\Gm}\right].
\end{equation}
First suppose that $\sg>0$. Given $\zeta>0$, let us define
\begin{equation}\label{PQdef2}
P^{\ast}=Q^{\ast}:=\TIF(\rr)+\zeta.
\end{equation}
Now we use (\ref{errorbound_brevity2}) to perform a union bound over all $\Gm_i$ such that $i\in\Tb$, writing $\xb_i$ for the minimum-norm solution on $\Gm_i$, giving
\begin{eqnarray}
&&\PP\left\{\exists\;\mbox{some}\;\Gm_i\;\mbox{such
    that}\;i\in\Tb\;\mbox{and}\;\|\xb_i-\xs\|>\sg\cdot\Psi\left[P^{\ast},Q^{\ast}\right]\right\}\nonumber\\
&=&\PP\left\{\bigcup_{i\in\Tb}\left(\|\xb_i-\xs\|>\sg\cdot\Psi\left[P^{\ast},Q^{\ast}\right]\right)\right\}\label{2necsplit0_NIHT}\\
&\le&\PP\left\{\bigcup_{i\in\Tb}\left(\|\xb_i-\xs\|>\sg\cdot\Psi\left[P_{\Gm_i},Q_{\Gm_i}\right]\right)\right\}\label{2necsplit1_NIHT}\\
&+&\PP\left\{\bigcup_{i\in\Tb}\left(\sg\cdot\Psi\left[P_{\Gm_i},Q_{\Gm_i}\right]\geq\sg\cdot\Psi\left[P^{\ast},Q^{\ast}\right]\right)\right\},\nonumber\\
&&\;\label{2necsplit2_NIHT}
\end{eqnarray}
since the event in (\ref{2necsplit0_NIHT}) lies in the union of the two
events in (\ref{2necsplit1_NIHT}) and (\ref{2necsplit2_NIHT}). It is an immediate consequence of
(\ref{errorbound_brevity2}) that the event in (\ref{2necsplit1_NIHT}) has probability
$0$. Taking limits of (\ref{2necsplit2_NIHT}) as $n\rightarrow\infty$, and cancelling $\sg$, we have
\begin{eqnarray}
&&\lim_{n\rightarrow\infty}\PP\left\{\exists\;\mbox{some}\;\Gm_i\;\mbox{such
    that}\;i\in\Tb\;\mbox{and}\;\|\xb_i-\xs\|>\sg\cdot\Psi\left[P^{\ast},Q^{\ast}\right]\right\}\nonumber\\
&\le&\lim_{n\rightarrow\infty}\PP\left\{\bigcup_{i\in\Tb}\left(\Psi\left[P_{\Gm_i},Q_{\Gm_i}\right]\geq\Psi\left[P^{\ast},Q^{\ast}\right]\right)\right\}\nonumber\\
&\le&\lim_{n\rightarrow\infty}\PP\left\{\cup_{i\in\Tb}(P_{\Gm_i}\geq P^{\ast})\right\}+\lim_{n\rightarrow\infty}\PP\left\{\cup_{i\in\Tb}(Q_{\Gm_i}\geq Q^{\ast})\right\},\label{2neclem_NIHT}
\end{eqnarray}
where we used the monotonicity of $\Psi$ with respect to $P$ and $Q$ in the last line. Since
  $\Tb\subseteq S_n$, and using (\ref{dist_def_NIHT}), we may apply
  Lemma~\ref{Fdist_tree} to (\ref{2neclem_NIHT}), yielding that each of the limits in the right-hand side of (\ref{2neclem_NIHT}) converges to zero exponentially in $n$, and so finally
$$\lim_{n\rightarrow\infty}\PP\left\{\exists\;\mbox{some}\;\Gm_i\;\mbox{such
    that}\;i\in\Tb\;\mbox{and}\;\|\xb_i-\xs\|>\sg\cdot\Psi\left[P^{\ast},Q^{\ast}\right]\right\}=0,$$
with convergence at a rate exponential in $n$ also by
    Lemma~\ref{Fdist_tree}. The same result also holds when $\sg=0$ by (\ref{errorbound_NIHT}). Since by Lemma~\ref{pseudo}, any stable point is necessarily a minimum-norm solution, and recalling the definition of $\Psi(P,Q)$ in (\ref{psi_def2}), and the definitions of $P^{\ast}$, $Q^{\ast}$ in (\ref{PQdef2}), we have
\begin{equation}\label{finalbound2}
\lim_{n\rightarrow\infty}\PP\left\{\begin{array}{ll}\exists\;\mbox{some $\au$-stable point $\xb_i$ on}\;\Gm_i\;\mbox{such
    that}\;i\in\Tb\;\mbox{and}\\
\;\|\xb_i-\xs\|>\sg\sqrt{\TIF(\rr)\left[1+a(\rr)+\zeta\right]^2+\left[a(\rr)+\zeta\right]^2}\end{array}
\right\}=0,
\end{equation}
with convergence exponential in $n$. Finally, by continuity,
$$\begin{array}{l}\|\xb_i-\xs\|>\sg\sqrt{\TIF(\rr)\left[1+a(\rr)\right]^2+1+\left[a(\rr)\right]^2}\\
\;\;\;\;\;\;\;\;\;\;\;\;\;\Longrightarrow\|\xb_i-\xs\|>\sg\sqrt{\TIF(\rr)\left[1+a(\rr)+\zeta\right]^2+\left[a(\rr)+\zeta\right]^2},\end{array}$$
for some $\zeta$ suitably small, and the result now follows from the definition of $\xi(\rr)$ in (\ref{xidef2_tree}).\hfill$\Box$\\
\\
It is now straightforward to prove the two main results for NITP.\\
\\
{\bf Proof of Theorem~\ref{recov2noise_tree}:} By Lemma~\ref{theta1lemNIHT}, there exists $\e>0$ such that N-IHT converges to an $\au(\dd,\rr;\e)$-stable point on some $\Gm_i$ such that $i\in\Tb$, and for this choice of $\e$, we can apply Lemma~\ref{theta2lemNIHT} to deduce the result.\hfill$\Box$\\
\\
{\bf Proof of Corollary~\ref{recov2noiseless_tree}:} The result follows by setting $\sg:= 0$ in Theorem~\ref{recov2noise_tree}.\hfill$\Box$

\section{Large deviations results in the tree-based setting}\label{large_dev}

This appendix develops large deviation bounds in the simplified proportional-growth asymptotic of Definition~\ref{propdimdef2} for various quantities related to Gaussian matrices, which are required to hold for all permissible support sets. 

In what follows, let the tree order $d$ to be some fixed integer with $d\geq 2$. We need to count $|\Tk|$, the number of permissible support sets in the $d$-ary tree-based framework, which is bounded above by $T(k)$, the total number of ordered, rooted $d$-ary trees of cardinality $k$. Recalling Lemma~\ref{tree_count}, we have 
$$T(k)=\frac{1}{(d-1)k+1}\binom{dk}{k}.$$

A similar result was proved in~\cite[Proposition 1]{modelbased} for the case of binary trees ($d=2$), though the result given above represents a generalization to any $d>2$, and in fact also gives a tightening of the result in~\cite{modelbased} in the case where $\log_2(N)>k$. Note also that we have an upper bound on $|\Tk|$ which is independent of $N$. This is in contrast to the total number of supports, which is $\binom{N}{k}$. However, $|\Tk|$ may not attain this upper bound if additional structure is imposed. In a typical wavelet tree model, for example, the root node has only $d-1$ children~\cite{exact}). In addition, the number of levels in a wavelet tree structure is typically limited to $J=\log_d(N)$, which represents a further restriction if $\log_d(N)<k$. It follows that, while it is possible to give an upper bound on $|\Tk|$ which is valid for any $N$, $|\Tk|$ does in general depend on both $k$ and $N$.

We will make use of the following limiting result for $T(k)$.

\begin{lem}[\textbf{Tree counting limit}]\label{treelimlem}
\begin{equation}\label{treelim}
\displaystyle\lim_{k\ra\infty}\frac{1}{k}\ln T(k)=d\cdot H(d^{-1}),
\end{equation}
where $H(\cdot)$ is defined in (\ref{shannon_def}).
\end{lem}

\textbf{Proof}:
$$\begin{array}{rcl}
\displaystyle\lim_{k\ra\infty}\frac{1}{k}\ln T(k)&=&\displaystyle\lim_{k\ra\infty}\frac{1}{k}\ln\left[\frac{1}{(d-1)k+1}\binom{dk}{k}\right]\\
&=&\displaystyle\lim_{k\ra\infty}\frac{1}{k}\ln\left[\frac{1}{(d-1)k+1}\right]+\displaystyle\lim_{k\ra\infty}\frac{1}{k}\ln\binom{dk}{k}\\
&=&0+\displaystyle\lim_{k\ra\infty}d\cdot\frac{1}{dk}\ln\binom{dk}{k}\\
&=&d\cdot H(d^{-1}),\end{array}$$
where the last step follows from Stirling's formula.\hfill$\Box$\\
\\
We proceed to proving the validity of the bounds on tree-based RIP constants for Gaussian matrices given in Definition~\ref{treeRIPdef}.\\
\\
\textbf{Proof of Lemma~\ref{RIPbounds}:} We may follow the proof of~\cite[Proposition 2.6]{lqphase}, replacing $\UU(\dd,\rr)$ with $\TU(\rr)$, and replacing $\lambda^{max}(\dd,\rr)$ with $\lambda^{max}(\rr)$, obtaining $$\PP\left[TU_k\geq\TU(\rr_n)+\e\right]\le 2|\Tk|\left[\lm^{max}(\rr_n)+\e\right]g_{max}\left[k,n;\lm^{max}(\rr_n)+\e\right],$$ 
and we may furthermore apply~\cite[Lemma 2.5]{lqphase} to give
\begin{equation}\label{beforelimits}
\PP\left[TU_k\geq\TU(\rr_n)+\e\right]\le 2|\Tk|\left[\lm^{max}(\rr_n)+\e\right]p_{max}\left[n,\lm^{max}(\rr_n)+\e\right]\exp\left\{n\cdot\psi_{max}\left(\lm^{max}(\rr_n)+\e,\rr\right)\right\},
\end{equation}
where $p_{max}(n,\lm)$ is  a polynomial in $n$ and $\lm$. Now we may take limits of both sides of (\ref{beforelimits}), using (\ref{treelim}), to deduce
$$\lim_{n\ra\infty}\frac{1}{n}\ln\PP\left[TU_k\geq\TU(\rr_n)+\e\right]\le d\rr\cdot H(d^{-1})+\psi_{max}\left(\lm^{max}(\rr)+\e,\rr\right),$$
which is strictly negative by Definition~\ref{RIPboundsdef}, from which (\ref{TU_result}) now follows using the same argument as in the proof of lemma~\ref{chisq_tree}. An analogous argument may be followed to prove (\ref{TL_result}) which we omit for the sake of brevity.\hfill$\Box$\\
\\
The large deviations bounds for the $\chi^2$ and $\FF$ distributions follow.\\
\\
\textbf{Proof of Lemma~\ref{chisq_tree}}: By~\cite[Lemma A.2]{stablepoint}, we have for all $i\in S_n$,
\begin{equation}\label{chisqtail}
\lim_{n\ra\infty}\frac{1}{n}\ln\PP(X_l^i\geq 1+\nu)\le -\frac{\gm}{2}[\nu-\ln(1+\nu)].
\end{equation}
Union bounding $\PP(X^i_l\geq 1+\nu)$ over all $i\in S_n$ gives
\begin{equation}\label{combsum}
\PP\{\cup_{i\in S_n}(X^i_l\geq 1+\nu)\}\le\sum_i\PP\left(X_l^i\geq 1+\nu\right)\le T(k)\cdot\PP(X_l^1\geq 1+\nu).
\end{equation}
Taking logarithms and limits of the right-hand side of (\ref{combsum}), using (\ref{chisqtail}) and (\ref{treelim}), we have 
$$\displaystyle\lim_{n\ra\infty}\frac{1}{n}\ln\left[T(k)\cdot\PP(X_l^1\geq 1+\nu)\right]=d\rr\cdot H(d^{-1})-\displaystyle\frac{\lm}{2}[\nu-\ln(1+\nu)],$$
and so (\ref{combsum}) implies that, for any $\eta>0$,
\begin{equation}\label{loglimit}
\frac{1}{n}\ln\PP\left\{\cup_{i\in S_n}(X_l^i\geq 1+\nu)\right\}\le d\rr\cdot H(d^{-1})-\displaystyle\frac{\lm}{2}[\nu-\ln(1+\nu)]+\eta,
\end{equation}
for all $n$ sufficiently large. By the definition of $\TIU(\rr,\lm)$ in (\ref{udef_tree}), and since $[\nu-\ln(1+\nu)]$ is strictly increasing on $\nu>0$, then, for any $\e>0$, setting $\nu:=\nu^{\ast}=\TIU(\rr,\lm)+\e$ and choosing $\eta$ sufficiently small in (\ref{loglimit}) ensures
$$\frac{1}{n}\ln\PP\left\{\cup_{i\in S_n}(X_l^i\geq 1+\nu^{\ast})\right\}\le-c_Q\;\;\;\;\mbox{for all $n$ sufficiently large},$$
where $c_Q$ is some positive constant, from which it follows that
$$\PP\left\{\cup_{i\in S_n}(X_l^i\geq 1+\nu^{\ast})\right\}\le e^{-c_Q\cdot n}\;\;\;\;\mbox{for all $n$ sufficiently large},$$
and (\ref{chisqresult1_tree}) follows. Combining the same union bound argument with the lower tail result of~\cite[Lemma A.2]{stablepoint} shows that, if we take $\nu^{\ast}=\TIL(\rr,\lm)+\e$ for some $\e>0$, then
$$\frac{1}{n}\ln\PP\left\{\cup_{i\in S_n}(X_l^i\le 1-\nu^{\ast})\right\}\le-c_P\;\;\;\;\mbox{for all $n$ sufficiently large},$$
where $c_P$ is some positive constant, and (\ref{chisqresult2_tree}) follows similarly to (\ref{chisqresult1_tree}).\hfill$\Box$\\
\\
\textbf{Proof of Lemma~\ref{Fdist_tree}}: By~\cite[Lemma A.5]{stablepoint}, we have for all $i\in S_n$,
\begin{equation}\label{Ftail} 
\lim_{n\rightarrow\infty}\frac{1}{n}\ln\PP(X_n^i\geq f)\le-\frac{1}{2}\left[\ln(1+f)-\rr\ln f-H(\rr)\right].
\end{equation}
Union bounding $\PP(X_n^i\geq f)$ over all $i\in S_n$ gives
\begin{equation}\label{combsum2}
\PP\left\{\cup_{i\in S_n}(X_n^i\geq f)\right\}\le\sum_{i\in S_n}\PP\left(X_n^i\geq f\right)=|S_n|\cdot\PP(X_n^1\geq f),
\end{equation}
Taking logarithms and limits of the right-hand side of (\ref{combsum2}), using (\ref{Ftail}) and (\ref{treelim}), we have
$$\displaystyle\lim_{n\ra\infty}\frac{1}{n}\ln\left[|S_n|\cdot\PP(X_n^1\geq f)\right]=d\rr\cdot H(d^{-1})-\displaystyle\frac{1}{2}\left[\ln(1+f)-\rr\ln f-H(\rr)\right],$$
which combines with (\ref{combsum2}) to imply that, for any $\eta>0$,
\begin{equation}\label{Feqn}
\frac{1}{n}\ln\PP\left\{\cup_{i\in S_n}(X_n^i\geq f)\right\}\le d\rr\cdot H(d^{-1})-\displaystyle\frac{1}{2}\left[\ln(1+f)-\rr\ln f-H(\rr)\right]+\eta,
\end{equation}
for all $n$ sufficiently large. By the definition of $\TIF(\rr)$ in (\ref{Fdef_tree}), and since the left-hand side of (\ref{Fdef_tree}) on $f>\displaystyle\frac{\rr}{1-\rr}$ is strictly increasing in $f$, then, for any $\e>0$, setting $f:=f^{\ast}=\TIF(\rr)+\e$ and choosing $\eta$ sufficiently small in (\ref{Feqn}) ensures
$$\frac{1}{n}\ln\PP\left\{\cup_{i\in S_n}(X_n^i\geq f^{\ast})\right\}\le-c_I\;\;\;\;\mbox{for all $n$ sufficiently large},$$ 
where $c_I$ is some positive constant, from which the result follows using the same argument as in the proof of lemma~\ref{chisq_tree}.\hfill$\Box$

\bibliographystyle{plain}
\bibliography{stablebib}

\begin{thebibliography}{10}

\bibitem{NIPS2010_3984}
A.~Agarwal, S.~Negahban, and M.~Wainwright.
\newblock Fast global convergence rates of gradient methods for
  high-dimensional statistical recovery.
\newblock In {\em Neural Information Processing Systems}, pages 37--45, 2010.

\bibitem{bach}
F.~Bach.
\newblock Learning with submodular functions: a convex optimization
  perspective.
\newblock {\em Foundations and Trends in Machine Learning}, 6(2-3):145--373,
  2013.

\bibitem{modelbased}
R.~Baraniuk, V.~Cevher, M.~Duarte, and C.~Hegde.
\newblock Model-based compressive sensing.
\newblock {\em IEEE Transactions on Information Theory}, 56:1982--2001, 2010.

\bibitem{CSSA}
R.~Baraniuk and D.~Jones.
\newblock A signal-dependent time-frequency representation: Fast algorithm for
  optimal kernel design.
\newblock {\em IEEE Transactions on Signal Processing}, 42(12):3530--3535,
  1994.

\bibitem{tractability}
B.~Bhan, L.~Baldassare, and V.~Cevher.
\newblock Tractability of interpretability via selection of group-sparse
  models.
\newblock In {\em International Symposium on Information Theory}, pages
  1037--1041, July 2013.

\bibitem{lqphase}
J.~Blanchard, C.~Cartis, and J.~Tanner.
\newblock Compressed sensing: How sharp is the restricted isometry property?
\newblock {\em SIAM Review}, 53(1):105--125, 2011.

\bibitem{greedy_tech}
J.~Blanchard, C.~Cartis, J.~Tanner, and A.~Thompson.
\newblock Phase transitions for greedy sparse approximation algorithms;
  extended technical report.
\newblock Technical Report ERGO 09-010, School of Mathematics, University of
  Edinburgh, 2009.

\bibitem{greedy}
J.~Blanchard, C.~Cartis, J.~Tanner, and A.~Thompson.
\newblock Phase transitions for greedy sparse approximation algorithms.
\newblock {\em Applied and Computational Harmonic Analysis}, 30(2):188--203,
  2011.

\bibitem{cgiht}
J.~Blanchard, J.~Tanner, and K.~Wei.
\newblock {CGIHT}: conjugate gradient iterative hard thresholding for
  compressed sensing and matrix completion.
\newblock {\em Information and Inference}, 4(4):289--327, 2015.

\bibitem{support_sizes}
J.~Blanchard and A.~Thompson.
\newblock On support sizes of restricted isometry constants.
\newblock {\em Applied and Computational Harmonic Analysis}, 29(3):382--390,
  2010.

\bibitem{thresh}
T.~Blumensath and M.~Davies.
\newblock Iterative thresholding for sparse approximations.
\newblock {\em Journal of Fourier Analysis and its Applications},
  14(5):629--654, 2008.

\bibitem{ihtCS}
T.~Blumensath and M.~Davies.
\newblock Iterative hard thresholding for compressed sensing.
\newblock {\em Applied and Computational Harmonic Analysis}, 27(3):265--274,
  2009.

\bibitem{union_subspaces}
T.~Blumensath and M.~Davies.
\newblock Sampling theorems for signals from the union of finite-dimensional
  linear subspaces.
\newblock {\em IEEE Transactions on Information Theory}, 55(4):1872--1882,
  2009.

\bibitem{normalized}
T.~Blumensath and M.~Davies.
\newblock Normalised iterative hard thresholding: guaranteed stability and
  performance.
\newblock {\em IEEE Journal of Selected Topics in Signal Processing},
  4(2):298--309, 2010.

\bibitem{candes1}
E.~Cand\`es, J.~Romberg, and T.~Tao.
\newblock Robust uncertainty principles: exact signal reconstruction form
  highly incomplete frequency information.
\newblock {\em IEEE Transactions on Information Theory}, 52(2):489--509, 2006.

\bibitem{candes}
E.~Cand\`es and T.~Tao.
\newblock Decoding by linear programming.
\newblock {\em IEEE Transactions on Information Theory}, 51(12):4203--4215,
  2005.

\bibitem{exact}
C.~Cartis and A.~Thompson.
\newblock An exact tree projection algorithm for wavelets.
\newblock {\em IEEE Signal Processing Letters}, 20(11):1028--1031, 2013.

\bibitem{stablepoint}
C.~Cartis and A.~Thompson.
\newblock A new and improved quantitative recovery analysis for iterative hard
  thresholding algorithms in compressed sensing.
\newblock {\em IEEE Transactions on Information Theory}, 61(4):1--24, 2013.

\bibitem{best_kterm}
A.~Cohen, W.~Dahmen, and R.~DeVore.
\newblock Compressed sensing and best $k$-term approximation.
\newblock {\em Journal of the American Mathematical Society}, 22:211--231,
  2009.

\bibitem{sp}
W.~Dai and O.~Milenkovic.
\newblock Subspace pursuit for compressive sensing signal reconstruction.
\newblock {\em IEEE Transactions on Information Theory}, 55(5):2230--2249,
  2008.

\bibitem{cartbob}
D.~Donoho.
\newblock {CART} and best ortho-basis: A connection.
\newblock {\em Annals of Statistics}, 25(5):1870--1911, 1997.

\bibitem{donoho}
D.~Donoho.
\newblock Compressed sensing.
\newblock {\em IEEE Transactions on Information Theory}, 52(4):1289--1306,
  2006.

\bibitem{neighborliness}
D.~Donoho.
\newblock High-dimensional centrosymmetric polytopes with neighborliness
  proportional to dimension.
\newblock {\em Discrete and Computational Geometry}, 35(4):617--652, 2006.

\bibitem{treeMP}
M.~Duarte, M.~Wakin, and R.~Baraniuk.
\newblock Fast reconstruction of piecewise smooth signals from random
  projections.
\newblock In {\em Signal Processing with Adaptive Sparse Structured
  Representations}, November 2005.

\bibitem{hidden_markov}
M.~Duarte, M.~Wakin, and R.~Baraniuk.
\newblock Wavelet-domain compressive signal reconstruction using a hidden
  markov tree model.
\newblock In {\em IEEE International Conference on Acoustics, Speech and Signal
  Processing}, pages 5137--5140, 2008.

\bibitem{HTP}
S.~Foucart.
\newblock Hard thresholding pursuit: an algorithm for compressive sensing.
\newblock {\em SIAM Journal on Numerical Analysis}, 49(6):2543--2563, 2011.

\bibitem{foucart_rauhut}
S.~Foucart and H.~Rauhut.
\newblock {\em A mathematical introduction to compressive sensing}.
\newblock Birkh\"{a}user, 2013.

\bibitem{concrete}
R.~Graham, D.~Knuth, and O.~Patashnik.
\newblock {\em Concrete mathematics: a foundation for computer science}.
\newblock Addison-Wesley, 1994.

\bibitem{fast_proj}
C.~Hegde, P.~Indyk, and L.~Schmidt.
\newblock A fast approximation algorithm for tree-sparse recovery.
\newblock In {\em IEEE International Symposium on Information Theory}, pages
  1842--1846, June 2014.

\bibitem{nearly_linear}
C.~Hegde, P.~Indyk, and L.~Schmidt.
\newblock Nearly linear-time model-based compressive sensing.
\newblock In {\em International Colloquium on Automata, Languages and
  Programming}, pages 588--599, July 2014.

\bibitem{approx_tol}
C.~Hegde, P.~Indyk, and L.~Schmidt.
\newblock Nearly linear-time model-based compressive sensing.
\newblock In {\em Symposium on Discrete Algorithms}, pages 1544--1561, January
  2014.

\bibitem{approx_alg}
C.~Hegde, P.~Indyk, and L.~Schmidt.
\newblock Approximation algorithms for model-based compressive sensing.
\newblock {\em IEEE Transactions on Information Theory}, 61(9):5129--5147,
  2015.

\bibitem{indyk_price}
P.~Indyk and E.~Price.
\newblock K-median clustering, model-based compressive sensing, and sparse
  recovery for earth mover distance.
\newblock In {\em 43rd Annual ACM Symposium on Theory of Computing}, pages
  627--636, June 2011.

\bibitem{sparse_regression}
P.~Jain, N.~Rao, and I.~Dhillon.
\newblock Structured sparse regression via greedy hard-thresholding.
\newblock \texttt{http://arxiv.org/abs/1602.06042}, 2016.

\bibitem{discrete_convex}
A.~Kyrillidis, L.~Baldasarre, M.~El~Halabi, Q.~Tran-Dinh, and V.~Cevher.
\newblock Structured sparsity: discrete and convex approaches.
\newblock In H.~Boche, R.~Calderbank, G.~Kutyniok, and J.~Vyb\'{i}ral, editors,
  {\em Compressed sensing and its applications}, Applied and Numerical Harmonic
  Analysis. Springer, 2015.

\bibitem{recipes}
A.~Kyrillidis and V.~Cevher.
\newblock Recipes on hard thresholding methods.
\newblock In {\em Computational Advances in Multi-Sensor Adaptive Processing},
  December 2011.

\bibitem{clash}
A.~Kyrillidis and V.~Cevher.
\newblock Combinatorial selection and least absolute shrinkage via the {CLASH}
  algorithm.
\newblock In {\em International Symposium on Information Theory}, pages
  2216--2220, July 2012.

\bibitem{treeOMP}
C.~La and M.~Do.
\newblock Signal reconstruction using sparse tree representation.
\newblock In {\em SPIE Optics and Photonics}, volume Wavelets XI, July 2005.

\bibitem{mallat}
S.~Mallat.
\newblock {\em A Wavelet Tour of Signal Processing: The Sparse Way}.
\newblock Academic Press, third edition, 2009.

\bibitem{cosamp}
D.~Needell and J.~Tropp.
\newblock {CoSaMP}: {I}terative signal recovery from incomplete and inaccurate
  samples.
\newblock {\em Applied and Computational Harmonic Analysis}, 26(3):301--321,
  2008.

\bibitem{yurii}
Y.~Nesterov.
\newblock {\em Introductory lectures on convex optimization : a basic course}.
\newblock Applied optimization. Kluwer Academic, Boston, Dordrecht, London,
  2004.

\bibitem{nocedal_wright}
J.~Nocedal and S.~Wright.
\newblock {\em Numerical Optimization}.
\newblock Springer, 1999.

\bibitem{thesis}
A.~Thompson.
\newblock {\em Quantitative analysis of algorithms for compressed signal
  recovery}.
\newblock PhD thesis, School of Mathematics, University of Edinburgh, 2012.

\end{thebibliography}

\end{document}